\documentclass[12pt]{article}
\usepackage[T1]{fontenc}
\usepackage[utf8]{inputenc}
\usepackage{amsmath,amssymb,slashed,upgreek,url}
\usepackage{amsfonts}
\usepackage{upgreek}
\usepackage{comment}

\numberwithin{equation}{section}

\usepackage[english]{babel}

\usepackage[mathscr]{eucal}
\usepackage{epsfig}
\usepackage{booktabs}
\usepackage{bbold}
\usepackage{slashed}

\DeclareMathAlphabet{\pxitfont}{OML}{pxmi}{m}{it}
\DeclareMathAlphabet{\pxitfontn}{U}{pxmia}{m}{it}

\DeclareFontFamily{U}{euc}{}%
\DeclareFontShape{U}{euc}{m}{n}{<-6>eurm5<6-8>eurm7<8->eurm10}{}%
\DeclareSymbolFont{AMSc}{U}{euc}{m}{n} %
\DeclareMathSymbol{\psitt}{\mathord}{AMSc}{"20}    
\DeclareMathSymbol{\chitt}{\mathord}{AMSc}{"1F}    

\def\sign{{\mathrm{sign}}}
\def\neg{\negthinspace}

\def\g{{\gamma}}
\def\cf{{\mathrm{cf}}}
\def\bg{\bar\gamma}
\def\com{{\mathrm{com}}}
\def\spinc{{\mathrm {spin}}_c}

\def\up{\uparrow}
\def\Pf{{{\mathrm{Pf}}}}
\def\etta{\upeta}
\def\sc{{\mathrm{sc}}}
\def\dn{\downarrow}
\def\h{\widehat}

\def\u{{\sf u}}

\def\Tsc{\slT{\neg_{\mathrm{sc}}}}
\def\top{{\mathrm{top}}}
\def\B{{\mathcal B}}
\def\sv{{\sf v}}

\def\k{{\sf k}}
\def\K{{\sf K}}
\def\KK{{\mathcal K}}
\def\pin{{\mathrm{pin}}}
\def\pinp{{\mathrm{pin}^+}}
\def\pinc{{\mathrm{pin}}_c}
\def\sj{{\sf j}}
\def\CP{{\mathbf {CP}}}
\def\j{\sj}
\def\v{\sv}

\def\bp{\begin{pmatrix}}
\def\ep{\end{pmatrix}}
\def\sfR{{\sf R}}
\def\sfC{{\sf C}}
\def\slC{ \text{\sffamily\slshape C\/}  }
\def\slT{\text{\sffamily\slshape T\/}}
\def\slCT{\text{\sffamily\slshape CT\/}}
\def\sfCR{{\sf {CR}}}

\def\sfCT{{\sf {CT}}}
\def\sfT{{\sf T}}
\def\D{{\mathfrak D}}
\def\q{{\sf q}}

\def\I{{\mathfrak I}}

\def\etta{\upeta}
\def\grav{{\mathrm{grav}}}
\newcommand{\bea}{\begin{array}}
\newcommand{\eea}{\end{array}}
\newcommand{\beq}{\begin{equation}}
\newcommand{\eeq}{\end{equation}}
\newcommand{\beqn}{\begin{eqnarray}}
\newcommand{\eeqn}{\end{eqnarray}}
\newcommand{\tr}{{\rm tr}}
\def\tra{{\mathrm{tr}}}
\newcommand{\Tr}{{\rm Tr}}

\def\B{{\mathcal B}}

\newcommand{\eps}{\epsilon}
\newcommand{\veps}{\varepsilon}

\renewcommand{\L}{{\mathcal L}}

\newcommand{\RR}{{\mathcal R}}

\newcommand{\Z}{\ZZ}

\newcommand{\CS}{{\mathrm {CS}}}

\def\sp{\mathfrak{sp}}

\def\sign{{\rm sign}\,}

\def\frak{\mathfrak}

\def\d{{\mathrm d}}
\def\ind{{\mathrm{ind}}}

\def\Yuk{{\mathrm{Yuk}}}

\def\bg{\overline\gamma}
\def\g{\gamma}

\font\teneusm=eusm10
\font\seveneusm=eusm7
\font\fiveeusm=eusm5
\newfam\eusmfam
\textfont\eusmfam=\teneusm \scriptfont\eusmfam=\seveneusm
\scriptscriptfont\eusmfam=\fiveeusm

\font\tencmmib=cmmib10 \skewchar\tencmmib='177
\font\sevencmmib=cmmib7 \skewchar\sevencmmib='177
\font\fivecmmib=cmmib5 \skewchar\fivecmmib='177
\newfam\cmmibfam
\textfont\cmmibfam=\tencmmib \scriptfont\cmmibfam=\sevencmmib
\scriptscriptfont\cmmibfam=\fivecmmib
\def\cmmib#1{{\fam\cmmibfam\relax#1}}

\font\teneurm=eurm10
\font\seveneurm=eurm7
\font\fiveeurm=eurm5
\newfam\eurmfam
\textfont\eurmfam=\teneurm \scriptfont\eurmfam=\seveneurm
\scriptscriptfont\eurmfam=\fiveeurm

\font\teneufm=eufm10
\font\seveneufm=eufm7
\font\fiveeufm=eufm5
\newfam\eufmfam
\textfont\eufmfam=\teneufm \scriptfont\eufmfam=\seveneufm
\scriptscriptfont\eufmfam=\fiveeufm

\def\H{{\mathcal H}}

\def\bar{\overline}

\def\Spin{{\mathrm{Spin}}}
\def\hat{\widehat}

\def\Z{{\mathbf Z}}
\def\i{{\mathrm i}}

\def\q{{\cmmib q}}

\def\d{{\mathrm d}}
\def\N{{\mathcal N}}

\def\tilde{\widetilde}
\def\t{\tilde}
\def\D{{\mathcal D}}
\def\u{{\frak u}}
\def\R{{\mathbf R}}
\def\2{{\bf 2}}
\def\1{{\bf 1}}
\def\0{{\bf 0}}

\def\bar{\overline}
\def\u{{\frak u}}

\def\O{{\mathcal O}}

\def\be{\begin{equation}}
\def\ee{\end{equation}}
\def\g{\gamma}
\def\sD{\slashed{D}}

\def\sp{\slashed{\partial}}

\begin{document}

\thispagestyle{empty}
\begin{flushright}\footnotesize
~

\vspace{2.1cm}
\end{flushright}

\begin{center}
{\Large\textbf{\mathversion{bold} Gapped Boundary Phases of Topological Insulators via Weak
Coupling}\par}

\vspace{2.1cm}

\textrm{ Nathan Seiberg and Edward Witten}

\vskip1cm

\textit{ School
of Natural Sciences, Institute for Advanced Study, Princeton, NJ 08540}\\
 \vspace{3mm}

\par\vspace{2cm}

\textbf{Abstract}
\end{center}
\noindent
The standard boundary state of a topological
insulator in 3+1 dimensions has gapless charged fermions.  We present model systems that
reproduce this standard gapless boundary state
 in one phase, but also have gapped phases with topological order.  Our models are weakly
 coupled and all the dynamics is explicit.
We rederive some known boundary states of topological insulators and construct
 new ones.   Consistency with the standard spin/charge relation of condensed matter physics
 places a nontrivial constraint on models.
 \vspace*{\fill}

\setcounter{page}{1}

\newpage

\tableofcontents

\vskip2cm
\setcounter{section}{0}
\section{Introduction}

\subsection{Anomalies And Boundary States Of A Topological Insulator}
The  topological insulator \cite{Kanef}-\cite{HM} is a fascinating and experimentally
accessible example of a symmetry-protected
topological (SPT) \cite{Pollmannetal}-\cite{WenEtAl} phase of matter.    In general, an SPT
phase seems almost trivial in bulk at long distances.
It is gapped and lacks long range topological order.  Yet subtle features of the bulk
physics force nontrivial behavior
on the boundary of a  material that is in an SPT phase. The boundary in general may
spontaneously break some of the symmetries,
have gapless excitations, or have topological order.

A topological insulator for our purposes is  a $3+1$-dimensional\footnote{$2+1$-dimensional
topological insulators fit in a similar
framework, although the relevant anomaly may seem more exotic \cite{WQ}.} time-reversal
invariant
system of
electrons coupled to background electromagnetism with gauge field $A$. (We write $\sfT$ for
time-reversal.) We will take $A$ to be a
classical background field, which does not fluctuate.  We model the system by relativistic
Dirac fermions with an
electron mass that we assume may be position-dependent.  Time-reversal symmetry
forces the electron mass parameter $m$ to be everywhere real. (The electron mass term is
thus $m\bar\Psi\Psi$ with real $m$, while a $\sfT$-violating contribution
proportional to $\bar \Psi \gamma^5\Psi$ is absent.)  A topological insulator can be modeled
as a system
 in which  $m$ is real and positive outside the material and negative inside.  Thus the mass
 parameter changes sign near the
 surface of the material.  As in \cite{JR}, such a sign change leads to the appearance  of
 massless fermions on
the interface between the two regions.  These gapless modes propagate along the boundary as
a $2+1$-dimensional massless Dirac
fermion, and are characteristic of the standard boundary state of the topological
insulator.

One could make a chiral rotation of the electron field inside the material by an angle $\pm
\pi$ to
reduce to the case that $m$ is positive inside
as well as outside the material.  The chiral anomaly
means that this rotation induces the interaction
\be\label{inducedtheta}I_\ind= \pi \int_X \left(\hat A(R) +{1\over 2}{F\over 2\pi} \wedge
{F\over 2\pi} \right), ~\ee
where $X$ is the worldvolume of the material, $F=\d A$ is the field strength of
electromagnetism, and $\h A(R)$
is a certain bilinear function of the Riemann tensor $R$, defined in eqn. (\ref{index}).
(The overall sign of $I_\ind$ depends
upon whether one makes a chiral rotation by an angle $+\pi$
or $-\pi$ to reverse the sign of the mass parameter.)
The second term in eqn. (\ref{inducedtheta}) represents a contribution to the theta-angle
of the
electromagnetic field, which \cite{thetapi,thetapitwo} is  $\theta=\pi$  inside
a topological insulator assuming that it  vanishes outside.
Under $\sfT$, $\theta=\pi$  is mapped to $\theta=-\pi$. The physics of a bulk system
is invariant under $\theta\to\theta+2\pi$, so $\theta=\pi$ is equivalent
in bulk to $\theta=-\pi$.   The same consideration appears to show that the gravitational
term
$\pi\h A(R)$ is irrelevant,\footnote{\label{optimistic} One may also question more generally
whether considerations
involving coupling to gravity, and thus to a general spacetime manifold,
are meaningful in condensed matter physics, which lacks a microscopic relativistic
symmetry.
Experience seems to show that such considerations do give valid results, possibly because
considerations
that can be seen by considering the behavior on a general spacetime manifold
could instead be extracted from entanglement properties of many-body
quantum wavefunctions.}
 as it
naively represents a gravitational theta-term with a coefficient $\theta_\grav=2\pi$,
which one might expect to be equivalent to $\theta_\grav=0$.
However, we will see that this term plays a role when considered together with the
spin/charge relation of condensed
matter physics.

The subtlety of the topological insulator comes from the fact that although the interaction
(\ref{inducedtheta}) is
$\sfT$-invariant on a spacetime
manifold without boundary, it is not $\sfT$-invariant by itself if the integral in
(\ref{inducedtheta}) runs only over
the interior $X$ of a topological insulator.
For one thing, the  proof of $\sfT$-invariance of a theory with the interaction $I_\ind$
depends on integrality of ${1\over 2}\int_X F\wedge F/(2\pi)^2$ (which ensures that
$I_\ind$, which is odd under $\sfT$,  is $\sfT$-invariant mod
$2\pi\Z$).  This integrality holds if $X$ is a compact spin
manifold without boundary, but not if it is the worldvolume of a topological insulator,
which in the real world
always has a nonempty boundary.  On the contrary,
if $X$ has boundary $W$, then $\pi\int_X  F\wedge F/(2\pi)^2$ can serve as the definition of
the
Chern-Simons function of a gauge field $A$ on $W$:
\be\label{zeb}\CS(A)=\pi\int_X\frac{F}{2\pi}\wedge \frac{F}{2\pi}.\ee
Another explanation of why merely including a term $I_\ind$ in the effective action violates
$\sfT$-invariance
is   that although a global chiral rotation by an angle $\pi$ everywhere in spacetime is
indeed equivalent to a global chiral rotation by
an angle $-\pi$,   a chiral rotation by a spatially-dependent angle that varies from $0$
outside the material to $+\pi$ inside is not equivalent
to one that varies from 0 to $-\pi$.

In a $\sfT$-invariant system that has the interaction $I_\ind$ in its interior,
something interesting must happen on the boundary  to maintain $\sfT$-invariance.
  The standard boundary state of the topological insulator achieves this via the presence of
  gapless charged
fermions on the boundary, whose appearance was briefly explained above.
  A purely $2+1$-dimensional system of massless charged electrons propagating on a
  three-manifold $W$ suffers from what is commonly
  called a parity anomaly \cite{Redlich,Semenoff, ADM,WQ}.  Classically, this system of
  fermions has time-reversal symmetry
and a global $U(1)_A$ symmetry to which $A$ couples.  In the quantum theory one has to give
up at least one
of these symmetries.   But if $W$ is the boundary of a four-manifold $X$ and the bulk
physics on $X$ includes the interaction $I_\ind$,
then $\sfT$ symmetry and $U(1)_A$ can be simultaneously preserved.
This can be understood as an example of  anomaly inflow \cite{CallanHarvey}, in fact a
rather subtle example as the relevant anomaly
 is a mod 2 effect.

This interplay between the physics in the bulk and on the surface, which is controlled by
symmetries and anomalies, makes
the system that we have just described robust.  Small symmetry-preserving deformations
cannot change the essential properties:
the existence of the interaction $I_\ind$ in bulk, and the presence of gapless charged
fermions on the boundary.   What about
large symmetry-preserving deformations?

The tight structure based on anomalies makes it challenging to gap the boundary while
preserving the symmetries.
Clearly, if the massless fermions of
the standard gapless boundary state are absent, something else must contribute the same
anomaly.  The boundary hence cannot
be a trivial gapped system.  But the boundary can support a $2+1$ dimensional topological
quantum field theory (TQFT)
that has the same anomalous realization of time-reversal symmetry $\sfT$ and global
$U(1)_A$
symmetry as the  free fermions of the standard gapless boundary state.  Examples have  been
described in
\cite{MKF}-\cite{Al}, following work on gapped boundary states of certain bosonic systems
\cite{VS}.  The goal of this paper is to further understand the constraints on these
systems, to give a description in which
all of the dynamics is completely explicit, and to find additional  symmetry-preserving
gapped boundary states.

The models we construct are similar to models constructed in \cite{MKF} and \cite{Wangetal}
for boundary states of a topological insulator and in \cite{WS} and \cite{MetlitskiTSC}
for boundary states of a topological superconductor.  In these papers, starting with the
standard gapless
boundary state of a topological insulator, the first step is to put the boundary in a
superconducting state, for example by proximity
to a slab of s-wave superconductor.  The boundary then supports quasiparticles -- vortices
-- that carry ordinary magnetic flux and obey exotic statistics.  The key
second step is to assume a process of ``vortex condensation'' after which the
electromagnetic gauge symmetry is restored and
 the vorticity is only conserved modulo some integer $n$.
But in the real world, the ordinary magnetic flux through a two-dimensional surface (such as
the surface of a
topological insulator) is an exactly conserved quantity, which cannot be reduced by any
boundary dynamics to a mod $n$ conservation law.
Related to this, the dynamics postulated in these papers is not very explicit.  To avoid
such issues, we reinterpret their construction so that the flux
carried by the vortices is the flux of an emergent gauge field. With suitable hypotheses on
symmetry breaking, this flux can perfectly well be conserved mod $n$ for some $n$.
All of the required dynamics is explicit and elementary, rather as in the ``composite
fermion'' derivation of the Moore-Read state \cite{MR,GR}.  Similarly, \cite{MKF} exhibits a
second
and simpler gapped boundary state obtained from the first using the process of ``anyon
condensation.'' Again, the assumed dynamics must be strongly coupled.  Instead, we construct
a weakly coupled system leading to the same topological quantum field theory.
Analogous remarks apply concerning the relation of our treatment to the treatment of a
topological superconductor in \cite{WS} and \cite{MetlitskiTSC}.

\subsection{Coupling To Background Fields}\label{coupling}

\begin{figure}
 \begin{center}
   \includegraphics[width=4in]{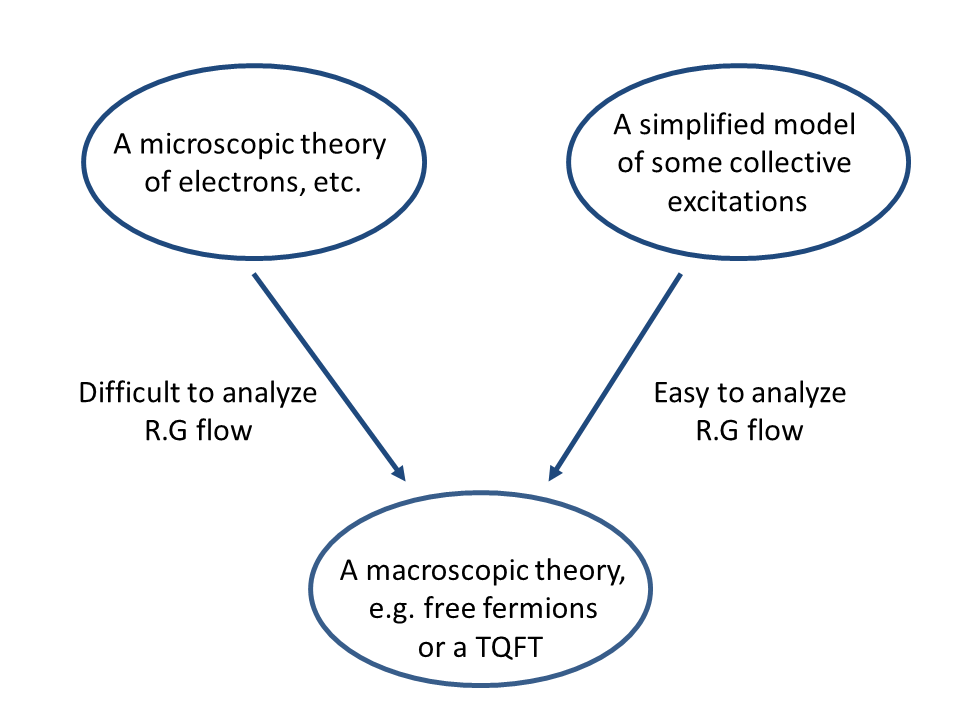}
 \end{center}
\caption{\small  We are interested in solving a complicated microscopic theory and finding
its macroscopic description.  Instead, we replace it with a simplified model that captures
some degrees of freedom.  This model is not supposed to be exact.  This model is weakly
coupled and allows us to find its long distance behavior easily.  The hope is that it is in
the same universality class as the original microscopic theory.  }
 \label{Threemodels}
\end{figure}

In condensed matter physics, one commonly discusses three distinct descriptions of a given
system (see fig.\ \ref{Threemodels}).
First, we have a microscopic Hamiltonian describing the electrons of the material.  This
system is quite complicated and typically
 cannot be analyzed explicitly.  Second, we consider a simplified model, which captures some
 collective excitations of the
microscopic system.  There is no claim that this model gives a complete and accurate
description of the system.
The idea is only that it is a simpler model that captures the essential properties of the
original microscopic theory;
they are in the same universality class.  We will refer to this model as a phenomenological
model.  Finally, we have the long-distance, macroscopic description of the physics.

Since it is often nontrivial to derive one of these descriptions from the others, it is
useful to constrain them
by finding appropriate consistency conditions. We have already mentioned that we impose
$\sfT$ and $U(1)_A$
symmetry.  Specifically, we impose that the $2+1$-dimensional boundary theory combined with
the bulk term $I_\ind$
respects these symmetries.

A refinement of these constraints follow from  what in particle physics is known as  't
Hooft
anomaly matching  \cite{tHooft}.  Instead of considering these three systems (the
microscopic system, the
simplified mode, and the long distance macroscopic description) as they are, we couple them
to a nontrivial electromagnetic
and gravitational background.  In the real original system of interest, the electromagnetic
vector potential
 $A$ is topologically trivial and the spacetime metric is flat.    But the microscopic
 Schrodinger equation is consistent in more
 generality.

For example, even though in nature, magnetic monopoles are not available to us, we can
entertain introducing a magnetic
monopole into our material. (In the current context this was done in
\cite{thetapi,FR,Metlitskipi}.)  This is consistent at the level of the microscopic
Schrodinger equation so it must
also be consistent in the other two descriptions.    Similarly, we can place the system on a
curved
spatial manifold, or, if we are optimistic (see footnote \ref{optimistic}), on a fairly
general curved spacetime.
Some conditions involving spin are needed, as discussed momentarily.
Time-reversal symmetry means that we can define the system even on a  spacetime that is not
necessarily
orientable.  When we do this, some properties must match between the microscopic and
macroscopic description.
As with the magnetic monopole, the real system of interest exists in flat spacetime, but
placing
it in a hypothetical curved spacetime turns out to be a powerful tool to learn about the
flat spacetime theory.

Finally, any system made from electrons only
(coupled to a background of nuclei that is inert except for lattice vibrations, and possibly
supplemented with photons) obeys what
we will call the spin/charge relation of condensed matter physics.  This asserts that -- in
a system of finite volume, which is
ultimately constructed from a finite number of electrons and nuclei --
states of odd electric charge have
half-integral spin and states of even electric charge have integral spin.\footnote{The
existence of neutrons means that the
spin/charge relation is not a fundamental property of Nature.  But it is valid in any
condensed matter system in which nuclear
spin plays no important role.}  If the spin/charge relation is satisfied microscopically,
then it must be satisfied in any phenomenological
description.

A useful way to incorporate the spin/charge relation, described in detail in section
\ref{spincharge}, is as follows.
Any system of fermions can be defined on a spin manifold.  But a system that obeys the
spin/charge
relation can be formulated on a more general class of manifolds known as $\spinc$ manifolds.
(For recent application of $\spinc$
manifolds in a condensed matter context, see \cite{M}.) If we start with a microscopic
system that obeys the spin/charge relation, then any more phenomenological description of
the same system should make
sense on a $\spinc$ manifold.  This gives constraints on phenomenological models that are
interesting and useful, even if we are really
not interested in the physics on $\spinc$ manifolds.

\subsection{Outline}

The outline of this paper is as follows.

 In section \ref{review}, we review the standard gapless boundary state of a topological
 insulator.
 We describe the action of free $2+1$-dimensional fermions and its symmetries.  We give a
 first description of the anomaly and we explain the spin/charge constraint and its relation
 to $\spinc$.

In section \ref{simple}, we present a simple one-parameter family of boundary
models for the topological insulator.  In one phase, these models describe the standard
gapless boundary
state of the topological insulator.  In another phase, a gap develops and a nontrivial
boundary TQFT is found.
The models are weakly coupled and the analysis is explicit and elementary.
The study of these models uncovers a new anomaly associated with the spin/charge relation.
We describe the collective excitations of these models and show that they exhibit
non-Abelian statistics.
(This should not be a surprise given previous literature \cite{MKF}-\cite{Al}.)
We also compute the partition functions of these models and clarify a number of subtle
issues.

Section \ref{TFTE} is devoted to the long distance macroscopic description of the system.
We
identify the low energy TQFT and explore its properties.  This motivates us to find
additional
models generalizing those in section \ref{simple}.  We recognize many of these TQFTs as
those found in \cite{MKF}.  More detail on these models, and an ``abelian'' model
that almost works as a boundary state of a topological insulator, can be found in  section
\ref{switched}.

Our main focus in this paper is on boundary states of a topological insulator, but the
constructions are also applicable to topological superconductors, as we discuss
in section \ref{app}.

A number of appendices present additional background information.  In Appendix \ref{norm},
 we summarize the proper normalization of some of the topological terms we use in three and
 four dimensions.
   In  Appendix \ref{calliastheorem}, we  review the Callias index theorem, which is useful
   in the analysis of vortices in these
   models.  In   Appendix \ref{CSGT}, we  summarize some useful information about
   Chern-Simons gauge theories,
   clarify some possibly confusing points in the literature, and present an explicit gauge
   theory Lagrangian
   for the TQFT of the Moore-Read state \cite{MR} and its generalizations.

\section{Review Of The Standard Gapless Boundary State}\label{review}
\subsection{Action And Symmetries}\label{actsym}

Let $X$ be the $3+1$-dimensional worldvolume of a topological insulator and $W$ its
$2+1$-dimensional boundary.   In general,
we assume that $X$ and $W$ are possibly curved Riemannian manifolds with a spin structure.
When  gauge fields are
present, we write $\sD_0$
for the Dirac operator for a fermion coupled to the geometry only (so on a flat manifold,
$\sD_0$ is simply the
flat space Dirac operator $\sp$) and $\sD$ for the full Dirac operator including the gauge
fields.

 A topological
insulator is gapped in bulk.
Its standard boundary state has a massless Dirac fermion $\Psi$ that propagates on $W$.  Its
action is
\be\label{thaction}I_\Psi=\int_W \d^3x\,\bar\Psi\i\sD\Psi=\int_W\d^3x \,\bar\Psi
\left(\i\sD_0+\slashed{A}\right)\Psi,\ee
where $A$ is the $U(1)$ gauge field of electromagnetism.
To describe in a simple way the action of the discrete symmetries $\sfT$ and $\sfC$
(time-reversal and charge conjugation), it is convenient
to use, in Lorentz signature, a basis of $2\times 2$ real gamma matrices $\g_\mu$,
$\mu=0,1,2$, that obey
\be\label{zonk}\{\g_\mu,\g_\nu\}=2\eta_{\mu\nu},~~\eta_{\mu\nu}=\mathrm{diag}(-1,1,1).\ee
With real gamma matrices, a Majorana fermion $\lambda$ is just a real 2-component fermion
field, which we think of as a column vector.
Then $\bar\lambda$ is defined simply to be $\lambda^\tra \g_0$, where $\tra$ denotes
transpose.  A Dirac
fermion is a complex 2-component fermion field $\Psi$; for a Dirac fermion $\Psi$, with
hermitian adjoint $\Psi^\dagger$ (both of which
are viewed as column vectors), one defines $\bar\Psi=(\Psi^\dagger)^\tr\g_0$.

 $\sfT$ is conventionally defined so that electric charge is even under $\sfT$ and magnetic
 vorticity is odd. Its action on $A$ is then
 \begin{align}\label{waction}\sfT (A_0(t,\vec x) )& =A_0(-t,\vec x) \cr
                                            \sfT (A_i(t,\vec x) )& =-A_i(-t,\vec x),
                                            ~~i=1,2. \end{align}
 It is clumsy to always write such a pair of equations, and we will usually abbreviate such
 a pair by writing
 \be\label{tuwaction}\sfT (A(t,\vec x))=-A(-t,\vec x),\ee
 where we write the transformation of the spatial components and it is understood that the
 time component transforms
   oppositely.\footnote{The motivation  comes from thinking of $A$ as a 1-form.  If
   $\tau:(t,\vec x)\to (-t,\vec x)$ is the time-reversal
   transformation of spacetime, then the two equations in (\ref{waction}) can be written
   $\sfT (A)=-\tau^*(A)$.  We omit the $\tau^*$ and just write $\sfT (A)=-A$.}
 $\sfC$ acts by
 \be\label{cconj}\sfC (A(t,\vec x))=-A(t,\vec x). \ee
 Therefore $\sfCT $ acts by
   \begin{align}\label{twaction}\sfCT (A_0(t,\vec x)) &=-A_0(-t,\vec x) \cr
                                            \sfCT (A_i(t,\vec x) )& =A_i(-t,\vec x), ~~i=1,2
                                            \end{align}
 or more briefly
 \be\label{twactionf}\sfCT (A(t,\vec x))  =A(-t,\vec x) ~.\ee
Often we write even more briefly $\sfT (A)=-A$ or $\sfCT (A)=A$.

To describe the action of the discrete symmetries on $\Psi$, it is most convenient to expand
$\Psi$ in terms of a pair of Majorana fermions  $\lambda_1,
\lambda_2$:   $\Psi=(\lambda_1+\i\lambda_2)/\sqrt 2$.  $\sfT$ acts by
\be\label{relsign}\sfT(\lambda_1(t,\vec x))=\g_0\lambda_1(-t,\vec x),~~~
\sfT(\lambda_2(t,\vec x))=-\g_0\lambda_2(-t,\vec x)    .\ee
Allowing for the fact that $\sfT$ is antiunitary so that $\i$ is odd under $\sfT$, this
definition implies that
\be\label{easysign}\sfT(\Psi(t,\vec x))=\g_0\Psi  (-t,\vec x), \ee
which somewhat deceptively makes it appear that $\Psi$ transforms linearly under $\sfT$.
$\sfC$ acts by
\be\label{cacts}\sfC(\lambda_1)=\lambda_1,~~\sfC(\lambda_2)=-\lambda_2,~~\sfC(\Psi)=\Psi^\dagger.\ee
The action of $\sfCT$ is accordingly
\be\label{ctacts}\sfCT(\lambda_i(t,\vec x))=\g_0\lambda_i(-t,\vec x),~~i=1,2,~~~~~~ \sfCT
(\Psi(t,\vec x))=\g_0\Psi^\dagger(-t,\vec x). \ee

In condensed matter physics, $\sfT$ can be a good microscopic symmetry, while $\sfC$ and
$\sfCT $ (which map electrons to positrons) are not.\footnote{
In particular, there is no particular constraint
on  the fermi energy for the massless
boundary fermions of a $3+1$-dimensional topological insulator except that it is within the
bulk energy gap.  So in relativistic terms, the boundary fermions are similar
to massless Dirac fermions with a nonzero chemical potential.  This chemical potential
violates $\sfC$ and $\sfCT$. Of course, $\sfC$ and $\sfCT$ are
also generically violated by all sorts of higher order terms in the Hamiltonian of the
massless fermions.}

Nevertheless, in this paper, we will keep track of all of these symmetries.   One reason for
this is that
all boundary states we construct do have $\sfC$ and
$\sfCT$ as well as $\sfT$ symmetry.   Also, many of the results have applications to
topological superconductors,
with $\sfCT$ playing the role of $\sfT$, and accordingly $\sfCT$ will be important in
section \ref{app}.

\subsection{Anomalies}\label{matching}

In constructing an alternative boundary state for the topological insulator, we want to
match certain anomalies of the standard boundary
state that we have just described.

The most basic and familiar anomaly is that the standard boundary state has anomalous $\sfT$
symmetry.  This means that if the action $I_\Psi$ of
eqn.\ (\ref{thaction}) is understood as the action of an abstract $2+1$-dimensional system,
then it cannot be quantized in a way that maintains
$\sfT$ symmetry (and also the $U(1)_A$ symmetry corresponding to conservation of electric
charge).  But it can be quantized in a $\sfT$-invariant
fashion if the three-manifold $W$ is the boundary of a four-manifold $X$ and the
electromagnetic theta-angle $\theta$ differs by $\pi$ between the vacuum and $X$.
Usually it is assumed that $\theta=0$ in vacuum and then anomalous $\sfT$ symmetry means
that $\theta=\pi$ inside the bulk of a topological insulator.

The $\sfT$ anomaly just described is commonly called a ``parity'' anomaly, but this
terminology is a little misleading because in $2+1$ dimensions,
parity, which acts on all spatial coordinates by $x^i\to -x^i$, $i=1,2$, is an element of
the connected part of the rotation group.  We prefer to call
this an anomaly in $\sfT$ and $\sfR$ symmetry, where $\sfR$ is a spatial reflection, for
example $(x_1,x_2)\to (-x_1,x_2)$.

The other anomalies to consider involve coupling to gravity, that is, they involve the
question of what happens if $W$ and $X$ are curved.   It is
not completely obvious that such anomalies are relevant in condensed matter physics, where
it is natural to consider a general curved spatial manifold,
but not a general curved spacetime.  However, experience -- notably \cite{HCR,WQ} with the
$3+1$-dimensional topological superconductor -- appears to show
that such anomalies  are relevant.  Probably this means that the results that can be deduced
from anomalies involving coupling to gravity
can alternatively be deduced by more subtle considerations of locality.   This has not yet
been demonstrated.

Here is one example of an anomaly involving gravity. Consider any $2+1$-dimensional
$\sfT$-invariant
state that is continuously connected to a system of free bose and fermi
fields.\footnote{Chern-Simons couplings of gauge fields
are quantized to have integer values
and so cannot be continuously turned off.  In general, gauge fields with Chern-Simons
couplings may contribute to the anomalies
that we are about to discuss.}  Assume that $\sfT^2=(-1)^F$,
where $(-1)^F$ is the operator that equals $-1$ for fermions and $+1$ for bosons.  Then the
action of $\sfT$ on fermions can be
diagonalized so that each Majorana fermion $\lambda$
transforms as $\sfT(\lambda(t,\vec x))=\pm \g_0\lambda(-t,\vec x)$, for some choice of
sign.
Let $n_+$  be the number of gapless Majorana fermions transforming under $\sfT$ with a $+$
sign and $n_-$ the number transforming with a $-$ sign.
A bare mass term $\i\bar\lambda \lambda'$ is $\sfT$-invariant if and only if $\lambda$ and
$\lambda'$ transform under
$\sfT$ with opposite signs, so only the difference $\nu_\sfT=n_+-n_-$ is an invariant when
$\sfT$-conserving bare masses are
turned on and off.  When interactions
are taken into account   \cite{FCV,WS,MetlitskiTSC,KitTwo},  $\nu_\sfT$ is no longer a
$\Z$-valued invariant, but is an invariant mod 16.
In fact, $\sfT$ symmetry can be
used to place any given $\sfT$-invariant theory on an unorientable manifold.   When this is
done, one obtains a theory whose partition function is in general
not well-defined; it has an anomaly that depends on the value of $\nu_\sfT$ mod 16.   This
was illustrated in an example
in \cite{HCR} and explained more systematically in
\cite{WQ}.   For the standard boundary state of the topological insulator, $n_+=n_-=1$ so
$\nu_\sfT=0$.

Similarly, in a $\sfCT$-invariant model,
 assuming that $(\sfCT)^2=(-1)^F$,  let $n_+'$  be the number of gapless Majorana fermions
 transforming under
$\sfCT$ as $\sfCT(\lambda(t,\vec x))=+\g_0\lambda(t,\vec x)$ and $n_-'$
the number transforming as $\sfCT(\lambda(t,\vec x))=-\g_0\lambda(t,\vec x)$.  Only the
difference $\nu_\sfCT=n_+'-n_-'$ is
invariant under turning $\sfCT$-conserving bare masses
on and off. When interactions are considered, $\nu_\sfCT$ is an invariant mod 16.  This can
again be understood
in terms of the anomaly that arises when $\sfCT$ symmetry is used to place a theory on an
unorientable 3-manifold.
For the standard boundary state of a topological insulator, $\nu_\sfCT=2$.

In condensed matter physics, as long as electric charge is conserved, $\nu_\sfT$ and
$\nu_\sfCT$ are highly constrained
by the spin/charge relation.  In Appendix \ref{unorientable}, we show that in a system that
obeys
the spin/charge relation,  $\nu_\sfCT$ is even and $\nu_\sfT$ does not give any information
beyond what one can learn on an orientable manifold.
(In the absence of the spin/charge relation, the only general
restriction is that $\nu_\sfT$ and $\nu_\sfCT$ are congruent mod 2.)

In condensed matter physics, $\sfCT$ is not really a symmetry so different boundary states
of a topological insulator can in general have different
values of $\nu_\sfCT$.  Hence in constructing boundary states of the topological insulator,
we will allow states with different values of $\nu_\sfCT$.
The distinction between states with different values of $\nu_\sfCT$, though not relevant for
a topological insulator, is relevant for topological
superconductors, as we discuss in section \ref{app}.

The reader may have noticed that we have not discussed what might appear to be the most
obvious anomaly involving  coupling of a $2+1$-dimensional
system to gravity:  the gravitational contribution to the anomaly in $\sfT$ symmetry, or
alternatively the jump in the gravitational theta-angle
in crossing from vacuum to the bulk of a topological insulator.   What should be meant by
the gravitational theta-angle depends on whether the spin/charge
relation is assumed.  For a theory that obeys the spin/charge relation, the appropriate
gravitational analog of $\theta=\pi$ is $\theta_2=\pi$, in the notation
of Appendix \ref{unorientable}.  The standard topological insulator has $\theta_2=0$, but in
Appendix \ref{unorientable}
and also in section \ref{TFTE}, we describe  symmetry-preserving boundary states (gapless or
gapped) for a hypothetical material of $\theta_2=\pi$.

\subsection{The Spin/Charge Relation And Spin$_c$ Manifolds}\label{spincharge}

The gravitational anomalies that were mentioned in section \ref{matching} are conveniently
studied by formulating a theory on a possibly curved four-manifold $X$, thus
 placing the boundary state on a possibly curved three-manifold $W$.  But what class of
 manifolds should be considered?

Since the microscopic systems of interest involve fermions, one might want to restrict $X$
to be a spin manifold.
But in a certain sense this is too restrictive.  In condensed matter physics, as long as we
consider systems in which electric charge
is conserved, there is a fundamental spin/charge relation that was already mentioned in
section \ref{coupling}.
A system that satisfies the spin/charge relation can be formulated not just on a spin
manifold but on a more general type of manifold
known as a $\spinc$ manifold.
On a possibly curved spin manifold, one has a covariant derivative $D_\mu^{(0)}$ that acts
on neutral spin 1/2 fermions coupled to gravity only.  In the presence of electromagnetism,
one additionally has a $U(1)$ gauge field $A$ and, for any integer $n$, a covariant
derivative $D_\mu^{(n)}=D_\mu^{(0)}+\i n A_\mu$ acting on spin 1/2 fermions of electric
charge $n$.
On a $\spinc$ manifold, $D_\mu^{(n)}$ is well-defined only for odd $n$; $D_\mu^{(0)}$ is not
well-defined by itself and $A_\mu$ is not well-defined globally as an Abelian gauge field.
Rather, $A_\mu$ is what we will call a $\spinc$ connection.   Though the notion of a
$\spinc$ manifold can be defined purely topologically, for our purposes a $\spinc$ manifold
always has a $\spinc$
connection $A_\mu$.

Concretely, $A_\mu$ is not well-defined globally as an Abelian gauge field because the field
strength $F=\d A$
 does not obey standard Dirac quantization of flux.   The standard quantization would say
 that if $
C\subset X$ is an oriented two-dimensional cycle, then $\int_C F/2\pi$ should be an integer.
The $\spinc$ condition says instead that
\be\label{fluxcond}\int_C\frac{F}{2\pi}=\frac{1}{2}\int_C w_2 ~~{\mathrm{mod}}~\Z.\ee
Here $w_2$ is the second Stieffel-Whitney class  of $X$.  All that matters for our purposes
is that in general $\int_C w_2$ is equal to 0 or 1 mod 2.
Eqn.\ (\ref{fluxcond}) shows that in general $F$ does not obey standard Dirac quantization,
but $2F$ does, since if we multiply by 2, the right hand side of (\ref{fluxcond}) becomes
integer-valued.  Accordingly, although a $\spinc$ connection $A_\mu$ is not well-defined as
an Abelian gauge field, $2A_\mu$ is an ordinary Abelian gauge field.

If we start with a microscopic system that can be placed on a general $\spinc$ manifold,
then any effective field theory that describes its low energy excitations and likewise
any topological field theory that describes a possible gapped phase must  make sense on a
general $\spinc$ manifold.  This will ensure that the states of the effective field theory
or topological field theory (in any sample of finite volume) will satisfy the spin/charge
relation.

In sections \ref{anomaly} and \ref{partfn}, we will see that even if a classical theory of
fermions satisfies the spin/charge relation, there may be a quantum anomaly in this
relation; equivalently, even if a classical theory can be formulated on a $\spinc$ manifold,
it may be that quantum mechanically a spin rather than $\spinc$ structure is required.
Models in which this happens cannot arise as effective field theories describing boundary
states of a topological insulator.  This will give a nontrivial constraint on our
constructions.

The ability to formulate a theory on a $\spinc$ manifold likewise gives a nontrivial
constraint on the topological field theories that can be used to describe gapped boundary
states at low energies.
We will describe this in the context of Chern-Simons gauge theories with Abelian gauge
group.  Here we will discuss
purely $2+1$-dimensional couplings (as opposed to couplings that are possible on the
boundary of a four-manifold).
For example, $A$ could be the electromagnetic gauge field interacting with the
$2+1$-dimensional
world-volume $W$ of a quantum Hall system. The Chern-Simons coupling $\CS(A)$ is supposed to
depend
mod $2\pi\Z$ only on the restriction of $A$ to $W$ (and possibly on the spin structure of
$W$).

We need a few preliminary facts, which we just state here, with explanations in Appendix
\ref{csc}.   It is relatively well-known that if $A$ is an Abelian gauge field, then on a
compact
three-dimensional spin manifold $W$,  the Chern-Simons
functional
\be\label{cscoupling}\CS(A)=\frac{1}{4\pi}\int_W A\d A \ee
is well-defined mod $2\pi \Z$, ensuring that it makes sense as a contribution to a quantum
effective action of a purely $2+1$-dimensional system.
  (In the absence of a chosen spin structure on $W$, $\CS(A)$ is only
well-defined mod $\pi\Z$ and must appear in the effective action with an even integer
coefficient.)   What if $A$ is a $\spinc$ connection rather than an ordinary $U(1)$ gauge
field?
Then a modified version of $\CS(A)$ with a certain gravitational correction is well-defined
mod $2\pi\Z$:
\be\label{corrcs}\CS(A,g)=\frac{1}{4\pi}\int_W A\d A +\dots.\ee
The ellipses refer to a purely gravitational contribution ($\Omega(g)$ in eqn.
(\ref{orelf}); here $g$ is the metric tensor of $W$).
  Now if $A$ is a $\spinc$ connection, and $a$ is an ordinary $U(1)$ gauge field,
then $A+a$ is a $\spinc$ connection.  So another expression that is well-defined mod
$2\pi\Z$ is
\be\label{another}\CS(A+a,g)=\frac{1}{4\pi}\int_W\left(a\d a+2 a\d A+A\d A\right)+\dots,\ee
with the same gravitational correction.
Subtracting the last two formulas, the purely gravitational term cancels out and we learn
that
\be\label{nother}\CS(A+a,g)-\CS(A,g) =\frac{1}{4\pi}\int_W a\d a+\frac{1}{2\pi}\int_W a\d A
\ee
is well-defined mod $2\pi \Z$.  On a spin manifold, with $a$ and $A$ being ordinary $U(1)$
gauge fields, the two terms on the right hand side of eqn.\ (\ref{nother}) would
be separately well-defined mod $2\pi\Z$.  In the more general case of a $\spinc$ manifold,
with $\spinc$ connection $A$,
the two terms are separately well-defined only mod $\pi\Z$ but their sum is well-defined mod
$2\pi\Z$.

Given these facts, let us consider a 3d theory with emergent $U(1)$ gauge fields
$a^1,a^2,\dots, a^n$, as well as the electromagnetic potential $A$.  Here $a^i$ are
dynamical 3d fields and $A$ is a classical background field.  (Later, $A$ will actually be
the restriction of a 4d field to 3d.)
We consider a theory with Chern-Simons action
\be\label{CSn}I_\CS= \int_W\left( {k_{ij} \over 4\pi}a^i\d  a^j + {q_i\over 2\pi }A\d
a^i\right).\ee
If $W$ is a spin manifold, then the condition for  this action to be well-defined mod
$2\pi\Z$ is simply that
the coefficients $k_{ij}$ and $q_i$ should all be integers.  (If $W$ is a purely bosonic
manifold
with no spin structure, then in addition the diagonal elements $k_{ii}$ must be even.)
However, suppose that $I_\CS$ is supposed to be part of the microscopic description of a
theory
that satisfies the usual spin/charge relation.  In this case, $I_\CS$ must be well-defined
mod $2\pi\Z$
on any $\spinc$ manifold $W$  with $\spinc$ connection $A$.  In view
of our discussion of eqn.\ (\ref{nother}),  the condition for this (beyond integrality of
the $k_{ij}$ and $q_i$) is
\be\label{CScon}q_i\cong k_{ii}~~~{\mathrm{mod}}~2. \ee

As a check, note that the condition (\ref{CScon}) is invariant under
field redefinitions   $a^i\to a^i+N^i A$ provided the coefficients $N^i$ are even, but not
otherwise.  This reflects the fact that $2A$ is
an ordinary $U(1)$ gauge field but $A$ is not, so if $a$ is a $U(1)$ gauge field, then a
redefinition $a\to a+2A$ makes sense but
$a\to a+A$ does not.  More generally, let us arrange $a^i$ as a column vector
$\begin{pmatrix}a^1\cr a^2\cr\vdots \cr a^n\end{pmatrix}$, and consider
an allowed\footnote{In making such a change of variables, it does not make sense to add to
$A$ a multiple of $a^i$, since $A$ is a classical background field while the $a^i$ are
dynamical fields.
(Moreover, the $a^i$ are purely three-dimensional while later $A$ will be the 3d restriction
of a 4d field.)  But conversely, there is no problem to add to $a^i$ a multiple of $A$.}
 change of variables
 \be\label{MNcon}a\to Ma +NA, \ee
with $M\in GL(n,\Z)$ and $N$ a column vector with even integer entries.
The reader can verify that this transforms $k_{ij}$ and $q_i$ to
 another set of integers, still satisfying (\ref{CScon}), and also adds to the action an
 integer multiple of $8\CS(A)$.

As explained above, on a $\spinc$ manifold, the basic Chern-Simons invariant $\CS(A)$,
normalized as in eqn.\ (\ref{cscoupling}), requires
a gravitational correction to make it well-defined mod $2\pi\Z$.  Is some multiple of
$\CS(A)$ well-defined mod $2\pi\Z$ without
a gravitational correction?  The answer is that the basic multiple of $\CS(A)$ with this
property is
\be\label{factoreight} 8\CS(A)=\frac{8}{4\pi}\int_W A\d A. \ee
That a factor of 8 is necessary here is explained in Appendix \ref{csc}.  That a factor of 8
is sufficient may be seen as follows.  On any three-manifold $W$
(not necessarily spin), for any $U(1)$ gauge field $b$,  $2\CS(b)$ is well-defined mod
$2\pi\Z$.  Setting $b=2A$, we learn that $8\CS(A)$ is well-defined
mod $2\pi\Z$.  A check is that the change of variable (\ref{MNcon}) shifts the action by a
multiple of $8\CS(A)$ and maintains its consistency.

The restriction (\ref{CScon}) has been previously discussed  \cite{MB} more along the
following lines. Consider the Wilson line operators
\be\label{tranl}\exp\left(\i n^ik_{ij} \oint a^j\right),\ee
with integers $n^i$.  These particular  line operators are ``transparent''; they have
trivial braidings and describe the worldlines of excitations that can move
to the bulk of the system.  Using standard formulas for Abelian Chern-Simons theory,
these operators have spins $n^ik_{ij}n^j/2$.  Their coupling to $A$ shows that their charges
are $q_in^i$.  Requiring that these
line operators should satisfy the usual spin/charge relation gives eqn.\ (\ref{CScon}).

\section{A Simple Class Of Boundary States}\label{simple}

\subsection{Some Models}\label{models}

In this section, we will describe a simple class of  boundary states for the topological
insulator that are gapped but nonetheless
have the same symmetries and anomalies as the
standard boundary state that we reviewed in section \ref{actsym}.  Boundary states with this
property inevitably have nontrivial topological
order,  and in our examples this will arise in a simple way.

In constructing models, we want to maintain the spin/charge relation of condensed matter
physics:  states of odd charge have half-integral
spin and states of even charge have integral spin.  To do so, we simply will take all fermi
fields to have odd electric charge and all bose fields
to have even electric charge.

In our construction, we assume that on the boundary $W$ of the topological insulator, there
propagates an emergent $U(1)$ gauge field $a$
as well as the $U(1)$ gauge field $A$ of electromagnetism.  We call the combined gauge group
$U(1)_A\times U(1)_a$.
 $\sfT$-invariance implies that the action has no Chern-Simons terms, so for $a$ we
assume a Maxwell-like action proportional to $f_{\mu\nu}f^{\mu\nu}$, where
$f_{\mu\nu}=\partial_\mu a_\nu-\partial_\nu a_\mu$.    We normalize
$a_\mu$ so that $f_{\mu\nu}$ has the standard Dirac flux quantum\footnote{In condensed
matter physics, an emergent Abelian gauge field
always has gauge group $U(1)$, not $\R$, and equivalently can have nontrivial magnetic flux
on a compact manifold.  Otherwise, there would
be no monopole operators (see section \ref{anomaly}) and the flux of $a$ would be an exactly
conserved quantity with no microscopic origin.} of $2\pi$.

We also add a boundary Dirac fermion $\chi$  that couples to $U(1)_A \times U(1)_a$ with
charges $(1,2s)$, for some integer $s$.   Assigning to $\chi$
odd charge under $U(1)_A$ is motivated by the spin/charge relation, as just explained.  In
addition, the fact that $\chi$ has odd charge under $A$
and even charge under $a$ means that $\chi$ generates the same $\sfT$-anomaly as the fermion
field $\Psi$ of the standard boundary state of eqn.\ (\ref{thaction}).    (This
$\sfT$-anomaly is a mod 2 effect, as it involves the question of whether the effective
theta-angle is an odd or even multiple of $\pi$.
So it only depends on the values of the $U(1)_A\times U(1)_a$ charges mod 2;
the $\chi$ field of  charges $(1,2s)$ therefore produces the same anomaly as the $\Psi$
field of charges
$(1,0)$.  See section \ref{partfn} for a fuller explanation.)
We assume that $a$ and $A$ transform in the same way under $\sfT$ and $\sfCT$, as described
in eqns.\ (\ref{waction}) and (\ref{twaction}).
We likewise assume that $\chi$ transforms like $\Psi$ under $\sfT$ and $\sfCT$ (eqns.\
(\ref{easysign}), (\ref{ctacts})).
These assumptions ensure that the kinetic energy
\be\label{ichi}I_\chi=\int_W\d^3x\sqrt g\,\i\bar\chi
\slashed{D}\chi=\int_W\d^3x\i\bar\chi\left(\slashed{D}_0+\slashed{A}+2s\slashed{a}\right)\chi\ee
is $\sfT$- and $\sfCT$-invariant.   They also ensure that the model has  $\nu_\sfCT=2$, just
like the standard boundary
state of a topological insulator.

Let us write $\q$ and $\k$ for the ordinary electric charge and the conserved charge of the
emergent gauge field.  $\sfT$ and $\sfCT$ are conventionally
defined so that $\q$ is even under $\sfT$ and odd under $\sfCT$.  Since we take $a$ to
transform like $A$ under the discrete symmetries, the same
is true of $\k$:
\be\label{samej}\sfT \k=\k\sfT,~~~~\sfCT\k=-\k\sfCT. \ee

In addition to $a$ and $\chi$, we add two complex scalar fields $w$ and $\phi$ of charges
$(0,1)$ and $(2,4s)$ under $U(1)_A\times U(1)_a$.
The purpose of adding these two fields is to make it possible to exhibit two different
phases.
In one phase, $w$ has an expectation value, while $\phi$ has zero expectation value and a
positive mass squared.
In this phase, $U(1)_A\times U(1)_a$ is spontaneously broken to $U(1)_A$.   Expanding around
this vacuum, the only gapless field is the Dirac fermion
$\chi$, which has charge 1 under the unbroken symmetry
$U(1)_A$.  At low energies, this phase is indistinguishable from the standard boundary
state
of the topological insulator.

We also want a phase in which the $\chi$ field undergoes $s$-wave pairing, with
$\epsilon_{ab}\chi^a\chi^b\not=0$ (here $a,b=1,2$ is
a spinor index).  We could introduce some sort of attractive interactions that would cause
such pairing to occur dynamically, but instead to get a simple
explicit model we introduce the scalar field $\phi$, whose vacuum expectation value
describes pairing.

To make possible a Yukawa coupling of $\phi$ to $\chi$, we take $\phi$ to transform under
the discrete symmetries as
\be\label{transphi}\sfC(\phi)=\bar\phi,~~~~\sfT(\phi)=-\phi. \ee
By contrast, for $w$ we take\footnote{We could also take $\sfT (w) =-w$ (as for $\phi$),
which differs from our choice by a $U(1)_a$ gauge transformation.  In that case, assuming
the expectation value of $w$ to be real,
 the unbroken time-reversal symmetry in the phase where $w$ Higgses $U(1)_a$ is a
 combination of $\sfT$ and a gauge transformation.  We prefer to call this combination
 $\sfT$ from the start, as in eqn.\ (\ref{transw}).}
\be\label{transw}\sfC (w)=\bar w,~~~~\sfT (w) =w. \ee
 If we expand $w$
in terms of real scalar fields as
$w=(w_1+\i w_2)/\sqrt 2$, then the above formulas imply
\be\label{ransw}\sfT (w_1)=w_1,~~~\sfT (w_2)=-w_2,\ee
along with
\be\label{answ}\sfC (w_1)=w_1,~~\sfC (w_2)=-w_2,~~~~\sfCT (w_i)=w_i, ~~i=1,2.\ee
These formulas have  obvious analogs for $\phi$.

The field $\chi \bar w^{2s}$ has the quantum numbers of a bulk electron, so we can include
in the action or the Hamiltonian a coupling of
the bulk electron field $e$ (restricted to the boundary) to $\chi \bar w^{2s}$:
\be\label{electron}I_e=\int \d^3x\sqrt g \left( \bar e\chi \bar
w^{2s}+\mathrm{h.c.}\right).\ee

The theory admits a gauge-invariant, Lorentz-invariant, and discrete symmetry preserving
Yukawa coupling of $\phi$ to $\chi$,
\be\label{Yuk}I_{\Yuk}=\int\d^3x\sqrt g\, h\left(\i\epsilon_{ab}\chi^a\chi^b\bar\phi+
\i\epsilon_{ab}\chi^{\dagger\,a}\chi^{\dagger\,b} \phi\right).\ee
(Here $a,b=1,2$ are spinor indices, $\epsilon_{ab}$ is the corresponding antisymmetric
tensor, and $h$ is a real coupling constant.)
Bearing in mind that for  Majorana fermions $\lambda$ and $\t\lambda$,
$\i\epsilon_{ab}\lambda^a\t\lambda^b$ is hermitian,
it is straightforward to verify that $I_\Yuk$
is hermitian.  To verify $\sfT$-invariance, which requires the minus sign in eqn.\
(\ref{transphi}), one uses the fact that $\i$ is odd under $\sfT$ but
(because the matrix $\g_0$ in the transformation law $\sfT(\chi(t,\vec x))=\g_0\chi(-t,\vec
x)$ has determinant $+1$) $\epsilon_{ab}\chi^a\chi^b$ is even.
Finally, $\sfC$ invariance is clear.

The phase with $\langle\phi\rangle\not=0$ preserves the discrete symmetries if we accompany
them by suitable gauge transformations.  If we take
$\langle\phi\rangle$ to be real (and, say, positive), then $\sfC$ needs no accompanying
gauge transformation, but $\sfT$ and $\sfCT$
must be accompanied by a $U(1)_a$ gauge transformation, which we can take to have  gauge
parameter $\exp(2\pi \i /8s)$.
In the phase with  $\langle\phi\rangle\not=0$, $U(1)_a$ is broken to a subgroup $\Z_{4s}$,
generated by a discrete gauge symmetry $\K=\exp(2\pi\i \k/4s)$; we also set $\K^{1/2}
=\exp(2\pi\i \k/8s)$.    The unbroken time-reversal symmetry of the system is $\slT=\sfT
\K^{1/2}$.
 Eqn.\ (\ref{samej}) implies that
 $\sfT \K^{1/2}=\K^{-1/2}\sfT$, so  the usual relation $\sfT^2=(-1)^F$ for free electrons
 is unchanged:
\be\label{sphi}\slT^2=(-1)^F . \ee
Likewise the effective $\sfCT$ transformation is \be\label{effct}\slCT=\sfCT \K^{1/2}.\ee
But since $\sfCT$ commutes with $\K^{1/2}$,
 the usual $(\sfCT)^2=(-1)^F$ becomes
\be\label{ctphi}(\slCT)^2=(-1)^F\K.\ee
Finally, the usual $\sfT(\sfCT)=(\sfCT)\sfT$ becomes
\be\label{transsign} (\slCT) \slT=\slT(\slCT )\K^{-1}.\ee
Individual quasiparticles can have $\K\not=1$ -- for example, $w$ corresponds to a
quasiparticle with $\K=\exp(2\pi i/4s)$.  But as $\K$ is a gauge
transformation in
the emergent gauge group, it leaves invariant any globally-defined state of a sample of
finite size.  So for any state of a finite sample, all quasiparticles together
always combine to a state of $\K=1$, and therefore to a state on which $\slT=\sfT$,
$\slCT=\sfCT$, and all the standard relations are obeyed.

The unbroken electric charge in the phase with $\langle\phi\rangle\not=0$ is a linear
combination of $\q$ and $\k$ under which $\phi$ is neutral.
This combination is
\be\label{elch}\q'=\q-\frac{\k}{2s}. \ee

In the phase with $\langle\phi\rangle\not=0$, all bosons and fermions have bare masses at
tree level.  So potentially, this is a gapped, symmetry-preserving
boundary state for the topological insulator.
However, a restriction on the value of $s$ is needed, as we will see in section
\ref{anomaly}.

 The model that we have described is analogous to the composite fermion model \cite{GR} of
 the Moore-Read state \cite{MR} of a fractional
 quantum Hall system.  In that context, the composite fermion is  coupled to $U(1)_A\times
 U(1)_a$
 where $a$ is an emergent Abelian gauge field.  And pairing of the composite fermion is a
 key ingredient.  There are  a   few differences.
 The Moore-Read state has no $\sfT$ symmetry, so    the gauge fields used in constructing it
 can and do have Chern-Simons couplings.  Also, the composite
 fermion of the Moore-Read state is spin-polarized (it has only one spin state of spin
 $+1/2$),
 and its pairing is $p$-wave pairing.  Finally, in the context of  the Moore-Read state,
there is no analog of Higgsing
 to a standard state, and hence there is  no analog of $w$.

The models that we have constructed turn out to be related to those of
\cite{MKF}.  The main difference is our use of the
 emergent gauge field $a$.  As we have already mentioned in the introduction, this allows us
 to analyze the models in a reliable way using standard weak coupling techniques.  This is
 what we will do below.

\subsection{Monopole Operators And The Anomaly}\label{anomaly}

In a theory constructed from electrons only, all local gauge-invariant operators must obey
the usual spin/charge relation.  Here ``gauge-invariant'' operators
are those that are invariant under any emergent gauge symmetries, in our case $U(1)_a$.  The
spin/charge relation says that operators of half-integral
spin carry odd electric charge, and those of integer spin carry even electric charge.

We have constructed our models so that the obvious $U(1)_a$-invariant local operators -- the
ones constructed as polynomials in the fields and their
derivatives -- obey the spin/charge relation.  But we must also consider monopole
operators.

A monopole operator (also known as an 't Hooft operator) is defined  by postulating a Dirac
monopole singularity at a specified point $p\in \R^3$
and performing a path integral in the presence of this singularity.  Very near $p$, all
fermi and bose fields that we have introduced can be treated
as free fields.  Moreover, the terms in the action that violate conformal invariance are all
irrelevant at short distances.  As in \cite{K}, which we follow
in the analysis below, the asymptotic conformal invariance near $p$ enables one to
determine the quantum numbers of monopole operators in radial quantization.  This amounts
to
 making a conformal mapping from $\R^3\backslash p$ (that is, $\R^3$ with the point $p$
omitted) to $\R\times S^2$.  We view the $\R$ direction as (Euclidean) ``time.''  To
construct the basic monopole operators with minimum charge,
we place on $S^2$ one unit of flux of $a$.   The quantum states obtained by quantization in
this situation correspond to the monopole operators.

In general, Chern-Simons couplings of gauge fields can generate nontrivial quantum numbers
for monopole operators.  However, because of $\sfT$ symmetry,
our models have no such couplings.  That being so, we get nontrivial quantum numbers only
from quantization of the fermion zero-modes.  The relevant
zero-modes are time-independent modes that are zero-modes of the Dirac operator on $S^2$.

The $\chi$ field, because it has $U(1)_a$ charge $2s$, has $2s$ zero-modes $\chi_i$,
$i=1,\dots,2s$.  These modes all have the same $2d$ chirality
and they transform with spin $J=(2s-1)/2$ under rotations of $S^2$.  The complex conjugates
of the $\chi_i$ are zero-modes
$\chi_i^*$ of the adjoint field $\chi^\dagger$.  These zero-modes can be normalized to give,
after quantization, a system of standard creation and annihilation operators,
\be\label{zendo}\{\chi_i,\chi_j\neg{}^\dagger\}=\delta_{ij},~~~\{\chi_i,\chi_j\}=\{\chi_i{}\neg^\dagger,\chi_j{}\neg^\dagger\}=0.
\ee
Because of the existence of these zero-modes, the lowest energy level in the monopole sector
is degenerate.  It contains a state $|\neg\downarrow\rangle$
that is annihilated by the $\chi_i$,
\be\label{wendo} \chi_i| \neg\downarrow\rangle=0.\ee
Acting repeatedly with the $\chi_i^\dagger$ gives a state that they annihilate:
\be\label{plendo}\chi^\dagger_i| \neg\uparrow\rangle=0,~~~|
\neg\uparrow\rangle=\chi_1^\dagger\chi_2^\dagger\dots \chi_{2s}^\dagger|  \downarrow\rangle.
\ee

Each time we act with one of the $\chi_i^\dagger$, we shift the $U(1)_A\times U(1)_a$
charges of a state by $(-1,-2s)$.  Since we do this $2s$
times to map $|  \neg\downarrow\rangle$ to $|  \neg\uparrow\rangle$, the charges of $|
\neg\uparrow\rangle$ exceed those of $|  \neg\downarrow\rangle$ by
$2s(-1,-2s)$.  On the other hand $\sfCT$ exchanges $|  \neg\downarrow\rangle$ with $|
\neg\uparrow\rangle$ and ensures that the charges of those
two states are equal and opposite.  Hence $|  \neg\downarrow\rangle$ has charges $(s,2s^2)$
and $|  \neg\uparrow\rangle$ has charges $(-s,-2s^2)$.

Let us construct monopole operators that are $U(1)_A\times U(1)_a$-invariant.  To construct
such an operator
from the Hilbert space that we have just described, one has to act on the state $|
\neg\downarrow\rangle$ with precisely $s$ creation operators.  The
resulting states are of the form
\be\label{middle}\chi^\dagger_{i_1}\chi^\dagger_{i_2}\dots \chi^\dagger_{i_s}|
\neg\downarrow\rangle.\ee

For any integer $s$, the spin $(2s-1)/2$ of the operators $\chi_i^\dagger$ is
half-integral.
The gauge-invariant states in eqn.\ (\ref{middle}) are obtained by acting on the spinless
state $|  \neg\downarrow\rangle$ with $s$ of these ``creation operators,''
so they have half-integer or integer spin depending on whether $s$ is odd or even.   If $\O$
is the monopole operator corresponding to a linear combination
of the states (\ref{middle}), then $\O$ is $U(1)_A\times U(1)_a$-invariant and its spin is
$s/2$ mod $\Z$.

Thus, such an $\O$ satisfies the spin/charge relation if and only if  $s$ is even.
  Henceforth, therefore, in investigating this class of models, we restrict  to even $s$.

  All  monopole operators of unit charge appear in the operator product
  expansion of the $\O$ constructed above
  with an ordinary local operator (a gauge-invariant polynomial in the fields and their
  derivatives).  So for even $s$, all charge 1 monopole operators
  satisfy the spin/charge relation.  Taking products of these operators, we learn that for
  even $s$, this relation is satisfied by monopole operators of any charge.

For odd $s$, with $\O(p)$ a monopole operator of charge 1,
 if $|  \Psi\rangle$ is any state that satisfies the spin/charge relation, then $\O(p)|
 \Psi\rangle$ is a state that violates it.
Since the operator $\O(p)$ is a source of one flux quantum of $f=\d a$, the states that
violate the spin/charge relation are simply those that
have an odd number of such flux quanta.

\subsection{Breaking A Symmetry}\label{addmonopole}

 In condensed matter physics, the only exact  symmetries are the ones that can be defined
microscopically (such as electric charge conservation and time-reversal symmetry).   There
might be additional emergent symmetries in the infrared,
but in a more precise description, they should all be explicitly broken, possibly by
operators that are irrelevant in the renormalization group sense.
For example, the Lorentz invariance and the $\sfC$ and $\sfCT$ symmetry of the models
introduced in section \ref{models} are all not natural in condensed matter physics
and should be explicitly broken, possibly by irrelevant interactions.  There is one more
symmetry of these models as presented so far that does not
correspond to any microscopic symmetry in condensed matter physics and so must be broken
explicitly in a more precise treatment.  This is the
symmetry generated by the conserved current
\be\label{yphi}j_\mu=\eps_{\mu\nu\lambda}\frac{f^{\nu\lambda}}{4\pi}.\ee
 We have normalized
$j_\mu$ so that -- with $a_\mu$ understood to obey standard Dirac quantization of flux --
the corresponding charge
$\u=\int\d^2x\, j^0=\int \d^2x \,f_{12}/2\pi$
has integer eigenvalues. $\u$ is simply the number of flux quanta.

The operators that violate conservation of $\u$ are simply the monopole operators that we
analyzed in section \ref{anomaly}.  We can eliminate an unphysical global conservation law
from the model
by adding to the action  a charge 1 monopole operator  with a small coefficient $\veps$.
For even $s$, a suitable linear combination of the states
(\ref{middle}) carries zero spin.    So if $\O$ is the corresponding monopole operator, with
hermitian conjugate $\O^\dagger$ of opposite monopole charge,
 we can eliminate
the conservation of $\u$, preserving all other symmetries, by adding $\veps(\O+\O^\dagger)$
to the action, or equivalently to the Hamiltonian.

In section \ref{anomaly}, we  have constructed the monopole operators at short distances
without worrying about what sort of states they act on.  This of course
depends on the physics at long distances.  In the phase that corresponds to the standard
boundary state of a topological insulator,
with $\langle w\rangle\not=0$, $U(1)_a$ is completely broken.  This leads to the existence
of vortices with any integer value of the vorticity $\u$.
The monopole operators create and annihilate these vortices, so adding a $\u=1$ monopole
operator to the action means
vortices can annihilate to quasiparticles that do not carry vorticity.

In the gapped phase with $\langle\phi\rangle\not=0$, matters are more interesting.  A
discrete subgroup of $U(1)_a$ is unbroken, so there
are stable quasiparticles with fractional $\u$.  They are discussed in section \ref{vortex}.
However, in a sample of finite size, the total $\u$
of all quasiparticles is always an integer, and in the presence of a perturbation
$\veps(\O+\O^\dagger)$, there is no conservation law associated to this
integer.

The phase with $\langle\phi\rangle\not=0$ is gapped at $\veps=0$, so adding
$\veps(\O+\O^\dagger)$
to the action with small $\veps$ does not produce any nontrivial dynamics.  The
 only purpose of this step is to explicitly break a symmetry, reducing the number of
 quasiparticle types.
In treatments of similar models that have appeared in the literature, the analogous step has
involved  ``vortex proliferation''
 or ``condensation of vortices,'' usually understood to involve some not completely explicit
 dynamics.

\subsection{The Vortex}\label{vortex}

The gapped phase with $\langle\phi\rangle\not=0$ has quasiparticles with fractional
vorticity.  In fact, since $\phi$ has charge $\k_\phi=4s$ under $U(1)_a$,
a vortex in the $\phi$ field, around which the phase of $\phi$ changes by $2\pi$, has flux
$\int_{\R^2} f=2\pi/\k_\phi=2\pi/4s$  and thus has $\u=1/4s$.

We prefer to define the integer-valued charge
\be\label{intcharge}\v=4s\u.\ee
The purpose of this is to describe the vorticity in a language analogous to what we use for
the unbroken $\Z_{4s}$ subgroup of $U(1)_a$.  This
subgroup is generated by $\K=\exp(2\pi \i\k/4s)$, where $\k$ is an integer-valued invariant
that is conserved mod $4s$.   So we likewise define an integer-valued
vorticity $\v$ that (once monopole operators are included, as in section \ref{addmonopole})
is conserved mod $4s$.

Vorticity transforms under the discrete symmetries oppositely to the electric charge.  Thus
the analog of eqn.\ (\ref{samej}) for the vorticity is
\be\label{torzo} \slT\v=-\v\slT,~~~~ \slCT\v=\v\slCT. \ee
We have written these formulas in terms of the discrete symmetries $\slT$ and $\slCT$ that
are unbroken in the phase under study.

 For the basic case of $\v=1$ (or $-1$), there is a well-known classical solution
that describes a vortex at rest.  This solution is rotation-invariant up to a gauge
transformation, and  (after suitable
gauge-fixing) it is unique  up to a spatial translation.  In particular, this standard
vortex solution is $\slCT$-invariant.  We will see that this fact is very useful.

In a purely bosonic theory with $U(1)_a$ broken to $\Z_{4s}$, and no other pertinent degrees
of freedom, quasiparticles would be labeled by the pair
$(\k,\v)$  and would satisfy Abelian statistics.  For example, in the simplest version of
$\Z_{4s}$ gauge theory, a quasiparticle of charges $(\k,\v)$,
when winding around another quasiparticle of charges $(\k',\v')$, would acquire a phase
$\exp(2\pi \i(\k\v'-\k'\v)/4s)$.

The main difference between the model under study here and a simple $\Z_{4s}$ gauge theory
results from the fact that in a vortex field with $\v=1$,
the $\chi$ field has a single zero-mode.  The phase of this zero-mode can be chosen to
make it $\slCT$-invariant, and this fixes the phase of the zero-mode up to sign.  But
actually, even without $\slCT$ symmetry, hermiticity of
the Hamiltonian and fermi statistics imply that
the space of fermion zero-modes always has a natural real structure,\footnote{In a real time
Hamiltonian framework, any
 fermion field can always be expanded in real components.  Upon doing so, the  general form
of a quadratic fermion Hamiltonian allowed by hermiticity and fermi statistics is
$H=\i\sum_{kl}h_{kl}\Psi_k\Psi_l$, where $h$ is a real
antisymmetric matrix and for clarity we write the formula in terms of a discrete sum over
modes.  As $h$ is real, the space of its zero-modes is a real
vector space.}  which in the case of a single zero-mode
fixes the mode (once we normalize it) up to sign.
Such an unpaired real
fermion  zero-mode is usually called a Majorana zero-mode.

The zero-mode  that arises in this problem in a vortex field
is somewhat analogous to fermion zero-modes that appear, for example, in the field of a
$1+1$-dimensional kink or a $3+1$-dimensional
monopole \cite{JR}. In all these
examples, when a fermion mass arises from symmetry breaking, the fermion acquires a
zero-mode in a field of suitable
topology. In a $2+1$-dimensional model
with the same relevant property (the context was a topological insulator with pairing
induced on its surface by proximity
to an $s$-wave superconductor),
this mode was found in \cite{KF}.  It was later interpreted \cite{CalliasAp} in terms of
the
Callias index theorem \cite{CalliasIndex}, following earlier work \cite{JRo,EWeinberg};
we review this interpretation in  Appendix \ref{calliastheorem}.  More recently it was also
investigated in the context of topological superconductors \cite{MetlitskiTSC}.
This zero-mode is very robust.  Even if we perturb away from the idealized models that were
introduced
in section \ref{models}, adding irrelevant interactions (for example, breaking the
symmetries $\sfC$ and $\slCT$ that are not natural in condensed
matter physics), a vortex field of this system with $\v=1$ still has an exact Majorana
zero-mode.  Local interactions can never lift a single
Majorana zero-mode.

As in the composite fermion approach to the Moore-Read state (where an analogous zero-mode
appears \cite{GR,Volovik} for somewhat different reasons),
the occurrence of a single Majorana zero-mode in the field of a vortex leads to non-Abelian
statistics.
Let us consider a system of $k$ widely separated vortices in $\R^2$.  To avoid extraneous
subtleties involving properties of the Clifford algebra,\footnote{For
example, if $k$ is odd, there is no completely natural quantization of $k$ fermion
zero-modes; if $k$ is congruent to 2 mod 4, the operator
$(-1)^F=\g_1\g_2\dots\g_k$ that plays an important role below squares to $-1$ rather than
$+1$.  Such issues are not relevant for our present purposes
and to avoid them, we take $k$ to be divisible by 8, which ensures that $k$ fermion
zero-modes can be represented by real matrices of rank $2^{k/2}$.}
we take $k$ to be divisible by  8.  (In any event, a globally defined state in a sample of
finite volume would have $k$ a multiple of $4s$, which is divisible
by 8 since $s$ is even.)

Simply because an isolated vortex would have
an exact fermion zero-mode and the theory is gapped, it follows that in the limit that the
vortices are widely separated, the field $\chi$ has $k$ modes that are
exponentially close to being zero-modes.  Whether these modes are exact zero-modes or are
only exponentially close to being zero-modes will
not be important in what follows.\footnote{With the aid of $\slCT$ symmetry, one can use the
Callias index theorem to show
that any state of vorticity $\v$ has $|  \v|  $ exact Majorana zero-modes.  (See Appendix
\ref{calliastheorem}.) With small explicit breaking of $\slCT$,
these modes are exponentially close to zero in the case of widely separated vortices.}

Let $\g_i$, $i=1,\dots,k$, be the Majorana zero-mode associated to the $i^{th}$ vortex.  The
$\g_i$ can be normalized so that after quantization they obey
a Clifford algebra:
\be\label{clifford}\{\g_i,\g_j\}=2\delta_{ij}. \ee
An irreducible representation $\RR$ of this algebra has $2^{k/2}$ states.  So this
$k$-vortex system has,
on the average, $\sqrt 2$ quantum states per vortex.  The nonintegrality of
this number is a symptom of non-Abelian statistics.

Suppose that we adiabatically move the vortices around, keeping them widely separated.
The quantum state varies according to the Berry connection
\cite{Berry}.   The monodromies around closed orbits\footnote{Vortices are identical objects
and a ``closed orbit'' may permute them.}
  in the configuration space of $k$ vortices give  a representation of the braid group $B_k$
  on the Hilbert space $\RR$.

It turns out that this representation can be easily determined using two elementary facts:

(1) Each vortex has its own zero-mode, which is uniquely determined up to sign.  So the
braid group can only act by permutations and sign changes
of the $k$ objects $\g_1,\g_2,\dots,\g_k$.

(2) The action of the braid group commutes with the operator $(-1)^F$, which counts the
number of fermions mod 2.  This operator anticommutes
with all of the fermion zero-modes $\g_1,\g_2,\dots,\g_k$.  Up to an irrelevant  sign, the
operator on the Hilbert space $\RR$ that has this property
(and squares to 1) is the chirality operator, the product of all the gamma matrices:
$(-1)^F=\g_1\g_2\dots\g_k$.

\begin{figure}
 \begin{center}
   \includegraphics[width=3in]{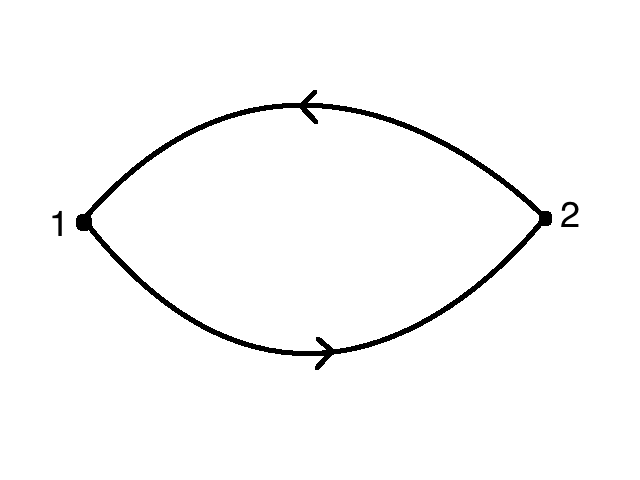}
 \end{center}
\caption{\small  Vortex 1 is initially to the left of vortex 2.  The two vortices are
exchanged via a motion in the counterclockwise direction.  }
 \label{counterclockwise}
\end{figure}

Now let us consider a basic operation: the exchange of two vortices, say labeled 1 and 2,
that move half way around each other in, say, the counterclockwise direction  (fig.
\ref{counterclockwise}).
How does this elementary move act on the gamma matrices?  One might expect that the answer
would involve a simple transposition
$(\g_1,\g_2)\to (\g_2,\g_1)$, but this is not possible as it would reverse the sign of the
chirality operator $(-1)^F$.   Rather, we must use the fact
that the $\g_i$ were naturally defined only up to sign.   The basic operation acts on the
gamma matrices with a sign change:
\be\label{toff} B_{12}:(\g_1,\g_2)\to (\g_2,-\g_1).\ee
Actually, we could equally
well take $(\g_1,\g_2)\to (-\g_2,\g_1)$.   The two operations differ by conjugation by
$(\g_1,\g_2)\to (\g_1,-\g_2)$ (or $(\g_1,\g_2)\to (-\g_1,\g_2)$).
The sign choice  in eqn.\ (\ref{toff}) fixes the relative signs of the $\g_i$, which have
been arbitrary until this point.

The crucial minus sign in eqn.\ (\ref{toff}) means that $B_{12}$ does not coincide with its
inverse, and so we get different results depending on whether
particle 1 moves around particle 2 in a clockwise or counterclockwise direction.  In fact,
$B_{12}^2\not=1$ but $B_{12}^4=1$.

The fact that $B_{12}^2\not=1$ means that braiding of vortices gives a representation of the
braid group that does not merely come from a representation
of the permutation group.  The fact that the operation (\ref{toff}), extended to a $k$
particle system, does give a representation of the braid group
can be  verified as follows.

Start with an initial configuration in which $k$ vortices, labeled $1,2,\dots,k$, are
arranged on the real axis in ascending order.  The braid group
$\B_k$ on $k$ letters is generated by elements $B_{i,i+1}$, $i=1,\dots,k-1$ that exchange
the $i^{th}$ and $i+1^{th}$ objects in a counterclockwise
direction.  As above, we take $B_{i,i+1}$ to act by $(\g_i,\g_{i+1})\to (\g_{i+1},-\g_i)$
(with trivial action on the other $\g_j$).   To get a representation
of $\B_k$, the only relation that we have to check is
\be\label{braidrel}B_{i,i+1}B_{i+1,i+2}B_{i,i+1}=B_{i+1,i+2}B_{i,i+1}B_{i+1,i+2}. \ee
Indeed, $B_{i,i+1}B_{i+1,i+2}B_{i,i+1}$ acts by
\be\label{firstone}(\g_i,\g_{i+1},\g_{i+2})\to (\g_{i+1},-\g_i,\g_{i+2})\to
(\g_{i+1},\g_{i+2},\g_i)\to (\g_{i+2},-\g_{i+1},\g_i),\ee
while $B_{i+1,i+2}B_{i,i+1}B_{i+1,i+2}$ acts by
\be\label{secondone}(\g_i,\g_{i+1},\g_{i+2})\to (\g_i,\g_{i+2},-\g_{i+1})\to
(\g_{i+2},-\g_i,-\g_{i+1})\to (\g_{i+2},-\g_{i+1},\g_i).\ee

What is the actual matrix $B_{i ,i+1}$ such that  $B_{i,i+1}^{-1}(\g_i,\g_{i+1})B_{i,i+1}=
(\g_{i+1},-\g_i)$, while $B_{i,i+1}$ commutes with the other gamma matrices?  It is
\be\label{uptosign}B_{i,i+1}=\pm \frac{1+\g_i\g_{i+1}}{\sqrt 2},\ee
where $\slCT$-invariance implies that the overall phase must be a sign $\pm 1$, and
 to satisfy eqn.\ (\ref{braidrel}), this sign must be independent of $i$.  The reason for
 this sign ambiguity is the following.  There is a 1-dimensional
representation of the braid group -- the representation associated to fermions -- in which
each
$B_{i,i+1}$ is equal to $-1$.  Tensoring with this representation
changes the sign in eqn.\ (\ref{uptosign}).  The two choices give representations that are
definitely inequivalent; the eigenvalues of $B_{i,i+1}$ are either $\i^{\pm 1/2}$ or
$\i^{\pm 3/2}$.
We will see in section \ref{quasi} that the basic vortex corresponds to the $+$ sign in
(\ref{uptosign}).
The opposite sign arises for another $\slCT$-invariant quasiparticle of the model (a state
of vorticity 3).

The representation of the braid group that we have just described is very close to one that
arises in
Ising conformal field theory in two dimensions, or in the corresponding Chern-Simons
topological field theory in three dimensions.   The conformal blocks of the spin fields of
the Ising model are
characterized by a representation of the braid group in which $B_{i,i+1}$ acts
as described above, up to an Abelian factor (that is, a phase factor)  that violates $\slCT$
symmetry.  (This representation was first found in the RCFT context and appeared explicitly
in \cite{Moore:1988qv}.  It was later used in the context of non-Abelian anyons in
\cite{NayakWilczek,Ivanov}.)
Clearly, since the models that we are studying are $\slCT$-conserving, they do not
generate this $\slCT$-violating phase.  We will see instead that the models considered here
can be constructed by
combining in a certain sense an Ising model with a theory (a version of
$\Z_{4s}$ gauge theory)  that cancels the phase factor. By itself, this latter theory would
generate Abelian statistics.

So far we have considered only vortices with $\v=1$.    What happens for $\v>1$?
Assuming $\slCT$ symmetry, a system of $\v$ widely separated vortices of vorticity 1 has
$\v$  zero-modes, all transforming the same way
under $\slCT$, since the individual vortices are identical (or since this is predicted by
the Callias index theorem).   We choose the sign of the operator $\slCT$ so that the
approximate zero-modes
all transform as $+1$ under $\slCT$.   This remains so when we bring the individual vortices
together to make a quasiparticle of vorticity $\v$.

We write $\g_j$, $j=1,\dots,\v$ for the fermion zero-modes in a quasiparticle of vorticity
$\v$.  If we add to the underlying Lagrangian or Hamiltonian irrelevant interactions that
explicitly break $\slCT$ symmetry,
with a small coefficient, this will induce an effective Hamiltonian for the vortex
\be\label{effham}\Delta H=\i \sum_{k,l}m_{kl}\g_k\g_l+\dots \ee
where $m_{kl}$ is a real, antisymmetric matrix.  Generically, this completely lifts all of
the fermion zero-modes if $\v$ is even, and leaves 1 zero-mode if $\v$ is odd.

In a gapped system, adding an irrelevant operator with a small coefficient cannot affect the
topological field theory that describes the braiding of quasiparticles.
So this braiding can be computed assuming that a vortex of vorticity $\v$ has 1 real fermion
zero-mode if $\v$ is odd and none if $\v$ is odd.   Accordingly, vortices of vorticity $\v$
have non-Abelian statistics as described above if $\v$ is odd, and have only Abelian
statistics if $\v$ is even.\footnote{The state selected by the $\slCT$-violating
perturbation
$\Delta H$ is not a $\slCT$ eigenstate.  Consequently, although eqn.\ (\ref{toff}) is valid
for any odd $\v$, it is not immediately clear
that  the phase in the formula (\ref{uptosign}) for $B_{12}$ is
just $\pm 1$.  But topological field theory is rigid in a way that makes this phase robust
against small $\slCT$-violating
perturbations.}

If instead we maintain $\slCT$ symmetry and take it into account, then the vortices have a
richer structure.  This will be more fully explored in section \ref{quasi}, but as a
preliminary,
we note that only the value of $\v$ mod 8 is important.  This is true for essentially the
same reason that there is a mod 8 periodicity in the theory of the Majorana chain
\cite{MajoranaChain}.
$\slCT$ symmetry does not allow a quadratic Hamiltonian as in eqn.\ (\ref{effham}), but it
does allow a quartic Hamiltonian
\be\label{neffham}\Delta H=\sum_{ijkl}p_{ijkl}\g_i\g_j\g_k\g_l,\ee
where $p_{ijkl}$ is real and completely antisymmetric.  A Hamiltonian of this form can
remove zero-modes in groups of 8.  The precise meaning of ``removing zero-modes in groups of
8''
is the following.  Quantization of 8 fermion zero-modes gives $2^{8/2}=16$ states.  A
quartic fermion Hamiltonian generically selects a single $\slCT$-invariant ground
state  in this 16-dimensional space (moreover, as explained in  Appendix
\ref{calliastheorem}, this can be done in such a way that the state in question is  bosonic
and has spin zero).  Thus in the presence of a generic $\slCT$-conserving Hamiltonian, real
fermion zero-modes can be omitted in groups of 8.

As a check on the claim that the statistics of a vortex of vorticity $\v$ only depend on the
value of $\v$ mod 2, we will consider the braiding of a vortex of vorticity 1 with one of
some arbitrary
vorticity $p$.
 The $\v=1$ vortex supports a single gamma matrix $\g_1$ and
the $\v=p$ vortex supports $p$ additional gamma matrices $\g_2,\g_3,\dots,\g_{p+1}$.
Braiding of the $\v=1$ vortex counterclockwise
around the composite $\v=p$ vortex
 is simply described by a product $B_{p,p+1}B_{p-1,p}\dots B_{1,2}$ of the basic moves
 described above.  The result is $(\g_1,\g_2,\dots,\g_{s+1})\to
 (\g_2,\g_3,\dots,\g_{p+1},(-1)^p\g_1)$.   For
 even $p$, there is no nontrivial sign, and the braiding involves a simple permutation of
 the gamma matrices.  For odd $p$, we get the same sign change that
 we found earlier for $p=1$, leading to the same non-Abelian statistics as before.   This
 reasoning can be straightforwardly extended to consider braiding of vortices with any two
 given
 vorticities $p$ and $p'$.

An alternative explanation of why nonabelian statistics arise precisely for odd vorticity is
as follows.  Consider a collection of $n$ widely separated quasiparticles of vorticity
 $\v_i$,  $i=1,\dots,n$,
with total vorticity $\v=\sum_i\v_i$.  Assume first that all $\v_i$ are even, $\v_i=2k_i$.
An irreducible representation of the $2k_i$ gamma matrices of the $i^{th}$ quasiparticle
gives a Hilbert space $\H_i$ of dimension $2^{\v_i/2}=2^{k_i}$.  The tensor
product\footnote{This
tensor product (and others discussed momentarily) must be taken in a $\Z_2$-graded sense,
which means that one includes
factors that ensure that the gamma matrices of one quasiparticle anticommute with those of
another, rather than commuting.}
 $\H=\otimes_i  \H_i$ of the individual $\H_i$ has dimension $2^{\sum_i k_i}=2^{\v/2}$
and gives an irreducible representation of the whole set of $\v$ gamma matrices.   Now
suppose instead that the $\v_i$ are all odd numbers $\v_i=2k_i+1$, and assume that $n$ is
even so that $\v=\sum_i\v_i$ is even.\footnote{In a condensed matter system of finite
volume, with compact boundary, $\v$ will always be divisible by $4s$.  In RCFT,
one would say that $\v$ must be even or the correlation functions all vanish and the space
of conformal blocks is zero. Similarly below we assume $n_{\mathrm o}$ to be
even.}   An irreducible representation of $2k_i+1$ gamma matrices has dimension $2^{k_i}$,
so this
is the dimension of the Hilbert space $\H_i$ that we might associate to the $i^{th}$
quasiparticle.  The tensor product $\otimes_i \H_i$ then has dimension $2^{\sum_ik_i}$,
while the Hilbert space of the whole system, which has to represent the whole set of
$\sum_i\v_i=2\sum_i k_i+n/2$ gamma matrices, will have dimension $2^{\sum_ik_i+n/2}$.
So the Hilbert space of the whose system is not simply $\otimes_i \H_i$; it is
$\H=Q\otimes_i \H_i$, where $Q$, known in RCFT as the space of conformal blocks, has
dimension
$2^{n/2}$.  The existence of a space of conformal blocks of dimension greater than 1 is the
sign of nonabelian statistics.  More generally, suppose that there are $n_{\mathrm e}$
quasiparticles of even vorticity and $n_{\mathrm o}$ of odd vorticity.  A counting as above
shows that the dimension of the space of conformal blocks is $2^{n_{ \mathrm o}/2}$.

\subsection{Quasiparticles}\label{quasi}

In this section, we describe the quasiparticles of the theory more systematically.

\subsubsection{Enumerating the Quasiparticles}\label{enum}

 First we consider  the ``elementary'' quasiparticles,
the ones that do not carry vorticity.   These are the fermion field $\chi$ and the boson
$w$.  (In the phase with $\langle\phi\rangle\not=0$,
there is no quasiparticle corresponding to $\phi$.)   $\chi$ is electrically neutral and has
spin 1/2.
In enumerating quasiparticles, we only keep
track of the spin mod 1, since integer spin can be converted to orbital angular momentum.
So we do not distinguish the two components of $\chi$, which have spin $\pm 1/2$.   $w$ has
spin 0 and electric charge $\q'=-1/2s$.
By fusion of $\chi$ with $w$ or $\bar w$, we can make additional quasiparticles $w^n$ and
$w^n \chi$, with the relations
\be\label{relations}\chi^2=1,~~~\bar w^{2s}\chi=e,\ee
which reflect the following.
  Since $\chi$ has a Majorana mass, a pair of $\chi$ excitations can be created or
  annihilated.  And the bulk coupling $I_e$
of eqn.\ (\ref{electron}), where $e$ is the electron field, means that a combination of
quasiparticles $\chi\bar w^{2s}$ can be converted to an electron $e$ and disappear into
the bulk; similarly $\chi w^{2s}$ is equivalent to a hole.

In describing the low energy physics, it is possible to consistently keep track of the
electric
charge only modulo 2, basically because a pair of electrons can carry integer angular
momentum, which is irrelevant in the low energy topological field theory.
We can thus consider $e^2$ and $\bar e^2$ to be trivial, and thus we can  set $w^{4s}=\bar
w^{4s}=1$ and identify $w^n$ with $\bar w^{4s-n}$.   In fact,
we have implicitly done that in enumerating the quasiparticles in the last paragraph.  If we
want to keep track of  electric charge as a real number (and not just its
reduction mod 2), then we have to distinguish $w^n$ from $\bar w^{4s-n}$.  In this case, the
full set of quasiparticles of zero vorticity would be $w^n$, $\bar w^n$,
$\chi w^n$, $\chi\bar w^n$, with the relations (\ref{relations}), along with $\chi
w^{2s}=\bar e$ and $\bar w w = 1$.  For brevity, in the following we use the smaller
set of quasiparticles.

The quasiparticles $w^n$ and $\chi w^n$ are mapped to themselves by the effective
time-reversal symmetry $\slT$ of the low energy theory.
Since $\slT^2=(-1)^F$, the states $w^n$ are Kramers singlets and $\chi w^n$ are Kramers
doublets.  $\slC$ maps $\chi$ to itself and $w^n$ to $\bar w^n$.
From this, the action of $\slCT$ follows.  Since $(\slCT)^2= (-1)^F\K$ (eqn.\ (\ref{ctphi}))
and $\chi$ is
invariant under $(-1)^F\K$, $\chi$ transforms as a Kramers singlet under $\slCT$.  $\slCT$
exchanges $w^n$ with $\bar w^n$ with phases that are determined
by the relation $(\slCT)^2=(-1)^F\K$.

If $\slCT$ symmetry is weakly broken by a perturbation such as that of eqn.\ (\ref{effham}),
then it is straightforward to enumerate the quasiparticles
for any $\v$.  For odd $\v$, there is a single real fermion zero-mode $\g_0$.  It obeys
$\g_0^2=1$ and can be represented as 1 (or $-1$)  in the vortex state (that is, in the Fock
vacuum obtained
by quantizing the nonzero-modes of $\chi$ in the field of the vortex).  In effect, $\g_0$
has an expectation value in the vortex state.
Moreover, $\g_0$ is a mode of $\chi$.  So acting with $\chi$ on the vortex
does not produce a new quasiparticle type.  We can still act with $ w^n$, $n=0,\dots,4s-1$,
producing $4s$ quasiparticle types for odd $\v$.

For even $\v$, with $\slCT$ symmetry weakly broken, there are no fermion zero-modes and,
just as for $\v=0$,
 we can make $2\times 4s=8s$ quasiparticle types by acting with $\chi$ and/or $ w^n$,
 $n=0,\dots,4s-1$.
 For both even and odd $\v$, it is a little tricky to determine the spins of these
 quasiparticles.  We will return to this shortly.

\subsubsection{Consequences of $\slCT$ Symmetry}

 $\slCT$ symmetry leads to a richer structure, though it does not change the list of
 quasiparticles.  As a preliminary, the gauge charges $\q'$ and $\k$ and the spin or angular
 momentum $\j$
 are all odd under $\slCT$.  So $\slCT$ symmetry ensures that the filled Dirac sea of
 negative energy modes of $\chi$  does not contribute to any of those quantum numbers.
 There is similarly
 no contribution from the bosons. The quasiparticle
 quantum numbers are therefore determined entirely by quantization of the fermion
 zero-modes.   What we are really interested in here are the fractional parts of the
 quasiparticle quantum numbers,
 which are observables of the topological field theory that governs quasiparticle
 interactions at low energies.  This topological field theory is not affected by weak
 $\slCT$-violating perturbations,
 so the determination below of the fractional parts of the quantum numbers is also valid in
 the absence of $\slCT$ symmetry.

Even in the presence of  $\slCT$ symmetry, the quasiparticle structure  in a sector of
vorticity $\v$ only depends on $\v$ mod 8, as explained in the discussion of eqn.\
(\ref{neffham}).
So we can limit ourselves to $0\leq \v\leq 7$.  But actually, even in this range, there is a
symmetry $\v\to 8-\v$.  In fact, $\slT$ maps $\v$ to $-\v$, which, since $\v$ is only
defined mod $4s$,
is equivalent to $4s-\v$.   Because of the anomaly described in section \ref{anomaly}, we
take $s$ to be even, so $4s$ is a multiple of 8 and hence $4s-\v$ is congruent mod 8  to
$8-\v$.
Combining these statements, we see that everything is determined by what happens for $0\leq
\v\leq 4$.   (Some subtlety is needed in applying this statement, because $\slT$, which was
used
in relating $\v$ to $8-\v$, does not commute with $\slCT$ (eqn.\ (\ref{transsign}).)

For $\v=1$, the vortex has just one $\chi$ zero-mode.  As explained in section \ref{enum},
this mode has an expectation value in the vortex state, as a result of which acting with
$\chi$ does not produce a new
quasiparticle.  But we obtain $4s$ quasiparticle types by acting with powers of $w$.  These
are mostly exchanged in pairs by $\slCT$, since
 $\slCT $ exchanges $w$ and $\bar w$.

To determine the spins of these quasiparticles, we need to take into account that in a
vortex sector, the conserved angular momentum $\j'$ is not
just the conventional angular momentum $\j$, but is a linear combination of $\j$ with the
emergent gauge charge $\k$:
\be\label{linearcom}\j'=\j+\v\k/4s. \ee
Since $w$ has $\j=0$, $\k=1$, it carries $\j'=\v/4s$, so it has spin $1/4s$ in a field of
$\v=1$.   Thus if $\Psi_1$ is the basic vortex of $\v=1$, the
quasiparticles $w^n\Psi_1$ have charge $\q'=-n/2s$ and spin $\j'=n/4s$.

For $\v=2$, there are two $\chi$ zero-modes.  As explained in Appendix \ref{calliastheorem},
these modes have $\j'=\pm 1/2$.  Quantizing those modes
gives two states of $\j'=\pm 1/4$.  These two states are electrically neutral, of course,
and are exchanged by the action of $\chi$.

The fact that the spins of the states obtained by quantizing the fermion zero-modes are $\pm
1/4$ enables us to answer a question that was left open in section
\ref{vortex}.  Starting with a rotation-symmetric classical state  of $\v=2$, separate the
two vortices in a way that preserves the symmetry under a $\pi$ rotation
that exchanges the two vortices
(but not, of course, the full rotation symmetry).  In this process, $\j'$ is not conserved,
but it is conserved mod 2, since the $\pi$ rotation symmetry is preserved.
So the possible angular momentum states of a system of two basic vortices of $\v=1$ are of
the form $2p\pm 1/4$, $p\in \Z$.
This means that the eigenvalues of a $\pi$ rotation are $\exp\left(\i\pi(2p\pm
1/4)\right)=\exp(\pm \i\pi/4)$.
This $\pi$
rotation is a way to describe the braiding of two vortices.    The eigenvalues in the
braiding of two vortices are thus $\exp(\pm \i\pi/4)$, which means that
we must use the $+$  sign in eqn.\ (\ref{uptosign}).

The remaining quasiparticles of $\v=2$ are obtained by acting with $w^n$ on these two states
of $\j'=\pm 1/4$.
For $\v=2$, $w$ has $\j'=1/2s$.  So the states obtained by acting with $w^n$  have spins
$\j'=\pm 1/4+n/2s$ and electric charge $\q'=-n/2s.$

For $\v=3$, the three $\chi$ zero-modes have spins $\j'=1,0,-1$, as found in Appendix
\ref{calliastheorem}.  Quantizing the states of  $\j'=\pm 1$
gives a pair of quantum states with $\j'=\pm 1/2$.  The additional $\chi$ zero-mode of
$\j'=0$ leads to non-Abelian statistics.  Acting with $w^n$,
we get quasiparticles with $\j'=\pm 1/2+3n/4s$, $\q'=-n/2s$.

Finally, for $\v=4$, the four $\chi$ zero-modes have spins $\j'=3/2,1/2,-1/2,-3/2$.  Their
quantization gives a pair of states of spins $\pm 1$ and a second
pair of spins $\pm 1/2$.  The states with spins $\pm 1$ are bosonic and, as we discuss
shortly, form a Kramers doublet under $\slCT$.
The states with spins $\pm 1/2$ are fermionic and form another Kramers doublet.  Acting with
$w^n$ gives spins $\pm 1+n/s$
and $\pm 1/2+n/s$, with $\q'=-n/2s$.

\subsubsection{Some More Delicate Questions}\label{delicate}

There remain some more delicate questions concerning the transformation of the
quasiparticles under symmetries.  In particular,
according to eqn.\ (\ref{ctphi}), the theory should admit symmetry operators  $\slCT$,
$(-1)^F$,
and $\K$ that should be be linked by
\be\label{lima}(\slCT)^2 = (-1)^F \K\ee
(and more trivial relations: $\K^{4s}=1$ and $\slCT,$ $(-1)^F$, and $\K$ all commute).   We
discuss below the action of $\slT$.

For quasiparticles of vorticity 0, one can use classical field theory to define symmetry
operators that obey all expected relations.
This is essentially how eqn.\ (\ref{lima}) was deduced in section \ref{models}.  Even when
one includes vortices, it remains true that
 for any state of the system
that can be defined in a compact sample,  the symmetry
operators can be defined and
obey the expected relations.  Concretely, this is because, in a compact sample, the total
vorticity is always a multiple of\footnote{Similarly, as  remarked after eqn.
(\ref{transsign}), in any state of a compact sample, the discrete gauge charges $\k_i$ of
all quasiparticles always add up to a multiple of $4s$, so that
globally $\K=1$.} $4s$
and therefore (since $s$ is even) is divisible by 8.  So the fermion zero-modes of the whole
system, taken together, generate a Clifford algebra
whose rank is a multiple of 8.  Such a Clifford algebra has a real representation in which
all symmetries are realized in the expected way.

By contrast, if $\v$ is not a multiple of 8, then the fermion zero-modes of a quasiparticle
of vorticity $\v$ generate a Clifford algebra whose
rank is not a multiple of 8, and in this case, in general the symmetries are not realized
``locally'' (i.e. on just one quasiparticle) in the expected way.
An analog is the theory of the Majorana chain \cite{MajoranaChain}.  A state of the whole
system realizes the symmetries (in that case, $\sfT$ and
$(-1)^F$) as expected, but if one attempts to factorize the Hilbert space as a tensor
product of a space associated to one end of the chain
and a space associated to the other, then the symmetries cannot be defined on the individual
factors or do not act in the expected way.

In the present problem, we find that  the symmetries act in a fairly normal way if $\v$ is
even, but not if $\v$ is odd.  There are several
ways to explain this.   Quasiparticles of even vorticity obey abelian statistics, so that,
as noted at the end of section \ref{vortex},
if one considers states with only such quasiparticles, the Hilbert space of the whole system
is the tensor product of subspaces associated
to individual quasiparticles.  This is enough to ensure that one can define all of the
expected operators for each quasiparticle,  but allows the possibility
that in doing so, one will run into a central extension of the expected symmetry group.  In
a sense, the occurrence in eqn. (\ref{lima}) of the operator
$\K$, which is trivial for a global state of the system but nontrivial for an individual
quasiparticle, represents such a central extension. Alternatively,
 as we explain at the end of section \ref{nonabelians}, by considering
a topological insulator interacting with magnetic monopoles, one can create a situation in
which a compact sample can have a state
of any even vorticity (not necessarily a multiple of $4s$),\footnote{In section
\ref{nonabelians}, we show that with unit magnetic charge inside the sample, one gets on the
surface a quasiparticle of vorticity $\pm 2$
together with $s$ quasiparticles of type $w$ or $\bar w$.  Since the symmetries can
certainly be defined for the classical fields $w$ and $\bar w$, and
they can also be defined for a global state of the whole system, it must be possible to
define the symmetries for a quasiparticle of vorticity $\pm 2$.}  so it must be possible to
define the symmetries for any even value of $\v$.

Let us begin by explaining the difficulty in realizing the symmetries when $\v$ is odd.  The
problem is particularly obvious for $\v=1$.
There is only one fermion zero-mode $\gamma_1$ and it has an expectation value in the vortex
state.   In eqn.\ (\ref{lima}), the operator $(-1)^F$
is supposed to anticommute with all fermion operators (and commute with bosons), and
likewise $\chi$ is odd under\footnote{$(-1)^F$ and $\K$ differ
because they act differently on $w$.}  $\K$.   So both $(-1)^F$ and $\K$ should anticommute
with $\g_1$. Clearly, an expectation value of $\gamma_1$ is not compatible with the
existence of a symmetry
$(-1)^F$ or $\K$ under which $\g_1$ is
odd.\footnote{The problem in defining K in the sector with odd v is reminiscent of the
problem in defining the color of magnetic monopoles in grand unified theories \cite{ColNe}.
In both cases we have an unbroken gauge theory, but the ``global gauge charge'' cannot be
defined in a topological nontrivial sector.}
  However, there is no obstruction to realizing the products $(-1)^F\K$ or $\K^2$, which
  commute with
$\g_1$.

More generally, for any odd $\v$, there are $\v$ zero-modes $\g_1,\dots,\g_\v$, and the
operator $\bg=\g_1\g_2\dots\g_\v$ commutes with them and so
is a $c$-number in an irreducible representation of the algebra.   But $\bg$ is odd under
$(-1)^F$ and under $\K$, so those operators cannot act
in an irreducible representation of the algebra.   Again, there is no problem with
$(-1)^F\K$ or $\K^2$.

Since $(-1)^F\K$ commutes with all the zero-modes,  it acts as a $c$-number in an
irreducible representation of the Clifford algebra.
To decide what this $c$-number should be, we return to eqn. (\ref{lima}), in which $\slCT$
is supposed to be an antiunitary symmetry.  We now
run into further subtleties, which depend on the properties of the rank $\v$ Clifford
algebra generated by the zero-modes.  For $\v=1$,
the Clifford algebra has a 1-dimensional representation, with the only gamma matrix being
$\g_1=1$.  On this space, we can define $\slCT=\KK$
(complex conjugation), obeying $(\slCT)^2=1$.  So we must define $(-1)^F\K=1$ on the vortex
ground state for $\v=1$.  $\K^2$ should commute with
$\slCT$, so it should act on this
state as\footnote{There seems to be no natural way to fix this sign.  In the language of
section \ref{nonabelians}, the signs depends on whether we
associate $\K^2$ with the one-form global symmetry $e^{2\i\oint c}$ or $W_\psi e^{2\i\oint
c}$.  These are exchanged by $\slCT$ so neither is preferred.}
 $\pm 1$.   For $\v=3$, there are three zero-modes,  which we can represent as $2\times 2$
 matrices $\g_i=\sigma_i$ (the Pauli matrices).
 Now we run into the fact that on this two-dimensional space, it is not possible to define
 $\slCT$, because there does not exist any antiunitary operator
  that commutes with the $\g_i$, as $\slCT$ is supposed to do.  There does exist the
  antiunitary operator $\KK \sigma_2$, which
 {\it anti}commutes with the $\g_i$, and obeys $(\KK\sigma_2)^2=-1$.  We cannot interpret
 this operator as $\slCT$, but we can interpret it as $\slCT\K$, which is supposed to
 anticommute with the $\g_i$.  As $(\slCT\K)^2=(-1)^F\K^3$, we learn that we must take
 $(-1)^F\K^3=-1$ for $\v=3$.  As for $\K^2$,
 having fixed its sign for $\v=1$, we can determine its sign for $\v=3$ by fusing three
 vortices of $\v=1$.

 More generally, for any odd value of $\v$, since there is no way to define $\K$, it is
 clear that we at most may be able to define $\slCT$ or $\slCT\K$,
 but not both.  Continuing in the above vein, one finds that for $\v=5$, one can define
 $\slCT$ (and it obeys $(\slCT)^2=-1$), while for $\v=7$,
 one can define $\slCT\K$ (and it obeys $(\slCT\K)^2=1$).

 Now we consider the case of even $\v$, which is more straightforward as explained above
 (but which still has some subtleties that we will not fully
 explore).  For $\v=2$, there are two gamma matrices, which can be given a real $2\times 2$
 representation.
  So the antilinear operator $\slCT$ that commutes with them is simply, up to an irrelevant
phase,  $\KK$.  In particular, it obeys $(\slCT)^2=1$, so the vortex states are Kramers
singlets of $\slCT$.
  Accordingly, we expect $(-1)^F\K=1$.  Because $(-1)^F$ and $\K$
should commute with $\slCT$ and anticommute with the $\g_i$, they must be, up to sign,
$(-1)^F=-\K =\bg=\g_1\g_2$.   As a check, the eigenvalues
of $\K=\exp(2\pi \i\k/4s)$ are $\pm \i$, corresponding to $\k=\pm s$.  These are the
expected values as the $\g_i$ (being modes of $\chi$)
carry $\k=2s$ or $\K=-1$, so that by the usual logic \cite{JR}  of charge fractionation, the
states that the $\g_i$ act on must have $\k=\pm s$.
 We should note, however, that the square of $(-1)^F$ is $-1$, not the expected $+1$.  We
 can regard this as representing
a (further) central extension of the algebra, a possibility allowed by the general remarks
above.

For $\v=4$, the gamma matrices $\g_1,\dots,\g_4$ can be represented in a four-dimensional
Hilbert space $\H$.
It is not possible to take them to be all real.  One can have three of the gamma matrices
real and the fourth imaginary, for example
$\g_1=\sigma_1\otimes 1, $ $\g_2=\sigma_3\otimes 1$, $\g_3=\sigma_2\otimes \sigma_2$,
$\g_4=\sigma_2\otimes \sigma_1$.   With this
choice, the $\slCT$ transformation that commutes with the $\g_i$ is uniquely determined, up
to an irrelevant phase,
to be $\slCT=\KK\g_{123}$, where $\g_{i_1i_2\dots i_k}$ is an abbreviation for
$\g_{i_1}\g_{i_2}\dots \g_{i_k}$.  This operator obeys $(\slCT)^2=-1$, so the states are all
in Kramers doublets, and we want $(-1)^F\K=-1$.
The operators $(-1)^F$ and $\K$ must anticommute with the $\g_i$ and commute with $\slCT$.
Up to sign, this forces $(-1)^F=-\K=\bg=\g_1\g_2\g_3\g_4$.

The fact that $\K=-(-1)^F$ for these states is a kind of anomaly.  We have found the vortex
states by quantizing the
theory of the $\chi$ field coupled to $\phi$ and $a$ (but with $w$ playing no role).  In a
restricted theory without the field $w$, at the classical level $\K=+(-1)^F$
for all fields, but we have found that in the $\v=4$ sector, the sign in this relation is
reversed.  This is closely related to the anomaly that was
described in section \ref{anomaly}.  To understand the relation, set $s=1$, so that a state
of $\v=4$ can be created by a monopole operator.
The anomaly says that the relation between spin and charge for monopole operators is the
opposite of the classical one, and this corresponds
to the minus sign in the relation $\K=-(-1)^F$.

The action of $\slT$ is largely determined by the relation $\slT^2=(-1)^F$ and the fact that
$\slT$ exchanges states of vorticity $\v$ with those of vorticity
$-\v$.  The nontrivial cases are the $\slT$-invariant values $\v=0$ and $\v=2s$.  For
$\v=0$, the action of $\slT$ is determined from the classical
considerations of section \ref{models}.  As for $\v=2s$, it is equivalent to $\v=0$ if $s$
is divisible by 4.  If  instead $s$ is congruent to 2 mod 4, then
the rank of the $\v=2s$ Clifford algebra is congruent to 4 mod 8, and the analysis of the
action of $\slT$ in this case is similar to what we said above
concerning the action of $\slCT$ for $\v=4$.

In section \ref{TFTE} we will rederive this spectrum of quasiparticles and the action of the
various symmetries on them using the low-energy TQFT description of this system.

\subsection{The Partition Function}\label{partfn}

Here we will examine in detail the partition function of the Dirac fermion $\chi$, coupled
to
$B=A+2s a$.  One goal is to show carefully that the models
we have studied have precisely the same anomalies as the standard gapless boundary state of
a topological insulator, and thus are fit as boundary states.    A second goal is to
describe
the partition function in a way that will motivate a topological field theory description
that we will give in section \ref{TFTE}.  We will consider only orientable
spacetimes.\footnote{The extension to an unorientable spacetime, which can be made using
either $\sfT$ or $\sfCT$ symmetry, can be analyzed
by using the Dai-Freed theorem \cite{DF} rather than the APS index theorem \cite{APS}. }

\subsubsection{Review Of The Parity Anomaly}\label{revparity}

Naively the fermion partition function is given by the determinant $\det \,\D$ of the
hermitian operator $\D=\i\slashed{D}$.   Because $\D$ has real eigenvalues, this determinant
is naturally
real.   But  infinitely many eigenvalues  are negative, and this leads to a problem in
defining the sign of $\det\,\D$.

Naively, $\det\,\D=(-1)^{n_-} |\det \neg \, \D|$ where $n_-$ is the number of negative
eigenvalues of $\D$. As $n_-$ is infinite, some regularization is needed.  Formally let
$n_+$ be the number of positive
eigenvalues of $\D$ and $N=n_++n_-$ the total number of eigenvalues.  In trying to
regularize $n_-$, it is better to first subtract the infinite constant $N/2$, thus replacing
$n_-$ by $n_--N/2=(n_--n_+)/2$.
The difference $n_+-n_-$ is still ill-defined, but any reasonable regularization of it
gives the same result, the $\etta$-invariant of Atiyah, Patodi, and Singer  \cite{APS}.
For example, one may define
\be\label{ruff}\etta=\lim_{\epsilon\to 0^+}\sum_i
\exp(-\epsilon|\lambda_i|)\sign(\lambda_i), \ee
where the sum runs over all eigenvalues $\lambda_i$ of the operator $\D$, and we define
  \be\label{uv}\sign(\lambda)=\begin{cases} 1 &  \text{if}~\lambda\geq 0\cr
  -1&\text{if}~\lambda<0. \end{cases}.\ee
 (A different but equivalent regularization was used originally \cite{APS}.)  Thus we
 regularize   $(n_--n_+)/2$ by replacing it with $-\etta/2$, and so we replace
 $(-1)^{n_-}=\exp(\pm \i \pi n_-)$
 with $\exp(\mp \i\pi\etta/2)$.  Finally, the regularized version of the path integral of
 the $\chi$ field is \cite{ADM}
 \be\label{mezzo}Z_\chi =|\neg\det \neg \,\D|\exp(\mp\i\pi\etta/2). \ee

  The parity anomaly is often described in terms of the Chern-Simons function rather than
  the $\etta$-invariant, but this is not quite correct.
   The Atiyah-Patodi-Singer theorem \cite{APS} would let us replace $\exp( \i\pi\etta)$ with
   $\exp(\i\CS)$, where $\CS$ is a Chern-Simons function that is well-defined mod $2\pi$.
   Here $\CS$ is $\CS(A,g)$ of eqn.\ (\ref{corrcs}), which includes
   the gravitational contribution $\Omega(g)$ of eqn.\ (\ref{orelf}).
 We cannot make this sort of replacement in eqn.\ (\ref{mezzo}).  Naively, we would want to
 replace $\exp(\i\pi\etta/2)$ with $\exp(\i\CS/2)$, but as $\CS$ is only well-defined mod
 $2\pi$,
 $\exp(\i\CS/2)$ has an ill-defined sign. The parity anomaly revolves around the overall
 sign of the partition function, so we cannot understand it properly via a formula with an
 ill-defined sign.

 Because $\etta$ is not an integer, the sign $\mp$ in the exponent in eqn.\ (\ref{mezzo})
 matters.  In an approach based on Pauli-Villars regularization, this sign comes
 from the sign of the regulator mass \cite{ADM,WQ}. This sign violates $\sfT$ and $\sfR$
 symmetry, which say that on an orientable manifold the path integral should be real.
  This violation is often loosely called the parity
 anomaly, though it is more precise to call it an anomaly in $\sfT$ and $\sfR$ symmetry.
 This anomaly is a mod 2 effect because the original problem only involved the sign of the
 path integral.  If we had two $\chi$ fields, the path integral would have been
 formally $\det^2\D$, which is naturally
 positive.  In terms of the above derivation, we could define the path integral for one
 $\chi$ field using a $+$ sign in eqn.\ (\ref{mezzo}), and that of the second using a $-$
 sign.  This amounts to using a $\sfT$- and $\sfR$-preserving regularization.
 The phase factors would cancel, leaving us with a real, positive partition function.
 (Alternatively, we could use a $\sfT$- and $\sfR$-violating regularization, such that the
 two $\chi$ fields contribute with the same sign, say $+$.  Then the partition function is
 proportional to $\exp(+\i\pi\etta)$, which violates $\sfT$ and $\sfR$.  But now we can add
 to the bare action a properly normalized Chern-Simons term making the phase
 $\exp(+\i\pi\etta-\i\CS)=1$, where the last step uses the Atiyah-Patodi-Singer theorem.
 The combination of the regularization and this bare term respects $\sfT$ and $\sfR$.)

To avoid a possible confusion, we would like to add a clarifying comment.  The free fermion
theory without the gauge field $A$ has both a global $U(1)$ symmetry and a time-reversal
symmetry $\sfT$.  It has a conserved $U(1)$ current $j_\mu$ and all the flat space
correlation functions of operators at separated points are $U(1)$- and time-reversal
invariant. The problem of the anomaly can be seen as a possible failure of these symmetries
at coincident points; specifically, the product $j_\mu(x) j_\nu(0)$ can include a
$\sfT$-violating contact term proportional to $\epsilon_{\mu\nu\rho}\partial^\rho
\delta(x)$, where $\delta(x)$ is a $2+1$-dimensional delta function.  Such contact terms can
be probed when the theory is coupled to a classical background gauge field $A$.  Even then,
if $A$ is infinitesimal, we can add to the Lagrangian a counterterm proportional to $A\d A$
to cancel the contact term and thus preserve $\sfT$.  The anomaly is the statement that this
cannot be done when $A$ is not infinitesimal and the gauge group is compact.  Then, the
allowed $A\d A$ counterterms must have properly quantized coefficients and cannot cancel
arbitrary fractional contact terms in $j_\mu(x)j_\nu(0)$. The situation here is similar to
the 't Hooft anomaly.  There a system has some global symmetries, but the currents of these
symmetries suffer from contact terms, which prevent us from coupling them to background
gauge fields.

Once the $U(1)$ symmetry is gauged, there  is no way to eliminate the anomaly in $\sfT$ and
$\sfR$ symmetry in a purely three-dimensional context. But
  those symmetries can potentially be restored, as we discuss next,  if the three-manifold
 $W$ on which the $\chi$ field is defined is the boundary of an oriented four-manifold $X$
 over which the $\spinc$ structure of $W$ extends. Physically, $X$ will be the worldvolume
 of a topological insulator.

 \subsubsection{The APS Index Theorem And $\sfT$-Invariance}\label{topins}

The path integral of the topological insulator can be analyzed \cite{MW,WQ}     by using
the APS
 theorem for the index\footnote{As reviewed in section 2.1.8 of \cite{WQ}, this index is
 defined using  on $W
 =\partial X$ the nonlocal APS boundary conditions that were introduced in \cite{APS}.}
  $\I$ of the Dirac operator on $X$:
  \be\label{indexformula}\I= \int_X\left(\h A(R)+\frac{1}{2} \frac{G}{2\pi}\wedge
  \frac{G}{2\pi}\right) -\frac{\etta}{2}. \ee
  Here
  \be\label{bcurv}G=\d B=\d A+2s\d a=F+2s f\ee
  is the curvature of $B=A+2s a$, and $\h A(R)$ is a certain quadratic expression in the
  Riemann tensor $R$ (eqn. (\ref{index})).  If $X$ has no boundary, then there is no
  ${\etta}/2$ term on the right hand side of eqn.
  (\ref{indexformula}), and this formula reduces to the standard Atiyah-Singer formula for
  the index of the Dirac operator on a compact four-dimensional spin manifold  without
  boundary.

   To restore $\sfT$ and $\sfR$ symmetry, we assume the existence on $X$ of the bulk
   ``topological'' couplings suggested by the index formula:\footnote{$I_\top$
   as defined here coincides with $I_{\mathrm {ind}}$ as defined in eqn.
   (\ref{inducedtheta}) of the introduction except for a factor of $\i$ that reflects
   the fact that we are now in Lorentz signature, and some terms that depend on $a$.  The
   following discussion will show that the terms that depend
   on $a$ are equivalent to properly normalized Chern-Simons  couplings that can be
   understood as part of the boundary action, rather than as bulk couplings,
   while the part of $I_\top$ that does not depend on $a$ can be considered to have the
   origin that was described in the introduction.}
  \be\label{itop}I_\top=\mp \i\pi \int_X\left(\h A(R)+\frac{1}{2} \frac{G}{2\pi}\wedge
  \frac{G}{2\pi}\right).\ee
  The combined boundary and bulk path integral, including a factor of $\exp(-I_\top)$ from
  the bulk, is then
  \be\label{combined}Z_\com= |\neg\det\,\D|\exp(\mp \i\pi\etta/2)\exp(-I_\top)= |\det   \neg
  \,\D|(-1)^\I. \ee
  In the last step, we used the APS index formula (\ref{indexformula}).

  Since the combined partition function $|\neg\det \neg \,\D|(-1)^\I$ is real, it respects
  $\sfT$ and $\sfR$ symmetry.  $\sfC$ (and therefore $\sfCT$) symmetry is also
  manifest, since both factors $|\neg\det\,\D|$ and $(-1)^\I$ in eqn.\ (\ref{combined}) are
  invariant under a sign change of $B=A+2sa$.

    Now let us discuss the physical meaning of the bulk couplings that we have had to add.
  We first review the standard boundary state of a topological insulator, so we set $s=0$,
  $B=A$, and $G=F$.   Also, first we consider the case that $X$ is a spin manifold and $A$
  is a $U(1)$ gauge
  field, and then the slightly more subtle case that $X$ is a $\spinc$ manifold with
  $\spinc$ connection $A$.

  If $X$ is spin, the bulk couplings $\pi \h A(R)$ and $(\pi/2) F\wedge F/(2\pi)^2$ are each
  $\sfT$-conserving.  The $F^2$ coupling corresponds to an electromagnetic $\theta$-angle of
  $\pi$.  This
  assertion depends on the fact that $\frac{1}{2}\int F\wedge F/(2\pi)^2$ is an integer on a
  spin manifold, but has no further divisibility properties.  (See Appendix  \ref{norm} for
  this and some further remarks in this paragraph.)  The two $\sfT$- and $\sfR$-conserving
  values of the electromagnetic theta-angle are 0 and $\pi$, and the $F^2$ coupling in
  $I_\top$ means that the topological insulator is the case $\theta=\pi$.     However, on a
  spin manifold, the bulk
  coupling $\pi\h A(R)$ is trivial.  This is because $\int_X \h A(R)$ is an even integer on
  a compact spin manifold $X$ without boundary, so $\pi\int_X \h A(R)$ always vanishes mod
  $2\pi$.  Thus on a spin manifold,
  the gravitational coupling $\pi\int_X \h A(R)$ does not influence the bulk physics, and
  accordingly, the effect of $I_\top$ in bulk can be summarized by saying that the
  electromagnetic
  theta-angle equals $\pi$.  In the $\spinc$ case, the combined gauge and gravitational
  coupling $I_\top$ is $\sfT$-invariant in bulk, but separately the gauge and gravitational
  couplings in $I_\top$ are not
  $\sfT$-invariant.  These matters are discussed in Appendix \ref{norm}.

  \subsubsection{Extension To $s\not=0$}\label{exts}

    Now let us consider the more general case with $s\not=0$ and $G=F+2s f$.  There is
    something non-trivial that we have to check in order to decide if the formula
    (\ref{combined}) is physically sensible.
  The emergent gauge field $a$ and its curvature $f=\d a$ are defined only on the boundary
  $W$, not on the four-manifold $X$.  But the formula (\ref{combined}) involves $G=F+2s f$,
  so it presumes an extension of
  $f$ over $X$.  Such an extension may not exist\footnote{The obstruction is that the first
  Chern class of the line bundle on which $a$ is a connection may not extend over
  $X$.} and if it exists it is not unique.

  To begin with,  let us assume that $a$ and therefore $f=\d a$ can be extended over $X$ and
  show that the choice of this extension does not matter.
   The dependence of $Z_\com$ on the choice of extension  comes from the part of the
   topological
  couplings $I_\top$ that depend on $f$.  This part leads to a factor
  \be\label{thefactor}U_X=\exp\left(\pm\i\pi \int_X\left(2s \frac{F}{2\pi}\wedge
  \frac{f}{2\pi}+2s^2 \frac{f}{2\pi}\wedge\frac{f}{2\pi}\right)\right).\ee
  We need to know whether $U_X$ depends on how $a$ and $f$ were extended over $X$.

  \begin{figure}
 \begin{center}
   \includegraphics[width=3in]{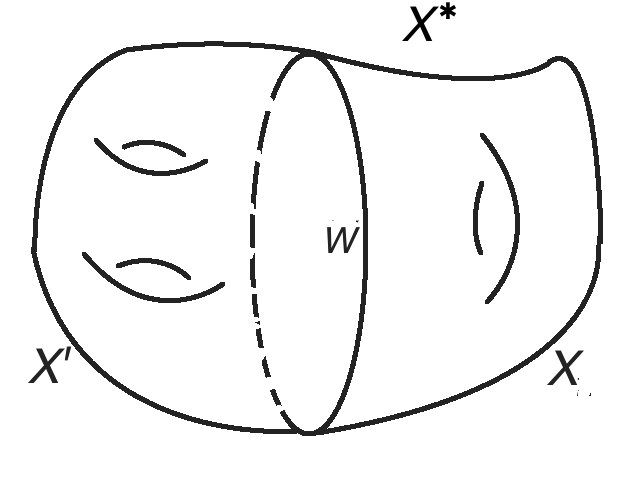}
 \end{center}
\caption{\small  Two manifolds $X$ and $X'$ with the same boundary $W$ are glued together
--
after reversing the orientation of $X'$ so that the orientations
are compatible -- to make a compact oriented manifold $X^*$ without boundary.  (With a view
to later generalizations, $X$ and $X'$ are depicted as being topologically different.) }
 \label{gluing}
\end{figure}

  There is a standard way to answer such a question.  Let $X'$ be another copy of $X$ (or in
  more general applications that we come to
  presently, some other four-manifold with the same boundary
  as $X$) with some possibly different extension of $a$ over $X'$.  By gluing together $X$
  and $X'$ along their common boundary, with a reversal of orientation of $X'$ so that
  their orientations agree along $W$, we make an oriented four-manifold $X^*$ without
  boundary (fig. \ref{gluing}).  The choices of extension of $a$ over $X$ and over $X'$ glue
  together to an extension over $X$.
  We want to know whether $U_X$ as defined in eqn.\ (\ref{thefactor}) is equal to $U_{X'}$,
  defined by the same formula evaluated on $X'$.  The ratio is
  \be\label{theratio}\frac{U_X}{U_{X'}}=  \exp\left(\pm\i\pi \int_{X^*}\left(2s
  \frac{F}{2\pi}\wedge \frac{f}{2\pi}+2s^2
  \frac{f}{2\pi}\wedge\frac{f}{2\pi}\right)\right).\ee
  We want to know whether this always equals 1.

  First suppose that $X^*$ is a spin manifold and $A$ is a $U(1)$ gauge field.  In this
  case, the cohomology classes $F/2\pi$ and $f/2\pi$ are integral,
  so the integrals $\int_{X^*} F\wedge f/(2\pi)^2$ and $\int_X f\wedge f/(2\pi)^2$ are both
  integers.  In the exponent in eqn.\ (\ref{theratio}), these integers multiply
  $2\pi\i s  $ or $2\pi \i s^2$, where $s$ is an integer, so the ratio $U_X/U_{X'}$ is
  indeed always equal to 1.

  If instead $X^*$ is a $\spinc$ manifold and $A$ is a $\spinc$ connection, then matters are
  different, and  we will rediscover the anomaly that was found
  in section \ref{anomaly}.  Indeed, in the $\spinc$ case, in general $F/2\pi$ is a
  half-integral class, so $p=\int_X F\wedge f/(2\pi)^2$ takes values in $\Z/2$.
  To ensure that $U_X/U_{X'}$ is always 1, we now need $\exp(2\pi\i s p)=1$ for all $p\in
  \Z/2$; in other words, we need $s$ to be even. This is the condition
  found in  section \ref{anomaly} for the model to make sense in the $\spinc$ context.

The analysis of the case that $a$ and $f$ cannot be extended over the worldvolume $X$ of the
topological insulator is a little technical, and the following paragraph could be omitted on
first reading. Even if $a$ and $f$ do not extend over $X$,
cobordism theory ensures that there is always some four-manifold $X'$ over which both $A$
and $a$, and therefore $F $ and $f$, as well as the spin structure of $W$,
  can be extended.  If we were studying a topological insulator with worldvolume $X'$ rather
  than $X$, then as in eqn.\ (\ref{combined}) the appropriate partition function would be
  $|\neg\det  \,\D|(-1)^{\I'}$, where now $\I'$ is the Dirac index on $X'$ (with APS
  boundary conditions) coupled to $A+2sa$.
  But of course, we really want to do physics  on $X$, not $X'$.
  What is wrong with the formula   $|\neg\det  \,\D|(-1)^{\I'}$ is that although, as above,
  it does not depend on how $a$ is defined on $X'$, it does depend on the
  unphysical choice of how
  $A$ is defined on $X'$.  To cancel this dependence, we need another factor.  An
  appropriate general formula is
  \be\label{werf}Z_\com=|\neg\det \neg \,\D|(-1)^{\I'(A+2sa)}(-1)^{\I^*(A)}. \ee
  Here $X^*$ is the four-manifold without boundary that is built as in fig. \ref{gluing} by
  gluing $X'$ to $X$, and $\I^*(A)$ (given in eqn.\ (\ref{zindex})) is the index
   on $X^*$ of a Dirac operator coupled to the $\spinc$ connection $A$.
   (The effect of the factor $(-1)^{\I^*(A)}$ is to remove the unphysical dependence on the
   restriction of
   $A$ to $X'$ and replace it with a dependence on the restriction of $A$ to the worldvolume
   $X$ of the topological insulator.  In
   writing eqn.\ (\ref{werf}), we have made explicit that $\I'$ is an index of a Dirac
   operator coupled to $A+2sa$ and $\I^*$ is an
   index of a Dirac operator coupled to $A$.)
   The formula (\ref{werf}) is manifestly real and thus $\sfT$-invariant.  $\sfC$-invariance
   is also manifest. The APS index theorem
  can be used to show that this formula does not depend on the choice of $X'$ or of the
  extensions of $a$ and $A$ over $X'$.

  In all of this, we have treated $\chi$ as a {\it massless} Dirac fermion coupled to the
  gauge field or $\spinc$ connection $A+2sa$.  What happens in the gapped phase
  in which $\chi$ acquires a mass by coupling to the Higgs field $\phi$?
  Since $\phi$ has charges $(2,4s)$,  when it has an expectation value,   $A+2sa$ is reduced
  at low energies to a $\Z_2$ gauge field (more
  precisely, the $\spinc$ structure that it defines reduces to a spin structure).
 $\chi$ then splits up as a pair of Majorana fermions coupled to the same spin structure or
 $\Z_2$ gauge field.
 They get equal and opposite masses $\pm m$ from the coupling to $\phi$, and this
  replaces $|\neg\det \neg \,\D|$ in the formulas for the partition function with the
  strictly positive quantity $\Pf(\D+\i m) \Pf(\D-\i m)=|\Pf(\D+\i m)|^2$ (here $\Pf$
  denotes the Pfaffian).
  The sign of the partition function is still as given in the formulas (\ref{combined}) or
  (\ref{werf}).

  If $B=A+2sa$ is generic, then the eigenvalues of the Dirac operator are generically
  nondegenerate, and the partition function should change sign when an eigenvalue
  passes through zero.  In terms of the formula (\ref{combined}) for the path integral of
  the system,\footnote{Similar remarks hold in the context of the more general formula
  (\ref{werf}).} there is of course no sign change of $|\neg\det\,\D|$ when an
  eigenvalue of $\D$ passes through zero, but when this happens $\etta$ jumps by $\pm 2$,
  causing a sign change in $\exp(\mp\i\pi\etta/2)$.  Likewise, $\I$ jumps by $\pm 1$,
  causing a sign change in $(-1)^\I$.  In the gapped phase, with the gauge group reduced at
  low energies to $\Z_2$, the eigenvalues of $\D$ have even multiplicity
  because of a version of Kramers doubling.  Since there are no gapless fermions, no sign
  change in the path integral is expected when a pair of eigenvalues passes through zero.
  Indeed, when that happens,
   $\etta$ jumps by $\pm 4$, so that $\exp(\mp\i\pi\etta/2)$ is a smoothly varying function,
   and  $\I$ jumps by $\pm 2$, so that $(-1)^\I$ is unchanged and is a topological
  invariant.  The topological invariance of the overall sign $(-1)^\I$ of the path integral
  is essential for the existence of the topological field theory interpretation that
  we construct in section \ref{TFTE}, and the fact that $\exp(-\i\pi\etta/2)$ is smoothly
  varying in the gapped phase is necessary for the relation to the Ising model
  that we discuss in section \ref{ising}.

The gapped phase also has vortices.  In the core of a vortex, the gauge symmetry is
restored.  The vortex represents a singularity in the low energy $\Z_2$ gauge
field or $\spinc$ connection $A+2sa$.  The singularity is associated to a monodromy of
$-1$,
familiar from the spin field of the Ising model, and from the Ramond sector in string
theory.
Analogs of eqns.\ (\ref{combined}) and (\ref{werf}) and of our statements above and also in
section \ref{ising} below hold in the presence of vortices.  But
one has to take into account the existence of a Majorana
zero-mode in the vortex core, and a detailed explanation is somewhat more technical.

  \subsubsection{Relation To The Ising Model}\label{ising}

 We make here a few remarks as preparation for section \ref{TFTE}, in which we will describe
 a topological field theory interpretation of the physics in the gapped phase.

 The combined partition function $Z_\com$ of eqn.\ (\ref{combined}) is the product of two
 phase factors, namely
 \be\label{firstfactor}\Phi_1=\exp(-\i\pi\etta/2)\ee
 (where now we make a choice of sign in the exponent) and
 \be\label{secondfactor}\Phi_2=\exp(-I_\top)=\exp\left( \i\pi \int_X\left(\h
 A(R)+\frac{1}{2} \frac{G}{2\pi}\wedge \frac{G}{2\pi}\right)\right).\ee
 $\Phi_1$ comes from integrating out the boundary fermions, and $\Phi_2$ from a bulk
 coupling.\footnote{Since $a$ and $f$ are only defined on $W$, the
 part of $\Phi_2$ that involves $a$ and $f$ is better written
 as a Chern-Simons coupling on $W$ than as an integral over $X$.  Equivalently, it can be
 written in a roundabout way along the lines of eqn.\ (\ref{werf}).}

 To interpret $\Phi_2$ in topological field theory, we will just write an Abelian
 Chern-Simons theory that reproduces this factor.
 But what is the topological field theory
 interpretation of $\Phi_1$?

 One of the simplest rational conformal field theories (RCFT's)
 in two dimensions is the Ising model.  From a chiral point of view, this is the theory of a
 Majorana-Weyl fermion $\psi$
 of dimension 1/2.  What is the chiral algebra of the Ising model?  There are two possible
 answers.  From one standard point of view, one considers the chiral algebra
 to consist of holomorphic fields of integer dimension.  From this point of view, the chiral
 algebra of the Ising model is simply generated by a stress tensor of central charge
  $c=1/2$.   This chiral algebra has three representations, usually denoted by $1$,
  $\sigma$, and $\psi$, whose dimensions are $0$, ${1\over 16}$, and $1\over 2$
  respectively. Alternatively, we can define a $\Z_2$-graded chiral algebra in which we
  allow holomorphic fields of half-integer spin.  With this definition, the chiral algebra
  of the Ising
  model is simply the $\Z_2$-graded algebra generated by the $\psi$ field itself.  We will
  say that the two points of view correspond respectively to RCFT's and to spin-RCFT's.

  RCFT's in two dimensions are associated to three-dimensional topological quantum field
  theories (TQFT's), and likewise spin-RCFT's are associated to three-dimensional
  spin topological
  field theories (spin-TQFT's).  A TQFT is defined on an oriented three-manifold, and
  similarly a spin-TQFT is defined on an oriented three-manifold with a choice of spin
  structure.  (In both cases, in general there is a framing anomaly.)
The Ising model viewed as an RCFT or as a spin-RCFT is associated to a 3d theory that we
will call the Ising TQFT or spin-TQFT.

Representations of the chiral algebra of an RCFT correspond to line operators -- or more
physically to quasiparticles -- in the
associated TQFT.  So the Ising TQFT has three line operators corresponding to
$1,\sigma,\psi$.
 The analogous statement in the case of a spin-RCFT has a subtlety that
 is explained in Appendix \ref{CSGT}. In this case, each representation of the
 $\Z_2$-graded
 chiral algebra leads to two distinct lines in the $2+1$-dimensional TQFT.  The spins of
 these two lines differ by ${1\over 2}\mod 1$ and each of them is associated with a
 different representation of the subalgebra of the chiral algebra consisting of fields of
 integer
 spin.  In the case of the Ising spin-TQFT, since there is only one representation, there
 are just
 two distinct lines, the trivial line with spin $0$ and  a line $W_\psi$ with spin $1\over
 2$.

  The Ising spin-TQFT will have a partition function on an oriented three-manifold $W$ with
  a choice of spin structure.  As for any TQFT or spin-TQFT, this partition
  function can be computed, starting with simple special cases, by applying a sequence of
  braiding and surgery operations.  However, in the case of the Ising spin-TQFT, these
  braiding and surgery operations all take place in a
  1-dimensional space, since the $\Z_2$-graded chiral algebra generated by $\psi$ has only
  one representation.
    Accordingly, the partition function of the Ising
  spin-TQFT will always be of modulus 1.

To illustrate some of these ideas,
  let us consider the simple example of the $S^2 \times S^1$ partition function of the Ising
  spin-TQFT.  It needs a spin structure around the $S^1$ factor, which we denote in an
  obvious way by $s=\pm 1$.  For $s=+1$ this partition function is $\Tr_{H_{S^2}} (-1)^F$
  and for $s=-1$ it is $\Tr_{H_{S^2}} 1$, where $H_{S^2}$ is the Hilbert space associated
  with $S^2$ (in the absence of any quasiparticles)
  and $(-1)^F$ is the usual operator that distinguishes bosons and fermions.  Since
  $H_{S^2}$ consists of a single bosonic state, the $S^2\times S^1$ partition function is
  equal to one for every $s$.  Now, the theory has a nontrivial line associated with the
  fermion $\psi$.  We can let this line pierce the $S^2$ at a point and wrap the $S^1$
  factor. The only change in the $S^2 \times S^1$ partition function calculation is that now
  $H_{S^2}$ should be replaced by $H_{S^2_\psi}$, the Hilbert space of the sphere pierced by
  the line.  This Hilbert space consists of a single fermionic state.  Therefore, the path
  integral
  in the presence of this line is $-s$.  As expected, the path integral is a phase and it
  depends on the spin structure.

  In general, we claim that the partition function of the Ising spin-TQFT (with no line
  operator insertions)
   is simply $\exp(-\i\pi\etta/2)$, where $\etta$ is the eta-invariant of a Majorana fermion
   on $W$, coupled to the relevant spin structure.
  This statement can be understood as a consequence of the Dai-Freed theorem \cite{DF}.
  Suppose that $W$ has a non-empty boundary $\Sigma$.  Then
  $\exp(-\i\pi\etta/2)$ is not well-defined by itself (its definition fails when the Dirac
  operator on $\Sigma$ acquires a zero-mode),
  but the Dai-Freed theorem says that the product of $\exp(-\i\pi\etta/2)$ with the path
  integral of a chiral fermion $\psi$
  on $\Sigma$ is smooth  and well-defined as a function of the metric on $W$.  This mirrors
  the standard relation between a TQFT or spin-TQFT in three dimensions
  and the corresponding RCFT or spin-RCFT in two dimensions:  $\exp(-\i\pi\etta/2)$ is the
  partition function of the bulk theory associated to the boundary theory of $\psi$.
  So $\exp(-\i\pi\etta/2)$ is the partition function of the Ising spin-TQFT.

  As a check, one can consider the dependence on the metric  of $W$.  If $\etta$ is the
  eta-invariant of a Majorana fermion coupled to a spin structure on $W$,
  then the dependence of $\exp(-\i\pi\etta/2)$ on the metric on $W$ is determined by the APS
  index theorem and corresponds to a framing anomaly with $c=\pm 1/2$ (where
  the sign depends on a choice of orientation),
  as expected for the Ising spin-TQFT.  In fact, the APS index theorem implies that the
  metric dependence of $\exp(-\i\pi\etta/2)$ is the same as that of
  \be\label{samefactor} \Theta=\exp\left(-\i\pi\int_X\h A(R)\right).\ee
  (Like $\exp(-\i\pi\etta/2)$, $\Theta$ really depends only on the metric of $W$, not $X$.)
  We note that $\Theta$ is the inverse of the gravitational factor in $\Phi_2$  (eqn.\
  (\ref{secondfactor})), so $\Theta$ cancels the framing anomaly of the Ising spin-TQFT.
   This cancellation is an inevitable consequence of the fact that the overall partition
   function is real,
  as is manifest in eqns.\ (\ref{combined}) and (\ref{werf}), so it cannot have a net
  framing anomaly.

  So in section \ref{TFTE}, we will interpret the phase factor $\Phi_2$ in terms of an
  Abelian Chern-Simons TQFT, and $\Phi_1$ in terms of the Ising
  spin-TQFT.  This does not mean that the spin-TQFT that governs the gapped phase of our
  models is  simply the product of those two theories.  The two theories are coupled
  because  the spin structure seen by the Ising spin-TQFT
  and entering in the definition of $\Phi_1=\exp(-\i\pi\etta/2)$
  is the one determined by $A+2sa$, which also appears in $\Phi_2$.

 To better understand this coupling, we need to understand how the Ising TQFT is related to
 the Ising spin-TQFT.\footnote{In Appendices \ref{utwo} and \ref{LSTQFT} we discuss the
 relation between these two theories when they are presented as Chern-Simons gauge
 theories.}  The partition
 function of the Ising TQFT is obtained from that of the Ising spin-TQFT by simply summing
 over spin structures.  In other words, the Ising spin-TQFT assigns
 a partition function to an oriented three-manifold $W$ with a choice of spin structure; if
 we sum this partition function over the choice of spin structure,
 we get the partition function of the Ising TQFT.  (This statement is equivalent to the
 statement
  that if the Ising TQFT is quantized on a two-manifold $\Sigma$, its physical states
   have a basis corresponding to spin structures on $\Sigma$.  But the general relation
   between 3d TQFT and 2d RCFT says that the
 physical states of the Ising TQFT are the conformal blocks of the Ising RCFT on the same
 surface.  And the conformal blocks of the Ising TQFT have
 a basis corresponding to spin structures on $\Sigma$.)

So far $a$ and $A$ have been classical background fields, but we still need to sum over the
quantum field $a$.  Since $A+2sa$ couples to $\chi$, which is a fermion, it might seem like
the sum over $a$ will amount to the sum over spin structures, making the spin-TQFT that
results from the sum an ordinary (non-spin) TQFT
that no longer depends on a particular choice of spin structure.  However, we should not
forget the $w$ quanta (and correspondingly the Wilson lines $e^{\i \oint a}$ in the low
energy TQFT).  These mean that the sum over $a$ still leaves dependence on the spin
structure.  One way to see that is to recall that the fundamental electron has the quantum
numbers of $\chi \bar w^{2s}$.  It is $U(1)_a$ invariant, but has spin $1\over 2$.
Therefore, it is sensitive to the spin structure even after the sum over $a$.  We conclude
that the low energy theory must be a spin-TQFT.

In section \ref{TFTE}, when we represent the two factors $\Phi_1$ and $\Phi_2$
 of the partition function  by the Ising spin-TQFT and  by an Abelian theory, we will have
 to take into
 account the fact that  these two factors couple to the same $\Z_2$ gauge field $A+2sa$. One
 way to describe how the two factors are coupled is the following.  We start with two
 decoupled theories and then gauge a common $\Z_2$ one-form symmetry \cite{KS,GKSW}.  We
 will describe that procedure in a concrete way in section \ref{nonabelians}.
 Alternatively, when the TQFT is presented as a Chern-Simons gauge theory, this one-form
 gauging is implemented by modding the gauge group by $\Z_2$.  We will do that in section
 \ref{ChernSimons}.  In our case, both the non-Abelian Ising TQFT and the Abelian TQFT are
 not spin.  But their $\Z_2$ quotient is a spin-TQFT.  In Appendix \ref{CSGT} we will review
 this modding of the gauge group and will demonstrate it in examples.  In particular, we
 will show how it can turn an ordinary TQFT to a spin-TQFT.

  The product $\Phi_1\Phi_2=(-1)^\I$ is $\sfT$-invariant, but the individual factors
  $\Phi_1$ and $\Phi_2$ are not.  An alternative and $\sfT$-conjugate factorization can
  be obtained by complex conjugating both $\Phi_1$ and $\Phi_2$ (or equivalently by
  reversing the sign in eqn.\ (\ref{mezzo})).  The construction that we will make based on
  factoring $(-1)^\I$ as $\Phi_1\Phi_2$ has the  advantage of giving a concise description
  of the low energy physics, but it has
  the drawback of somewhat hiding the  $\sfT$  and $\sfCT$ symmetry.

\section{A Topological Field Theory For The Low Energy Physics}\label{TFTE}

In this section we study the low energy behavior of our model of section \ref{models} and
then we generalize it.
Clearly, since the system is gapped, the low energy description is a topological field
theory.
This theory should exhibit all the symmetries of the system.  These include time-reversal
and charge conjugation
 as well as $U(1)_A$.  Also, the topological theory should exhibit the same anomalies as the
 microscopic theory.
 Finally, since the microscopic theory can be placed on a $\spinc$ manifold, the low energy
 theory should have the same property.

As we will see, it will be convenient to describe the topological field theory as having two
sectors, coupled in a certain way.
This separation into two sectors makes it easy to analyze the model, though it obscures some
of the global symmetries,
in particular $\slT $ and $\slCT $.

One way to understand the two sectors is to look at the partition function (\ref{combined})
and its discussion in
section \ref{ising}. One sector, which we will discuss in section \ref{Abelians}, leads to
$\Phi_2=\exp(-I_\top)=
\exp\left( \i\pi \int_X\left(\h A(R)+\frac{1}{2} \frac{G}{2\pi}\wedge
\frac{G}{2\pi}\right)\right)$ with
$G=F+2sf =\d(A+2sa)$.  The other sector, which we will add in section \ref{nonabelians},
leads to $\Phi_1=\exp(-\i\pi\etta/2)$.

In this section we will describe the TQFT of the system.  We will suppress the gravitational
interactions and will focus on the gauge theory.

\subsection{The Abelian Sector}\label{Abelians}

In the models of section \ref{models}, the
 expectation value of $\phi$ breaks the $U(1)_a$ gauge symmetry to $\Z_{4s}$.  Therefore,
 our low energy theory should include a $\Z_{4s}$ gauge theory.  One way to describe
 $\Z_{4s}$ gauge theory
  is to replace the Higgs mechanism of $\phi$ by a Lagrange multiplier $U(1)$ gauge field
  $c$ with the coupling \cite{MMS,BaS}
\be\label{Zfsg} {1\over 2\pi}c \d(4s a + 2A) ~.\ee
The equation of motion of $c$ states that $4sa+2A$ is a trivial gauge field. Its field
strength $\d(4s a + 2A)$ vanishes and its periods are trivial.  This does not mean that $a$
is a trivial gauge field.  Instead, we can shift $a$ by a $\Z_{4s}$ gauge field and still
satisfy these constraints.  Therefore, the dynamical part of $a$ is a $\Z_{4s}$ gauge
field.\footnote{The Lagrangian of eqn.\ (\ref{Zfsg}) can also be understood by starting with
the microscopic field $\phi=|  \phi|  e^{\i \varphi}$ and considering the effective
Lagrangian of the gauge mode $\varphi$.  The field $c$ is dual to the compact scalar
$\varphi$.}

Next, we integrate out $\chi$. As we said, we are going to separate the expressions in
section \ref{partfn} into $\Phi_1=\exp(-\i\pi\etta/2)$ (which we will discuss in section
\ref{nonabelians}) and
$\Phi_2=\exp(-I_\top)=\exp\left( \i\pi \int_X\left(\h A(R)+\frac{1}{2} \frac{G}{2\pi}\wedge
\frac{G}{2\pi}\right)\right)$. Ignoring the gravitational term $\h A(R)$, $\Phi_2$
corresponds to Abelian Chern-Simons couplings:
\be\label{singchi}{1\over 8\pi}(2sa+  A)\d(2sa+A) = {2s^2 \over 4\pi} a\d a + {s\over 2\pi}
A \d a  +{1\over 8\pi} A\d A + \d(\dots) ~.\ee
As expected, the $A\d A$ term is not properly normalized to be a purely three-dimensional
term.  This reflects the need to attach the system to the $3+1$-dimensional bulk.
The $a\d a$ term deforms the $\Z_{4s}$ gauge theory in eqn.\ (\ref{Zfsg}).  It makes it a
Dijkgraaf-Witten theory \cite{DW} with $k=2s^2$ (see Appendix \ref{zn}). The $a\d A$ term is
properly normalized.  But since $k$ is always even, the model
 satisfies the spin/charge relation only for even $s$.\footnote{\label{anotherway}
 Here we see another perspective on the spin/charge anomaly we discussed in section
 \ref{anomaly}.  When $s$ is odd we can restore the spin/charge relation in the low energy
 theory by adding to it an odd multiple of ${1\over 2\pi}adA$.  One way to do it is to add
 this term to the microscopic theory with $\chi$.  This would restore the relation and would
 allow $A$ to be a $\spinc$ connection.  But since this term violates $\slT $, we do not do
 that.  This is the hallmark of an anomaly.  The classical theory preserves $\slT$ and
 satisfies the spin/charge relation, but the quantum theory cannot have both.  It either
 preserves $\slT$, or satisfies the spin/charge relation.}

Let us analyze the theory based on the sum of equations (\ref{Zfsg}) and (\ref{singchi}):
\be\label{Zfsgs} {1\over 2\pi}c\d(4s a + 2A) +{1\over 8\pi}(2sa+  A)\d(2sa+A)= {1\over
8\pi}(8c+ 2sa+  A)\d(2sa+A) ~.\ee
The observables in this theory are Wilson lines
\be\label{Wilsonl}W_{n_a,n_c}=e^{\i  n_a \oint a + \i n_c \oint c} ~.\ee
They are generated by
\be\label{Wilsong}W_{1,0}=e^{\i  \oint a } \qquad , \qquad W_{0,1}=e^{\i  \oint c }~.\ee
The equations of motion of $a$ and $c$ lead to the relations
\be\label{Wilsonr}\begin{aligned}
&W_{1,0}^{4s}=e^{4\i s \oint a } =e^{-2\i \oint A}\\
& W_{0,1}^{4s}=e^{4\i s \oint c }=1~,\end{aligned}\ee
where we used the fact that $s$ is even.  This means that we can restrict to $n_a,n_c=0,\pm
1,\dots,\pm(2s-1),2s$.  It is also straightforward to find the spins of these lines
\be\label{Wilsons}S_{n_a,n_c} = \left({n_a n_c \over 4s} - { n_c^2 \over 16}\right) \mod 1
~.\ee

\subsection{Adding The Non-Abelian Sector}\label{nonabelians}

The Abelian model of eqn.\ (\ref{Zfsgs}) cannot be the whole story.  It ignores the factor $
\Phi_1=\exp(-\i\pi\etta/2)$ in the partition function and it does not capture the results
about the vortices and the non-Abelian statistics of the quasiparticles that we discussed in
sections \ref{vortex} and \ref{quasi}.

Here we add another sector to account for these issues.  As explained in section
\ref{ising}, this should be a topological theory, which we can refer to as the ``Ising
system.'' It has three line observables $W_1$, $W_\sigma$, and $W_\psi $, whose spins are
$0$, $1\over 16$, and $1\over 2$ respectively.  These lines have the standard non-Abelian
statistics of the Ising model.

As a first attempt in coupling the two systems, we do the following.  The Abelian theory
(\ref{Zfsgs}) is a $\Z_{4s}$ gauge theory.  This means that $2sa+ A$ is a $\Z_2$ gauge
field.  (More precisely, it can be a $\spinc$ connection.)  It is subject to the
Dijkgraaf-Witten term and the coupling to $A$ (\ref{singchi}).  The same $\Z_2$ gauge field
appears in the argument of $\etta$.  So our action is the sum of the Abelian $\Z_{4s}$ gauge
theory and $ -{\pi\over 2}\etta(2sa+ A)$.

Let us first assume that the two factors are decoupled and consider the theory of each of
them separately.  Above we describes the Abelian sector in terms of two $U(1)$ gauge fields
$a$ and $c$ (\ref{Zfsgs}).  We could try to do it for also for $ -{\pi\over 2}\etta(2sa+
A)$, introducing two $U(1)$ gauge fields $x,y$ with ``action''
\be\label{first} -\frac{\pi }{2}\etta(x) + \frac{1}{2\pi} \int 2y \d x~ .\ee
This is not really a sensible action in the usual sense, because $\frac{\pi}{2}\etta(x)$ is
not continuous mod $2\pi$ when its argument is a $U(1)$ rather than  $\Z_2$ gauge field.
However, this ``action'' does lead to a sensible path integral which moreover reproduces the
Ising TQFT, because after performing the integral over $y$, $x$ is constrained
to be a $\Z_2$ gauge field, and then the integral over $x$ reduces to the sum over spin
structures of $\exp(-\i\pi \etta/2)$.

Next, we consider the sum of equations (\ref{Zfsgs}) and (\ref{first})
\be\label{second} -\frac{\pi }{2}\etta(x) + \frac{1}{2\pi} \int 2y \d x + {1\over 8\pi} \int
(8 c + 2sa+ A)\d (2sa+ A)~. \ee
In order to identify $x=2sa+A$, we mod out the $U(1)_c \times U(1)_y$ gauge group by a
$\Z_2$ that acts on the two factors.  The quotient can be chosen so
 that $y$ and $c$ are no longer good gauge fields but $\tilde c = 2c$ and $\tilde y = y+c$
 are good $U(1)$ gauge fields.  In terms of them, the action becomes
\be\label{third} -\frac{\pi }{2}\etta(x) + \frac{1}{2\pi} \int (2\tilde y - \tilde c) \d x +
{1\over 8\pi} \int (4 \tilde c + 2sa+ A)\d (2sa+ A)~. \ee
Now, the equation of motion of $\tilde c$ sets $x=2sa+A $ (up to a gauge transformation) and
we end up with the desired result
\be\label{fourth} -\frac{\pi }{2}\etta(2sa+A)  + {1\over 8\pi} \int (8 \tilde y + 2sa+ A)\d
(2sa+ A), \ee
with the same gauge field $2sa+A$ in both the Ising and Abelian sectors.
In section \ref{ChernSimons} we will present another action for this system in terms of a
Chern-Simons theory of a continuous group without the problem of the first term in
(\ref{fourth}) not being a good action off-shell.  But for now, we take a simple lesson from
this exercise.  The model that we are trying to understand
can be constructed by coupling the Abelian sector to an Ising system and  dividing by a
$\Z_2$ which acts simultaneously on $c$ and on the Ising sector.

The operation of taking a $\Z_2$ quotient of the product of the Abelian sector and the Ising
system can be described in many different ways.
 From the RCFT point of view, taking the quotient
 amounts \cite{MSN} to extending the chiral algebra  by the operator associated with
 $E=W_{2s,0}W_\psi$.  From the TQFT point of view, this can be described as gauging a $\Z_2$
 global one-form symmetry that acts on the two sectors \cite{KS,GKSW}.  From the
 Chern-Simons gauge theory perspective the quotient is obtained \cite{MST} by dividing the
 gauge group by $\Z_2$, as we have done in the above derivation.

Gauging the one-form symmetry amounts to projecting on lines that have trivial braiding with
$E=W_{2s,0}W_\psi$, thus making it a ``transparent line.''  (The line operators
with trivial braiding with $E$ are the ones that in the above derivation can be written in
terms of $\tilde c$ and $\tilde y$ rather than $c$ and $y$.) The line observables of the
system are restricted to be
\be\label{AbeI}\begin{aligned}
W_{n_a,n_c} \quad &n_c ~ {\rm even}\\
W_{n_a,n_c}W_\psi\quad &n_c ~ {\rm even}\\
W_{n_a,n_c}W_\sigma \quad &n_c ~{\rm odd} ,\end{aligned}\ee
with spins
\be\label{AbeIS}\begin{aligned}
 \left({n_a n_c \over 4s} - { n_c^2 \over 16}\right) &\mod 1 \\
 \left({n_a n_c \over 4s} - { n_c^2 \over 16}+{1\over 2} \right) &\mod 1 \\
 \left({n_a n_c \over 4s} - { n_c^2 \over 16}+{1\over 16} \right)& \mod 1,
 \end{aligned}\ee
respectively.

Every one of these lines represents a quasiparticle of our system.  Let us compare this
description
of the quasiparticles to the discussion in section \ref{quasi}.  First, it is clear that the
line $W_{1,0}$ represents the elementary $w$ quanta.  More generally, the integer $n_a$
represents the $U(1)_a$ charge denoted by $\k$.  The relation $W_{1,0}^{4s}= e^{-2\i \oint
A}$ of eqn.\ (\ref{Wilsonr}) means that $\k$ is conserved modulo $4s$.  Also, the classical
object $e^{-2\i \oint A}$  in the right hand side has a simple physical interpretation.  The
field $\phi$ has $U(1)_A\times U(1)_a$ charges $(2,4s)$.  Its condensation means that the
corresponding line $W_{1,0}^{4s} e^{2\i \oint A}$ has trivial correlation functions and can
be replaced by the unit operator.

The fundamental fermion $\chi $ is represented by $ W_\psi$.  The Ising relation
$W_\psi^2=1$ corresponds to $\chi^2=1$ in eqn.\ (\ref{relations}).  It is consistent with
the fact that $\chi^2$ has the same charges as $\phi$, which condensed.  The transparent
line $E=W_{2s,0}W_\psi$, which we used in the $\Z_2$ quotient, is related to the underlying
electron (compare with the relation $\bar w^{2s}\chi=e$ in eqn.\ (\ref{relations})).  Note
that this line cannot be set to 1.  As it represents the electron, it has half-integer spin
and as such it is not trivial.  Yet, its trivial braiding reflects the fact that the
electron can leave the $2+1$-dimensional boundary and move into the $3+1$-dimensional bulk.

Next, we identify the vorticity as $\v=n_c$. A justification of this is that in the theory
(\ref{Zfsgs}), an insertion of the line operator $\exp\left(\i\v\oint  c\right)$ causes a
monodromy of $a$
that would be expected for  vorticity $\v$.  For nonzero values of $\v=n_c$ we recover the
spectrum of quasiparticles above.  For example, the basic vortex corresponds to $W_{0,1}
W_\sigma$, whose spin vanishes (see eqn.\ (\ref{AbeIS})). Two such vortices can fuse in two
possible channels, namely $W_{0,2}$ and  $W_{0,2}W_\psi$,
and therefore we have non-Abelian statistics.  These two lines represent $\v=2$ vortices.
Since the spins of these two lines are $\pm{1\over 4}$, the eigenvalues of the braiding
matrix are $e^{\i \pi(2\cdot 0 \pm {1\over 4})}=e^{\pm \i \pi/4}$, as we found in section
\ref{quasi}.  Similarly, the vorticity $\v=3$ line $W_{0,3}W_\sigma$ has spin ${1\over
2}\mod 1$.  Two such vortices can fuse to the two quasiparticles $W_{0,6}$ and
$W_{0,6}W_\psi$, whose spins are $\pm {1\over 4}$.  Therefore, the eigenvalues of the
braiding matrix of two $\v=3$ vortices are $e^{\i \pi(2\cdot {1\over 2} \pm {1\over
4})}=e^{\pm 3\i \pi/4}$.

We can also add elementary quanta to the vortices.  Adding $\chi$ is represented by
attaching $W_\psi$ to the lines.  For odd $\v$ this is trivial, because of the Ising fusion
rule $\psi\times\sigma=\sigma$.
This is consistent with the statement above and in section \ref{quasi}
that for odd $\v$ there is only a single state.  For even $\v$, this maps
$W_{0,\v}\leftrightarrow W_{0,\v}W_\psi$, i.e.\ it maps between the two states of the
vortex.  We can also add $\k$ $w$ quanta to the vortex by changing $n_a \to n_a+\k$.  This
shifts the spin of the line by $\k\v/4s \mod 1$ (see eqn.\ (\ref{AbeIS})), exactly as
expected.

Next, we would like to identify the action of the unbroken $\Z_{4s}$ gauge theory on the
lines.  Naively, we might expect the $\Z_{4s}$ generator $\K$ to be a one-form global
symmetry and we can attempt to identify it with $e^{\i\oint c}$.  This cannot be right,
because the line $e^{\i\oint c}$ was projected out by the quotient and the line $W_\sigma
e^{\i\oint c}$ cannot be used as a one-form generator, because its fusion is non-Abelian.
Although we can identify $\K^2= e^{2\i\oint c}$ or $\K^2=W_\psi e^{2\i\oint c}$, it seems
that we cannot identify $\K$  in the TQFT.  This should not be too surprising, since we have
seen in section \ref{quasi} that $\K$ can be defined in the even vorticity sectors, but only
even powers of $\K$ can be defined in the odd vorticity sectors.

Finally, let us discuss the action of the discrete symmetries.  The charge conjugation
symmetry $\sfC$ acts as
\be\label{Cmodel} \sfC( a)= -a \qquad,\qquad \sfC( c)= -c\qquad , \qquad \sfC(A)= -A \ee
and it is manifestly a symmetry of our system.  Time-reversal is more subtle.  From the
discussion in section \ref{simple}, it is clear that time-reversal changes the direction of
time but also acts on the gauge fields as\footnote{\label{shorthand}
We use an abbreviated notation for the action time-reversal that was described in section
\ref{actsym}.}
\be\label{Tmodel}\slT(a)= -a \qquad, \qquad \slT( c)=c\qquad,\qquad \slT( A)= - A~. \ee
This is a symmetry of the $c\,\d a$ term in the action, and it is also a symmetry of the
remaining Abelian part of the action together
with the eta-invariant (the product of those terms, or rather of their exponentials, is the
time-reversal invariant expression $(-1)^\I$).

Because of the choice of sign that went into factoring the model in terms of an Abelian
sector times a non-Abelian one,
time-reversal symmetry is not manifest in our presentation.  But the spectrum of
quasiparticles that we have found is clearly $\slT$-invariant.

Given that $s$ should be even,
the simplest TQFT that arises from this construction  has $s=2$.  Precisely this model was
constructed\footnote{If one sets $b=c+a$, then the abelian action
 (\ref{Zfsgs}) for $s=2$ becomes $\frac{1}{4\pi}(8b\d b-8c\d c)+\frac{1}{2\pi}(2A\d b)$,
 which is $U(1)_8\times U(1)_{-8}$ with a specific coupling of $A$, as in \cite{MKF}.} in
 \cite{MKF}, where it was called $T_{96}$.
Using a procedure known as ``anyon condensation,'' these authors also constructed another
model from it, with fewer quasiparticles.  This procedure does not result from any
weakly coupled dynamics.  In section \ref{moregeneral}, we will present a weakly coupled
theory that leads at long distances to that model.  But for the time being we would like to
describe their procedure using another language.

The ``anyon condensation'' procedure amounts to modding out the gauge group by a discrete
subgroup.  In our case, the Abelian sector has a $U(1)_a\times U(1)_c$ gauge symmetry and we
turn it into $\Big(U(1)_a\times U(1)_c\Big)/\Z_2$.  We implement the quotient by expressing
the Lagrangian in terms of the fields $a'=2a$ and $c'=c+a$ and view them as good $U(1)$
gauge fields, while $a$ and $c$ are not.  This turns equation (\ref{Zfsgs}) with $s=2$ into
\be\label{Zfsgst} {1\over 8\pi}(8c'-2a'+  A)\d(2a'+A) ~.\ee
This theory has $k=\begin{pmatrix}0&4\\ 4 &-2\end{pmatrix} $ and $ q=\begin{pmatrix}2\\
0\end{pmatrix} $.  This is a $\Z_4$ gauge theory with $k=-2$ and it satisfies the
spin/charge relation.  This is precisely the minimal
model of \cite{MKF} and \cite{Wangetal}.\footnote{One way of thinking about this quotient is
by identifying a one-form global symmetry of the original theory and then gauging it.  What
we did here is to gauge the global symmetry that shifts $a \to a+ \zeta$ simultaneously with
$c\to c+ \zeta$ with the same $\Z_2$ connection $\zeta$.  The theory also has a $\Z_2$
one-form global symmetry that shifts only $a\to a +\zeta$.  Gauging this symmetry leads to
the theory with $s=1$ in equation (\ref{Zfsgs}), but as we remarked there, this would
violate the spin/charge relation.  The conflict between gauging this $\Z_2$ one-form
symmetry (as opposed to the one that acts both on $a$ and on $c$) and the spin/charge
relation can be interpreted as a mixed 't Hooft anomaly between them.}

\begin{figure}
 \begin{center}
   \includegraphics[width=4in]{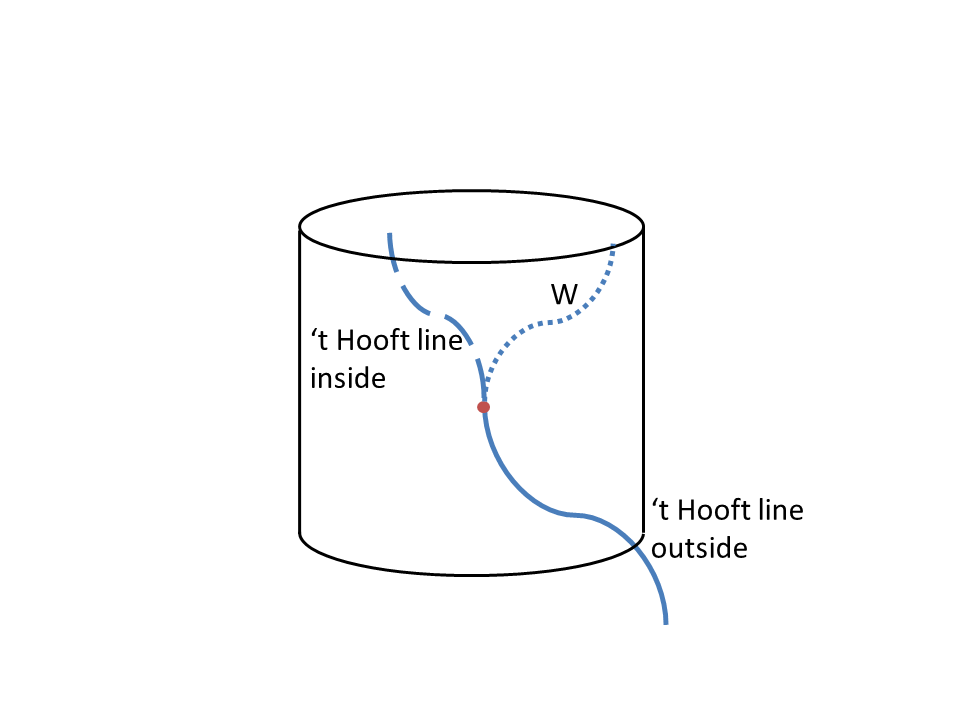}
 \end{center}
\caption{\small A magnetic monopole penetrates a topological insulator.  Its world-line is
represented by an 't Hooft line (the solid line). It penetrates the topological insulator at
the red point.  It continues as an 't Hooft line (the dashed line) inside the material, now
with electric charge $\pm 1/2$,
 and leaves behind a quasiparticle described by the dotted line on the surface.  Here time
 flows upward and we depicted only 2 out of the 3 dimensions of the bulk.  }
 \label{monopoleline}
\end{figure}

We will end this subsection by describing an interesting phenomenon.  As in
\cite{thetapi,FR,Metlitskipi}, let us consider taking a magnetic monopole of unit magnetic
charge
from outside the material through the boundary into the material.  Outside the material, the
monopole has vanishing electric charge; inside the material, because of the Witten effect,
it has electric charge $\pm{1\over 2}$.  It is expected that the charge is compensated by
leaving charge $\mp{1\over 2}$ on the surface of the material.  In a gapless phase, such
electric charge can be deposited on the surface using the charged massless fermions there
(this depends on zero-modes that
the fermions develop in the presence of a unit of magnetic flux).  But how does it happen in
the gapped phase?  The world-line of the magnetic monopole is an 't Hooft line.  It is
characterized by having nonzero flux through an $S^2$ that surrounds it: $\int_{S^2} F =
2\pi$.  This fact has interesting consequences at the point $p$ where the 't Hooft line
penetrates the material (the red dot in figure \ref{monopoleline}).  Momentarily ignoring
possible production of quasiparticles, the equation of motion of $c$ sets on the boundary
$2s\d a+\d A =0$, forcing $\d a$ to be nonzero near $p$.  Consider a Euclidean $S^2$ in the
surface that surrounds the point $p$. Assuming that $2s\d a+\d A=0$ on the  $S^2$, we have
$\int_{S^2} \d a =-{1\over 2s}\int_{S^2}\d A =-{2\pi \over 2 s}$, which is not properly
quantized.  This means that there must be a quasiparticle worldline  intersecting the
surface and
contributing  a singularity $\d a = {2\pi \over 2 s}\delta $, with $\delta$ a delta
function.  With this delta function, the integral $\int_{S^2}\d a$ vanishes.  The
interpretation is that a quasiparticle is emitted from $p$. Its world-line is described by a
line operator that ends at $p$.  More explicitly, this singularity is produced by  a line
operator whose dependence on $c$ is
 a factor of $\exp(-2\i \int c)$; i.e.\ this line operator describes a quasiparticle with
 vorticity $\v=-2$.  We would also like the flux integral $\int_{S^2}\d c$ to vanish.  Using
 the equations of motion, this means that the line $W$ should include a factor of $\exp(-s\i
 \int a)$, i.e.\ the vortex has $s$ quanta of $\bar w$. Since $\bar w$ has electric charge
 $1/2s$, this accounts for the expected deposition in the surface of electric charge $1/2$.
 We conclude that $W=W_{-s,-2}$ in the notation of eqn.\ (\ref{AbeI}); the red dot in figure
 \ref{monopoleline} represents an operator in the TQFT on which such a line can end.  We can
 also discuss the $\sfCT$ image of this picture.  Here the emitted line is $W_{s,-2}
 W_\psi$.  It differs from the previous case by an electron line $E=W_{2s,0}W_\psi$.  This
 electron line continues in the bulk along the monopole line (the dashed line) and changes
 the sign of its electric charge.

\subsection{More General Models}\label{moregeneral}

More general models can be studied in a similar way.  For example, let us
 generalize the  model introduced in section \ref{models}
by adding  pairs of fermions $\chi'_i$ and $\chi''_i$ with $U(1)_A\times U(1)_a$ charges
$(1,2s+n_i) $ and $(1,2s-n_i) $ with $n_i\in \Z$.
Without loss of generality, we take all $n_i> 0$.
These charges have been chosen to be compatible with the spin/charge relation and to allow
$\slT $- and $\slCT $-preserving couplings of the form
\be\label{chitildechi}
\bar\phi \chi'_i\chi''_i +\bar w^{n_i}\bar \chi\chi'_i + w^{n_i}\bar \chi\chi''_i +\mathrm{
h.c}. \ee
(see section \ref{more} for details on this and various statements that follow).
In the phase with  $\langle w\rangle\not=0,$  $\langle \phi\rangle=0$,  all fermions except
one linear combination  are lifted by the  coupling to $w$
 and we remain with the standard boundary state of a topological insulator.  We are
 interested in the other phase with $\langle w\rangle=0$ and $\langle \phi\rangle\not=0$.

As above, let us first analyze the Abelian sector.  Now, eqn.\ (\ref{Zfsgs}) is replaced by
\begin{align}\label{Zfsgsg}\notag
&{1\over 2\pi} c\d(4sa+ 2 A) + {1\over 8\pi}(2sa+  A)\d(2sa+  A)  \\ \notag
&+
{1\over 8\pi} \sum_i \Big[\left((2s+n_i)a + A\right)\d\left((2s+n_i)a + A\right) -
\left((2s-n_i)a + A\right)\d\left((2s-n_i)a + A\right) \Big]\\
&\quad={1\over 8\pi}\Big(8c+2(s+2n)a+A \Big) \d(2sa+  A)+ \d(\dots).\end{align}
On the left hand side, the first term describes the unbroken $\Z_{4s}$ gauge theory, the
second term represents the contribution of $\chi$, and the terms with the sum represent the
contribution of the other fermions.
(We explain in section \ref{more} that the minus sign in the sum is needed if one wants a
formalism without additional
eta-invariants.)
  The right hand side depends only on $n=\sum_i n_i$.  This Abelian topological field theory
  is characterized by a $k$-matrix and charge
vector
\be\label{kmatrix}k=\begin{pmatrix}0&4s\\ 4s &2s(s+2n)\end{pmatrix} \qquad, \qquad
q=\begin{pmatrix}2\\ s+n\end{pmatrix} ~.\ee

We again find a spin/charge anomaly.  The classical system satisfies this relation, but the
quantum theory satisfies it only when $s+n$ is even.

As in the original model with $\chi$ only, we should add the non-Abelian sector associated
with $\etta$, which can be represented by the Ising topological field theory and a $\Z_2$
quotient.

As above, $\sfC$ acts as in eqn.\ (\ref{Cmodel}) and it is a  manifest symmetry of our
system.  As usual, time-reversal is more subtle.  We still take $\slT$ to reverse the time
and to act as $a\to -a$, $A\to -A$, but the action on $c$ should be
\be\label{Taction}\slT:\quad  c\to c +na ~.\ee
To see that this is a symmetry, recall from section \ref{nonabelians} that the Ising sector
together with the Lagrangian of equation ({\ref{Zfsgs}) are invariant under a time-reversal
transformation that acts as $c\to c$.   The action of the more general model
 eqn.\ (\ref{Zfsgsg}) differs from that of the special case (\ref{Zfsgs}) only by an extra
  term proportional to $na$; the term $na$ in eqn.\ (\ref{Taction})
 was chosen to compensate for the transformation under $\slT$ of that term.

In general, different
 microscopic theories can lead to the same macroscopic topological theory.  We have already
 seen that a macroscopic description of this class of models
  depends only on $n$ and not on the individual $n_i$.  Additional identifications can be
  found using the freedom to shift $c$.  Since we need $n+s$ to be even, we can shift $c\to
  c-{n+s\over 2} a$ and write equations (\ref{Zfsgsg}) and (\ref{kmatrix}) as
\be\label{generalAb}\begin{aligned}
&{1\over 8\pi}\Big(8c-2s a+A \Big) \d(2sa+  A)\\
&k=\begin{pmatrix}0&4s\\ 4s &-2s^2\end{pmatrix}, \qquad q=\begin{pmatrix}2\\
0\end{pmatrix}.\end{aligned}\ee
For even $s$, this effective action is equivalent to that of
 equation (\ref{Zfsgs}).  (Shift $c \to c-{s\over 2} a$ in equation (\ref{Zfsgs}) to find
 equation (\ref{generalAb}).)  But here we can find a model also with odd $s$.  In these
 variables the $\slT$ symmetry transformation (\ref{Taction}) becomes
\be\label{Tactiong}\slT(c)= c-sa ~.\ee

Note that a microscopic theory with even $n$ (including the original
 model without $\chi'_i$ and $ \chi''_i$) needs even $s$ and a microscopic theory with odd
 $n$ needs odd $s$.  Yet the resulting macroscopic theories can be described uniformly
 (eqn.\ \ref{generalAb}).

Can there be other models that lead to additional low energy topological field theories?  We
can start with $U(1)_A$ and some other emergent gauge symmetry with some fermions and
scalars.  For simplicity, let us limit the search to models associated with a $U(1)_A\times
U(1)_a$ gauge symmetry (i.e., with only a single emergent gauge field $U(1)_a$) and $\phi$
with charges $(2l,m)$, where the $U(1)_A$ charge of $\phi$  has to be even in order to
preserve the spin/charge relation.  We also allow arbitrary fermions with various charges
such there is no anomaly associated with $U(1)_a$ and the spin/charge relation of $U(1)_A$
is satisfied. The most general Abelian sector resulting
from such a theory can be written as a sum of ${1\over 2\pi} c \d(m a+2lA) $ and a linear
combination of terms of the form $ a\d a $, $ A\d a $, and $ A\d A$.

Next we impose time-reversal.  For that we need to know which $\Z_2$ gauge field the Ising
sector couples to.  So far, it has been $A+2sa$.  Here we take
it to be more generally   $lA+{m\over 2}a$ (and hence $m$ must be even).  The motivation for
this choice is that an expectation value of $\phi$ turns this
combination into a $\Z_2$ gauge field.  Clearly, other options are also possible. We
generalize the action of
$\slT$ to include $c \to c + r a + 2p A$ with $r,p \in \Z$. Using the fact that the Ising
system plus equation (\ref{Zfsgs}) is $\slT$-invariant with $c\to c$, we learn that $m=4s$
with $s\in \Z$, and that  the Lagrangian should be
\be\label{mostgen}
{1\over 8\pi} (8c + 4(ra+2pA)+ 2sa+lA)\d(2s a+lA) ~.\ee

If we want anomalous time-reversal symmetry (which means that the coefficient of $A\d
A/4\pi$ should be a half-integer), then $l$ should be odd.  Using this fact, the spin/charge
relation sets $r=s\mod 2$.  This allows us to redefine $c\to c - {r+s\over 2} a$ and to turn
eqn.\ (\ref{mostgen}) to
\be\label{generalAbb}\begin{aligned}
&{1\over 8\pi} (8c -  2sa+(l+8p)A)\d(2s a+lA) \\
&k=\begin{pmatrix}0&4s\\ 4s &-2s^2\end{pmatrix}, \qquad \qquad q=\begin{pmatrix}2l\\
4sp\end{pmatrix},\qquad \\
&\slT: ~c\to c -sa+2pA.\end{aligned} \ee
By shifting $c$ and $a$ by  integer multiples of $2 A$ and possibly reversing the signs of
$c$ and $a$, we can limit ourselves to
\be\label{plvalues} p=0,1,\qquad l=1,3,\dots,2s-1,~\qquad s>0.\ee

The $A\d A$ interaction is
\be\label{ada}\frac{k_c}{4\pi}A\d A,~~~~ \qquad  k_c=4pl+{1\over 2} l^2.\ee
Since $l$ is odd, it follows that
\be\label{nada} k_c\cong \frac{1}{2}~~{\mathrm{mod}}~4. \ee
This conclusion does not depend on the specific set of fields that we used, since as
 observed in section \ref{spincharge}, $k_c$ is invariant mod 8 if $a$ and $c$ are shifted
 by even integer multiples of $A$.

The congruence (\ref{nada}) can actually be understood based on matters that are explained
in section \ref{spincharge} and Appendices \ref{thetangles} and \ref{csc}.
Let us first consider a conventional topological insulator, whose bulk action, in the
notation of Appendix \ref{thetangles}, has $\theta_1=\pi$ and $\theta_2=0$.
Thus the bulk action is $\pi I_1= \pi \int_X\left(\hat A(R)+\frac{1}{2}\frac{F\wedge
F}{(2\pi)^2}\right)$.  In addition to this bulk action, there may be boundary
contributions in the effective action.
These boundary terms  must be properly normalized three-dimensional Chern-Simons couplings.
The coupling $\pi I_1$ has a metric-dependence because of the $\h A(R)$
term.  However, this metric dependence is just right to cancel the metric dependence of the
term $-\pi\etta/2$ in the effective action; this is ensured by the APS index
theorem. This cancellation is a consequence of time-reversal symmetry (without which the
partition function of a gapped boundary
state can depend on the  boundary metric, via the framing anomaly of 3d TQFT).
Accordingly, any additional boundary contributions to the effective action must be
independent of the metric and can only depend on $A$ and $a$.  In particular,
as explained in section \ref{spincharge} and Appendix \ref{csc}, for a $\spinc$ connection
$A$, a properly normalized Chern-Simons coupling of the form $A\d A$
must be an integer multiple of $(8/4\pi)\int A\d A$; in other words it must contribute to
$k_c$ an integer multiple of $8$.

Based on this, it might seem that $k_c$ should be congruent to $1/2$ mod 8.  However, we
should also consider the possibility described in Appendix \ref{thetangles}
of a material with $\theta_1=\theta_2=\pi$.  Such a material is not a standard topological
insulator, but something more exotic.  The coupling $\pi I_2$ contributes
4 to the effective value of $k_c$.  The boundary Chern-Simons couplings will still
contribute a multiple of 8.  So a material with $\theta_1=\theta_2=\pi$ will have
a boundary state with $k_c\cong 9/2$ mod 8.

This reasoning explains the result found in eqn.\ (\ref{nada}), and also enables us to
understand which of our models are boundary states for a conventional topological
insulator and which are appropriate for an exotic material with $\theta_1=\theta_2=\pi$.
For $(p,l)=(0,\pm1\mod8) $ or $(1, \pm 3 \mod 8)$, we get $k_c\cong 1/2$ mod 8, so these
models correspond to the conventional topological
insulator.    For $(p,l)=(0,\pm 3 \mod 8)$ or $ (1,\pm1\mod8)$, we get $k_c\cong 9/2$ mod 8,
so these correspond  to the more exotic case $(\theta_1,\theta_2)=(\pi,\pi)$.

A microscopic construction of the models with $(p,l)=(0,1)$ was given above.  Other values
of $(p,l)$ can also be found.  For example, rescaling all the $U(1)_A$ charges in the models
above by $l$, we can construct the low energy theory with $p=0$ and arbitrary $l$.  Another
option is to start with $\phi$ and $\chi$ with $U(1)_A\times U(1)_a$ charges $(2l,4 s)$,
$(3l,6s)$ with coupling $\bar\phi^3\chi\chi$.  For $s$ even this leads to the effective
theory (\ref{generalAbb}) with $p=1$ and arbitrary $l$.

We conclude that a large class of microscopic models leads at long distances to
the same theory (\ref{generalAbb}), which is labeled by $s\in \Z$ and the integers $(p,l)$
as in
eqn.\ (\ref{plvalues}).  As we said above, by enlarging the emergent gauge symmetry $U(1)_a$
and by
adding several $\phi$-like fields, many additional models can be constructed.

Finally, we would like to show  that (as remarked at the end of section \ref{nonabelians})
our simplest topological field theory parameterized by
$s=l=1$, $p=0$ describes the minimal model of \cite{MKF} and \cite{Wangetal}.  To see that,
substitute $s=l=1$, $p=0$,
and $a=b + 2c$ in eqn.\ (\ref{generalAbb}) to find\footnote{Since the coefficient for $c$ in
this
redefinition is even, this change of variables does not affect the $\Z_2$ quotient in
section \ref{nonabelians}.}
\be\label{MKFm}\begin{aligned}
&{8\over 4\pi} c\d c -{2\over 4\pi}b\d b +{2\over 2\pi} A\d c +  {1\over 8\pi} A\d A \\
&\slT(b)= -b\\
&\slT(c)= c+b~.\end{aligned} \ee
This is a $U(1)_8\times U(1)_{-2}$ theory with only the $U(1)_8$ field $c$ coupled to $A$,
as in \cite{MKF} and \cite{Wangetal}.
Surprisingly, our simplest microscopic
models with a single $\chi$ field lead to more complicated low-energy theories (or at least
low-energy theories with more quasiparticles), because
they need even $s$.  And we derive the simplest topological theory (\ref{MKFm}) using a more
complicated microscopic model with at least three fermions.

For $s=1$, the model just described is the only one in the family (\ref{generalAbb}) that
obeys the appropriate
congruence conditions for a standard topological insulator. For
$s=2$, there are two possibilities:  $l=1$, $p=0$, or $l=3$, $p=1$.   These models both have
abelian sectors $U(1)_8\times U(1)_{-8}$
(with different couplings to $A$) and as noted in section \ref{nonabelians}, the $l=1$ model
is the model $T_{96}$ of \cite{MKF}.

\subsection{An Explicit Chern-Simons Lagrangian for the Whole System}\label{ChernSimons}

Here we give an explicit local Lagrangian that combines the Abelian and non-Abelian sectors
in the above analysis
and describes the whole system. The Abelian $\Z_{4s}$ sector is represented by the
$U(1)_a\times U(1)_c$ gauge theory (\ref{generalAbb}).
The Ising system can be viewed as a coset theory $SU(2)_2 \over U(1)_4$ \cite{GKO}.  Such  a
coset can be described \cite{MST}
by an $SU(2)_2 \times U(1)_{-4} \over \Z_2$ Chern-Simons gauge theory with action
\be\label{LagCSt}{2 \over 4\pi} \Tr\, (u\d u+{2\over 3} u^3) - {2\over 4\pi}(\Tr\, u
)\d(\Tr\, u) ~,\ee
where $u$ is a gauge field of $U(2)=(SU(2)\times U(1))/\Z_2$.  (See Appendix \ref{CSGT} for
more details.) The overall sign of these terms is related to the orientation of the Ising
system and was set to agree with the discussion of our microscopic models.  Note that even
though eqn.\ (\ref{LagCSt}) describes the chiral Ising model, i.e.\ the theory of a free 2d
fermion, this Chern-Simons theory does not require a spin structure (see
the discussion of eqn.\ (\ref{rela})).
In the language of  section \ref{ising}, it describes the Ising TQFT, not the Ising
spin-TQFT.

The $\Z_2$ quotient of the $U(1)_a\times U(1)_c$ gauge group in eqn.\ (\ref{generalAbb}) and
the $U(2)$ gauge theory in eqn.\ (\ref{LagCSt}) acts on the gauge fields $\Tr\,u$ and $c$.
Instead of considering the quotient with its additional bundles, we can change variables to
\be\label{changev}\begin{aligned}
&\tilde u=u-c {\bf I}\\
&e=(\Tr\, u ) - 2sa -(l-1)A~.\end{aligned}\ee
Here $\tilde u$ and $e$ are standard $U(2)$ and $U(1)$ gauge fields with no additional
quotient.  The shift by the terms linear in $a$ and $2A$ is for convenience.  This turns the
sum of equations (\ref{generalAbb}) and (\ref{LagCSt}) to
\be\label{LagCStrsf}\begin{aligned}
&{2 \over 4\pi} \Tr\, (\tilde u  \d\tilde u   +{2\over 3} \tilde  u^3) - {1\over 4\pi}(\Tr\,
\tilde u   )\d(\Tr\, \tilde u  )-{1\over 4\pi} ( e-A) \d ( e -A) \cr
 &\qquad + {1\over 8\pi}\Big(-4\Tr\, \tilde u +(8p+3l) A +2 sa \Big)\d(2s a
 +lA)~.\end{aligned}\ee

The field $ e$ describes a decoupled trivial $U(1)_{-1}$ theory and we included $A$ in its
coupling to make the spin/charge relation manifest.  It is easy to check that the first two
terms in eqn.\ (\ref{LagCStrsf}) depend only on the traceless part of $\tilde u$; i.e.\ the
field $\Tr\, \tilde u $ does not have a quadratic kinetic term.  This part of the theory can
be viewed as $U(2)_{2,0}$.  (Note that this theory is spin, but the coupling to $A$ is such
that $A$ can be a $\spinc$ connection.)  $\Tr\, \tilde u$ is a Lagrange multiplier
implementing $d(2sa +lA)=0$.  It leaves an unbroken $\Z_{4s}\subset U(1)$, whose gauge field
is $a$.  This $\Z_{4s}$ gauge theory has $k=2s^2$.  Since $ e$ decouples, the whole system
is effectively $\Big(SU(2)_2\times (\Z_{4s})_{k=2s^2}\Big)/\Z_2$.  It is straightforward to
check that this simpler theory leads to the correct spectrum of lines.

\section{Some Additional Topics}\label{switched}

In this section, we describe  a very simple $\sfT$-invariant topological field theory and
provide some more detail on models were described
in section \ref{TFTE}.

\subsection{$U(1)_2\times U(1)_{-1}$}\label{verysimple}

We consider a $U(1)_2\times U(1)_{-1}$ Chern-Simons gauge theory with the Lagrangian
\be\label{uonuon}\L={2\over 4\pi} a\d a -{1\over 4\pi} b \d b ~.\ee
This theory gives an explicit realization of the semion-fermion theory of \cite{Fidkowski},
denoted SF.  It has four non-trivial lines generated by $W_{1,0}=\exp(\i \oint a)$ and
$W_{0,1}=\exp(\i \oint b)$ (see Appendix \ref{CSGT}), whose spins modulo $1$ are $1\over 4$
and $1\over2$, respectively. In the language of \cite{Fidkowski}, $W_{1,0}$ is the semion
$s$ and $W_{0,1}$
is the fermion $f$.  The lines satisfy $W_{1,0}^2=W_{0,1}^2=1$.  It is easy to check that
this theory is $\sfT$-invariant under
\be\label{uoneuonet} \begin{aligned}
\sfT( a) &= a-b\\
\sfT( b )&= -b+2a ~.\end{aligned} \ee
The action on the lines is
\be\label{timea} \sfT:\quad W_{1,0}   \longleftrightarrow W_{1,0} W_{0,1} ~,\ee
which is consistent with the reversal of the spin under $\sfT$.  (Recall that in a TQFT only
the spin modulo $1$ is meaningful.)  In the semion-fermion language, eqn. (\ref{timea})
means that $\sfT$
exchanges $s$ and $sf$.

The square of the transformation (\ref{uoneuonet}) is the charge conjugation transformation
$a\to -a$ and $b\to -b$.  In that sense, this is not like ordinary time-reversal.  But
on-shell,
since $a \sim -a$ and $b \sim -b$, this fact is insignificant and $\sfT^2=1$.  (The fact
that $\sfT^2\not=1$ off-shell means that we do not know how to formulate this theory on an
unorientable three-manifold, which raises the possibility that there might be an anomaly in
doing so.  See section \ref{varying}.)

We  couple the system to a background $U(1)_A$ gauge field $A$ via
\be\label{uonuonA}\L={2\over 4\pi} a\d a -{1\over 4\pi} b \d b +{1\over 2\pi} A\d a ~.\ee
Then $W_{1,0}$ represents the world line of a quasiparticle with $U(1)_A$ charge $1\over 2$.
With nonzero $A$, we replace  the transformations (\ref{uoneuonet}) by
{\be\label{uoneuonetA} \begin{aligned}
\sfT( a)& = a-b +A \\
\sfT( b)&= -b+2a +A~.\end{aligned} \ee
We combine this with the usual $\sfT(A)=-A$.
Under this time-reversal transformation, the theory based on (\ref{uonuonA}) is invariant,
provided we add to the Lagrangian ${1\over 8\pi} A\d A$.  As in our previous examples, we
see that this theory is $\sfT$-invariant with the same anomaly as a topological insulator;
it is $\sfT$-invariant when coupled to a $3+1$-dimensional bulk with $\theta=\pi$.

Both the coupling of $A$ in the Lagrangian (\ref{uonuonA}) and the transformation laws
(\ref{uoneuonetA}) are incompatible with the spin/charge relation.  Therefore, the theory
needs a spin structure. $A$ must be a $U(1)$ gauge field rather than a $\spinc$ connection.
More generally, the
original theory (\ref{uonuon}) needs a spin structure.  It can be coupled to a $\spinc$
connection $A$ by adding to it ${1\over 2\pi } A\d b$ (rather than
${1\over 2\pi}A\d a$, as in eqn. (\ref{uonuonA})), but then the $\sfT$ transformation
(\ref{uoneuonet}) cannot be extended to nonzero $A$.

To explore this theory further, let us set $A=0$ and place it on a spatial torus $T^2$ and
choose a basis of one-cycles labeled by $a$, $b$ and a spin structure labeled in an obvious
way by $s_{a,b}= \pm 1$.  The Hilbert space $\H$ of our theory is two-dimensional, since
$U(1)_2$ has two states on $T^2$ and $U(1)_{-1}$ (for
any spin structure) has only one.  A line operator wrapping a 1-cycle in $T^2$ gives an
operator that acts on $\H$, and in a suitable basis
we have
\be\label{linemat}\begin{aligned}
&W_{1,0}^a=\begin{pmatrix}1& 0 \\ 0&-1 \end{pmatrix}\qquad\qquad\quad
W_{1,0}^b=\begin{pmatrix}0&1\\ 1&0 \end{pmatrix} \\
&W_{0,1}^a=-s_a \begin{pmatrix}1&0\cr 0&1\end{pmatrix} \qquad\qquad
W_{0,1}^b=-s_b \begin{pmatrix}1&0\cr 0&1\end{pmatrix} ~.\end{aligned}\ee  This action of
$W_{1,0}$ is standard and follows (up to a choice of basis)
from the fact that $W_{1,0}^2=1$ and $\{W_{1,0}^a,W_{1,0}^b\}=0$.
The form of $W_{0,1}$ around the two cycles was found by recalling that $W_{0,1}$ is the
worldline of a spinor and  is otherwise trivial.

Time-reversal should act as a $2\times 2$ matrix ${\cal T}$ combined with complex
conjugation.  Because of eqn.\ (\ref{timea}), we should have
\be\label{Trel}\begin{aligned}
&{\cal T}^{-1} {\cal T}^*  =1\\
&{\cal T}^{-1} W_{1,0}^a {\cal T}^*=-s_a W_{1,0}^a\\
&{\cal T}^{-1} W_{1,0}^b {\cal T}^*=-s_b W_{1,0}^b ~.
\end{aligned}\ee
We solve these (up to an overall sign)
by
\be\label{Tsol}
{\cal T}=\begin{cases} \begin{pmatrix}1&0\\ 0&-s_b\end{pmatrix} &\mbox{for  }  s_a=-1 \\
\begin{pmatrix}0&-s_b\\ 1&0\end{pmatrix} & \mbox{for  } s_a=1~. \end{cases}\ee
This satisfies
\be\label{Tsqu} {\cal T}^2=\begin{cases} {\bf 1} & \mbox{for  } s_a=-1 \\
-s_b {\bf 1} & \mbox{for  } s_a=1\end{cases}=\begin{cases}{\bf 1} & \mbox{for even spin
structures} \\
- {\bf 1} & \mbox{for odd spin structure} ~.\end{cases}\ee
We conclude that the two states are Kramers singlets in the case of an even spin structure
(either $s_a$, or $s_b$, or both are $-1$),
but form a Kramers doublet for the odd spin structure ($s_a=s_b=1$).

Despite the difficulty with the spin/charge relation, this model has various potential
applications.
 In eqn. (\ref{MKFm}), we have already encountered $U(1)_2$ (with opposite orientation and
 without the coupling to $A$) as a factor in a $\sfT$-invariant
system that satisfies the spin/charge relation.   The semion-fermion theory SF has been used
in \cite{Fidkowski} in studying topological superconductors and we will encounter it in that
guise in section
\ref{prelims}.
As another possible application, by multiplying the action (\ref{uonuonA}) by 2 (or by any
even integer), we get a $\sfT$-invariant theory that no longer
needs a spin structure and  might be a boundary state for
the bosonic topological insulators of \cite{VS}.  (After rescaling the action, $\sfT^2$ is
still the operation $(a,b)\to (-a,-b)$, but this
is now a nontrivial global symmetry.)

\subsection{More On Some Models From Section \ref{TFTE}}\label{more}

In section \ref{moregeneral}, we described models that can be constructed by adding
additional pairs of Dirac fermion fields to the basic model of section \ref{models}.
Here we will supply some more detail.  For brevity we consider the case of adding to the
original $\chi$ field of $U(1)_A\times U(1)_a$ charges $(1,2s)$
a single additional pair $\chi'$ and $\chi''$ of charges $(1,2s+n)$ and $(1,2s-n)$.

The first question that we want to address is the $\sfT$ and $\sfCT$
transformations\footnote{We study these symmetries only at the classical
level and do not analyze quantum anomalies on an unorientable manifold.} of $\chi$, $\chi'$,
and $\chi''$ and the Yukawa couplings that make possible
both Higgsing to the standard gapless boundary state and also to the gapped boundary state
described in section \ref{moregeneral}.

In general, if $\hat\chi$ is
 a charged Dirac fermion  coupled to $A$ and/or $a$, then given that $A$ and $a$ transform
 under $\sfT$ as in eqn.\ (\ref{waction}), the transformation of $\hat\chi$
under $\sfT$ is essentially uniquely determined.
When $\hat\chi$ is expanded as a sum of two Majorana fermions, one will have to transform
under $\sfT$ with a factor of $+\g_0$ and
one with a factor of $-\g_0$, as in eqn.\ (\ref{relsign});
after rotating to the right basis, the transformation of $\hat\chi$ under $\sfT$ will then
be as in eqn.\ (\ref{easysign}).
By contrast, the two Majorana components of $\hat\chi$ will transform the same way under
$\sfCT$, and there is a nontrivial choice of whether they will both transform as $+\g_0$ (as
in eqn.\ (\ref{ctacts})) or as $-\g_0$.  This choice will determine whether $\hat\chi$
contributes
$+2$ or $-2$ to $\nu_\sfCT$.  The contribution of $\hat\chi$ to $\nu_\sfT$ is always 0.

We would like to choose the signs in the $\sfCT$ transformations of $\chi,\chi'$, and
$\chi''$ so that Higgsing to the conventional boundary
state of a topological insulator will be possible.  This means that the signs must be such
that the net contribution to $\nu_\sfCT $ will be $+2$.
One might think that the ``new'' fields $\chi'$ and $\chi''$ should make canceling
contributions to $\nu_\sfCT$, and that the $+2$ comes from the
original $\chi$ field.  It turns out, however, that to make possible the Higgsing that we
want, $\chi$ should contribute $-2$ and $\chi'$, $\chi''$ contribute
$+2$ each.

In one phase, the scalar field $w$ of charges $(0,1)$ has an expectation value, breaking
$U(1)_A\times U(1)_a$ to $U(1)_A$.  After this breaking,
$\chi$, $\chi'$, and $\chi''$ are all simply Dirac fermions of ordinary electric charge 1.
To reduce to the standard gapless boundary state of a topological
insulator, we want $w$ to couple to these fermions in such a way that two linear
combinations become massive, leaving only one massless Dirac fermion.
In general, let $\chi_1$ and $\chi_2$ be Dirac fermions of charges $(1,p_1)$ and $(1,p_2)$,
transforming in the standard way under $\sfT$.  For definiteness
suppose that $p_1\geq p_2$.  Then in the phase with $\langle w\rangle\not=0$, $\chi_1$ and
$\chi_2$ can get a mass from a $\sfT$-invariant coupling
\be\label{zebbo}\int\d^3x\sqrt  g\left(w^{p_1-p_2}\bar\chi_1\chi_2  -  \bar
w^{p_1-p_2}\bar\chi_2\chi_1    \right).\ee
This is hermitian,\footnote{The difference from the situation described in footnote
\ref{zeldo} is that as $\chi_1$ and $\chi_2$
are assumed to transform with opposite signs under $\sfCT$, this expression is
$\sfCT$-conserving. }  and each term is separately $\sfT$-invariant.  The two terms are
exchanged by $\sfCT$, which is a symmetry if and only if $\chi_1$ and $\chi_2$
transform with opposite signs under $\sfCT$.

In the phase with $\langle\phi\rangle\not=0$, we want all fermions to get masses by Yukawa
couplings to $\phi$.  The only such couplings allowed
by $U(1)_A\times U(1)_a$ gauge symmetry are $\phi\chi\chi+\mathrm{h.c.}$ and
$\phi\chi'\chi'' +\mathrm{h.c.}$  We have already discussed the
$\sfT$, $\sfC$, and therefore also $\sfCT$ invariance of the former coupling in discussing
eqn.\ (3.8).   The same reasoning applies to  a similar coupling
$\phi\chi'\chi''+\mathrm{h.c.}$ and shows that it is $\sfCT$-invariant if and only if
$\chi'$ and $\chi''$ transform the same way under $\sfCT$.

So we must choose $\chi'$ and $\chi''$ to each contribute $+2$ to $\nu_\sfCT$, and hence
$\chi$ to contribute $-2$.  The allowed couplings involving
$w$ are, for $n>0$, $\bar w^n\bar\chi\chi'+\mathrm{h.c.}$ and
$w^n\bar\chi\chi''+\mathrm{h.c.}$   Such couplings yield a mass matrix of rank 2,
leaving one massless Dirac fermion in the phase with $\langle w\rangle\not=0$.

Next, we would like to justify more precisely the treatment of the model that was given in
section \ref{moregeneral}.  We recall from section
\ref{partfn} that the partition function of the original $\chi$ field of charges $(1,2s)$
can be factored as
\be\label{zelb} \exp\left(\pm \frac{\i\pi}{2}\etta(B)\right)\exp(\mp \i I_\top(B)), \ee
for a certain topological action $I_\top$.  Here $\etta$ is the eta-invariant of a Dirac
operator coupled to $B=A+2s$ and likewise  $I_\top$ is a function of $B$.
 Regardless of which sign we choose in the factorization of eqn.\ (\ref{zelb}), we get a
 description in terms of an Ising sector, coming from the
 factor $\exp(\pm \i\pi\etta(B)/2)$, coupled to an Abelian Chern-Simons theory, coming from
 $\exp(\mp\i I_\top)$.

 The partition function of the fields $\chi'$ and $\chi''$ can likewise be factored as
 \be\label{elb} \exp\left(\pm \frac{\i\pi}{2}\etta(B+na)\right)\exp(\mp \i I_\top(B+na))
 \ee
 and
 \be\label{welb} \exp\left(\pm \frac{\i\pi}{2}\etta(B-na)\right)\exp(\mp \i I_\top(B-na)).
 \ee
 Now we observe that at low energies, $B$ is a $\Z_2$ gauge field, so that $B$ is gauge
 equivalent to $-B$ and $B+na$ is gauge equivalent to $-(B-na)$.  But $\etta$
 is invariant under charge conjugation, so $\etta(B+na)=\etta(B-na)$.  Accordingly, if we
 choose a $+$ sign in eqn.\ (\ref{elb})
 and a $-$ sign in eqn.\ (\ref{welb}), or vice-versa, then the $\etta$-invariants will
 cancel, and we can write the partition function of the
 combined $\chi'$-$\chi''$ system simply as
 \be\label{telb}\exp\biggl(\pm\i\bigl(I_\top(B+na)-I_\top(B-na)\bigr)\biggr). \ee
 This description was used in section \ref{moregeneral}.

 Finally, we will discuss the quantization of the monopole operators, to recover the result
 found in section \ref{moregeneral}:
 the model obeys the spin/charge relation if and only if $s+n$ is even.   The discussion of
 eqns. (\ref{zendo})-(\ref{plendo}) is a useful starting point.
 There we found that in quantization on $S^2$ in the presence of a single flux quantum of
 $f=\d a$, the original $\chi$ field has a state $|\neg\neg\uparrow\rangle$
 of spin 0 and charges $(-s,-2s^2)$.  The fields $\chi'$ and $\chi''$ have analogous states
 $|\neg\neg\uparrow'\rangle $ and $|\neg\neg\uparrow''\rangle$
 of respective charges $(-(s+n/2),-2(s+n/2)^2)$ and $(-(s-n/2),-2(s-n/2)^2)$.
 Adding these values, the combined system in the presence of one unit of flux has a state
 $|\neg\uparrow\uparrow'\uparrow''\rangle$
 of charges $(-3s,-6s^2-n^2)$.  Starting from this state, there are many
 ways to construct a state that is invariant under the emergent gauge symmetry $U(1)_a$, and
 hence will correspond to a gauge-invariant monopole operator.
 One simple choice (not the choice that we made in section \ref{anomaly}) is to act with
 $6s^2+n^2$ powers of $w$, which has charges $(0,1)$.  Here
 we have to remember that $w$ has half-integral spin in the field of a magnetic monopole of
 $U(1)_a$.  Hence the state $w^{6s^2+n^2}|\neg\uparrow\uparrow'\uparrow''\rangle$ has
 electric charge $-3s$ and has spin $(6s^2+n^2)/2$ mod 1, or equivalently $n/2$ mod 1.  This
 state obeys the spin/charge relation if and only
 if $s+n$ is even.  If this one gauge-invariant state obeys the spin/charge relation, then
 so do all
 gauge-invariant states in this sector, since they can be obtained by acting on this
 state by a product of gauge-invariant products of elementary fields.

\section{Application To Topological Superconductors}\label{app}

\subsection{Preliminaries }\label{prelims}

With a few simple twists, the models constructed in this paper can be interpreted as gapped
symmetry-preserving boundary states of a topological superconductor.

The main difference, for our purposes,  between a superconductor and an insulator is that
pairing occurs in the superconductor, reducing the gauge group of electromagnetism
from $U(1)$ to $\Z_2$.  This means that in a superconductor, the electromagnetic gauge field
$A$ is gauge-equivalent to $-A$.  Accordingly, there is no natural role for
a symmetry $\sfC$ that reverses the sign of $A$, and no natural {\it a priori} distinction
between the time-reversal symmetries that we have called $\sfT$ and $\sfCT$.

Apart from the residual $\Z_2$ of electromagnetism, the important symmetry in a topological
superconductor is time-reversal.   As in our discussion of the
parity anomaly in section \ref{revparity}, here the flat space theory is $\sfT$-invariant,
but there is an anomaly after coupling to a background field.
In this case the background field is the metric and the anomaly is visible only when the
theory is placed on a non-orientable manifold.}

 In adapting the models of this paper
to a topological superconductor, either $\sfT$ or $\sfCT$ can play the role of
time-reversal.  However, the more interesting case is that the time-reversal symmetry of
the topological superconductor is identified with what until now has been called $\sfCT$.
The reason for this is simply that the models of this paper have $\nu_\sfT=0$,
which means that if they are interpreted as boundary states of a superconductor with $\sfT$
as the time-reversal symmetry, then this is a topologically trivial superconductor.
However, the same models have $\nu_\sfCT=2$, meaning that if they are interpreted as
boundary states of a superconductor with $\sfCT$ as the time-reversal symmetry,
then this is a topologically non-trivial superconductor with $\nu_\sc=2$.

Hence, in applications to a topological superconductor, we will take $\sfCT$ as the
time-reversal transformation.  However, the name $\sfCT$ is rather misleading
in this context, since there is no natural notion of $\sfC$ in a superconductor, so we will
rename $\sfCT$ as $\sfT_\sc$.  (It might cause too much confusion with the
rest of the paper if we rename $\sfCT$ as $\sfT$ in this section only.)

In our analysis so far of boundary states of topological insulators, the spin/charge
relation of condensed matter physics has provided an important constraint.
By contrast, for models of the class considered in this paper, there is no such constraint
for a topological superconductor.  We will give two explanations of this fact.

First of all, in any of our  models of topological insulator  boundary states that violate
the spin/charge relation, that relation could be restored (as discussed
in footnote \ref{anotherway}) by adding an
additional coupling
\be\label{addcoup}\int A\wedge \frac{\d a}{2\pi}. \ee
This coupling shifts by 1 the electric charge of any state with a unit flux of $\d a/2\pi$,
so it restores the spin/charge relation, if that relation is not otherwise satisfied.
But the coupling (\ref{addcoup}) is $\sfT$-violating, so it is forbidden in the context of a
topological insulator.  However, in a superconductor, since $A$ is gauge-equivalent
to $-A$, the coupling of eqn.\ (\ref{addcoup}) is actually $\sfT_\sc$-conserving.  By adding
this coupling, if necessary, one can eliminate any problem in the
spin/charge relation.

For a second explanation, consider a spin manifold with some chosen spin structure and with
a corresponding
spin connection $D_0$.  Let $A$ be a $\Z_2$ gauge field.   Then $D_0+A$ is the spin
connection of some
other spin structure, which we will call the effective spin structure.  So any boundary
theory that makes sense on a general spin manifold
can be interpreted as a boundary state of a superconductor, simply by coupling it to the
effective spin structure.   The spin/charge relation is manifestly
obeyed, since the boundary degrees of freedom couple only to the effective spin structure.

Thus, in the original model with a single $\chi$ field, for applications to a topological
superconductor, we can  consider odd as well as even values of $s$.   For a minimal
example, we return to eqn. (\ref{Zfsgs}), where now we set $s=1$ and omit the coupling to
$A$ (since $A$ is now absorbed in the effective
spin structure that is used in defining the Chern-Simons action and the $\etta$-invariant).
Setting $b=a+2c$, we can rewrite (\ref{Zfsgs}) as
\be\label{rewrite}\frac{1}{4\pi}\left(-8c\d c+2b\d b\right),\ee
corresponding to $U(1)_{-8}\times U(1)_2$.  Including the Ising sector, the model becomes
$(U(1)_{-8}\times \mathrm{Ising})/\Z_2\times U(1)_2$.

The model   $(U(1)_{-8}\times \mathrm{Ising})/\Z_2$ is known as T-Pfaffian
\cite{TPfaffian,TPfaffiantwo}.
The product theory $(U(1)_{-8}\times \mathrm{Ising})/\Z_2\times U(1)_2$
is\footnote{\label{remark} In section \ref{verysimple}, we described SF as $U(1)_2\times
U(1)_{-1}$.
The role of $U(1)_{-1}$ was to convert $U(1)_2$ into a spin theory (thus providing the
``transparent'' fermion of the semion-fermion theory SF)
while also canceling the framing anomaly of $U(1)_2$.  In the present context, T-Pfaffian is
already a spin theory, so we do not need a $U(1)_{-1}$ factor
to provide a transparent fermion, and moreover the microscopic construction with $\chi$
makes manifest that the model has no framing anomaly.  However, the fact
that we are getting the transparent fermion of SF from T-Pfaffian means that it may be
slightly oversimplified to identify our theory $(U(1)_{-8}\times\mathrm{Ising})/\Z_2\times
U(1)_2$ as T-Pfaffian$\times$SF.}     T-Pfaffian$\times$SF in the language of
\cite{Fidkowski}.

\subsection{Varying $\nu_\sc$}\label{varying}

In short, all models so far studied in this paper, including models that violate the
spin/charge relation in the context of a topological insulator, can be interpreted
as boundary states for a topological superconductor with $\nu_\sc=2$.

We can easily make boundary states with any even value of $\nu_\sc$ just by taking tensor
products of a number of  $\nu_\sc=2$ models.
(The methods of this paper do not suffice to make boundary states of a topological
superconductor with odd $\nu_\sc$.)

We instead will consider the following simple variant of that idea.  We generalize the
models of section \ref{models} with a single $\chi$ field simply
by introducing $r$ identical fermion fields $\chi_i$, $i=1,\dots,r$,
with exactly the same charges $(1,2s)$ under $U(1)_A\times U(1)_a$ and exactly the same
transformation under $\sfT$ and $\sfCT$ as
in section \ref{models}.  This will give a model with $\nu_\sc=2r$.
We assume that the $\chi_i$ all have the same coupling to $\phi$ as in the familiar model
with $r=1$:
\be\label{Yuka}I_{\Yuk}=\sum_{i=1}^r\int\d^3x\sqrt g\,
h\left(\i\epsilon_{ab}\chi_i^a\chi_i^b\bar\phi+
\i\epsilon_{ab}\chi_i^{\dagger\,a}\chi_i^{\dagger\,b} \phi\right).\ee
Then in the phase with $\langle\phi\rangle\not=0$, the model is gapped.   (This model is not
the same as the tensor product of $r$ copies of the original
model, because the various $\chi_i$ all couple to the same $a$, $w$, and $\phi$.)

Our main interest is to apply this model to a topological superconductor (in which case we
care about only $\sfT_\sc=\sfCT$ and not $\sfT$),
but if $r$ is odd, the model also provides a possible boundary state for a topological
insulator with $\theta=\pi$.  So we will start the analysis
keeping track of all symmetries, and only later concentrate on the application to the
topological superconductor.

For any $r$, the partition function of this model, on an orientable manifold, for a given
choice of the $\Z_2$ gauge field $2sa+A$,
 can be computed without further ado.  It is simply the $r^{th}$ power of what we computed
 in
eqn.\ (\ref{combined}) (or its generalization (\ref{werf})).  In the gapped phase,  the
partition function with a single $\chi$ field is simply
$(-1)^\I$, so if the number of  $\chi$ fields is $r$, the partition function  is
$(-1)^{r\I}$.  This depends only on the value of $r$ mod 2,
strongly suggesting that the topological field theory describing the model depends on $r$
only mod $2$
 at least if we restrict to orientable manifolds.  For odd $r$, this topological field
 theory should thus be the one that we studied in section \ref{TFTE}; for even $r$, since
 the partition function is identically 1,
 the topological field theory is simply (on orientable manifolds) a trivial version of
 $\Z_{4s}$ gauge theory with no Dijkgraaf-Witten term.

There is actually a simple way to prove that the topological field theory derived from this
model only depends on the value of $r$ mod 2.
While maintaining $\sfT$-invariance, it is possible to add a bare mass term for two
of the $\chi_i$ that preserves all symmetries except $\sfCT$.  For example, we can take
\be\label{baremass}I_{\mathrm{bare}}=m\int\d^3x\sqrt g
\,\left(\bar\chi_{r-1}\chi_r-\bar\chi_r\chi_{r-1}\right), \ee
with $m$ real.\footnote{\label{zeldo} Reality of $m$ and the relative minus sign between the
two terms are needed
 for $\sfT$-invariance and hermiticity.   With our definitions, if $\chi$ is a Majorana
 fermion
then $\bar\chi\chi=\epsilon_{ab}\chi^a\chi^b$ is $\sfT$-invariant but antihermitian.  The
relative minus sign in eqn.\ (\ref{baremass})
makes $I_{\mathrm{bare}}$ hermitian but $\sfCT$-violating.}
Once  the system is gapped  for $\langle \phi\rangle\not=0$ using the Yukawa couplings
(\ref{Yuka}), it remains gapped if the bare
mass term $I_{\mathrm{bare}}$ is turned on. This is true for all $m$.   For $m\to\infty$,
$\chi_{r-1}$ and $\chi_r$ decouple and we get the same model again
but with $r$ replaced by $r-2$.    But in general, the topological field theory that
describes a gapped system at low energies is invariant under
deformation of parameters as long as the system remains gapped.
Hence, as suggested by the discussion of the partition function in the last paragraph, the
low energy
topological field theory is the same for any odd number $r$ of $\chi$ fields.

Since the trajectory in theory space by which we interpolated between $r$ and $r-2$ is
$\sfCT$-violating, it is natural
to suspect that the theories with different values of $r$ are in different universality
classes and can actually be distinguished by
some observables that are sensitive to $\sfCT$ symmetry.  Such an observable is actually the
number of fermion zero-modes mod 8 in a vortex
field, which determines subtle properties of the quasiparticles.  In a basic $\v=1$ vortex,
each of the $\chi_i$ has a single Majorana
zero-mode, making a total of $r$ zero-modes, all transforming the same way under $\sfCT$.
In the absence of $\sfCT$ symmetry, generic
perturbations can add to the Hamiltonian
a term bilinear in the zero-modes, lifting the  zero-modes
in pairs.  So without $\sfCT$ symmetry, a universality class is only characterized by the
value of $r$ mod 2.  ($\sfT$ relates $\v=1$ to $\v=-1$,
and so places no constraint on what happens for
$\v=1$.)  But in the $\sfCT$-conserving case, just as in the theory
of the Majorana chain \cite{MajoranaChain}, a perturbation to the Hamiltonian can only lift
zero-modes in groups of 8. Therefore
the number mod 8 of zero-modes in a vortex field is an invariant of a $\sfCT$-conserving
universality class. It is equal to the value of $r$ mod 8.

The anomaly that arises if $\sfCT$ is used to place a theory on an unorientable manifold
depends on $\nu_\sfCT$ mod 16.  Since $\nu_\sfCT=2r$, it follows that
on an unorientable manifold, the partition function (and even its anomaly) depends on $r$
mod 8.  Presumably, if we understood
how to generalize the low energy topological field theory to an unorientable manifold, it
would capture this information.
The arguments given above claiming to show that the low energy theory depends only on $r$
mod 2  involve either computing the partition function
on an orientable manifold or making a deformation that violates $\sfCT$.  So these arguments
do not tell us about the low energy description
when $\sfCT$ is used to go to an unorientable manifold.

Now let us apply all this to a topological superconductor, meaning that we set
$\sfT_\sc=\sfCT$, and take $A$ to be of order 2.
For a basic example, we also set
 $s=1$, leading (if $r=1$) to T-Pfaffian$\times$SF, as described in section \ref{prelims}.
 If $r$ is even, we get a model with abelian statistics that can
be a boundary state of a topological insulator with $\nu_\sc=2r$ divisible by 4.  Here we
will focus on the case of odd $r$, corresponding to $\nu_\sc$ congruent to 2 mod 4.

The associated topological field theory is always T-Pfaffian$\times$SF, but with four
possible actions of $\sfT_\sc$, depending on whether we take $r=1,3,5$, or $7$, and
always with $\nu_\sc=2r$.
Indeed, it has been claimed in \cite{Fidkowski} that
there are two possible actions of time-reversal on T-Pfaffian and two on SF, combining to
four possible actions  on T-Pfaffian$\times$SF.
The four possibilities have been denoted
T-Pfaffian$_\pm\times$SF$_\pm$.  Moreover, these four cases have been claimed to correspond
to $\nu_\sc=2,6,10$, and 14 -- in other
words, to $\nu_\sc=2r$ with  $r=1,3,5,7$.

Let us compare our construction in more detail to the assertions in \cite{Fidkowski}.
First of all, the theories T-Pfaffian$_+$ and T-Pfaffian$_-$ are supposed
to correspond respectively to the case that the basic vortex of vorticity $\v=1$ is a
Kramers singlet or a Kramers doublet under time-reversal.  In our approach, this
vortex has $r$ fermion zero-modes (all of angular momentum zero).  Quantizing those
zero-modes gives the spinor representation of $\Spin(r)$.
This representation is real for $r=1,7$ and pseudoreal if $r=3,5$.  The real and pseudoreal
cases allow respectively an antiunitary symmetry of square 1 or $-1$.
So the first factor of T-Pfaffian$_\pm\times$SF$_\pm$ is T-Pfaffian$_+$ if $r=1,7$ and
T-Pfaffian$_-$ if $r=3,5$.    This seems to agree with what is claimed in \cite{Fidkowski}.
However, we should note the following subtlety.  Similarly to what is explained in section
\ref{delicate}, the antiunitary symmetry that can be defined in a sector
of $\v=1$ actually depends on the value of $r$ mod 4; it is really $\Tsc$ if $r=1,5$, and
$\Tsc\K$ if $r=3,7$.  (In each of these statements, $\Tsc$
is the effective time-reversal symmetry of the low energy phase and obeys
$\Tsc^2=(-1)^F\K$.)

On the other hand, the distinction between SF$_+$ and SF$_-$ is supposed to be as follows.
Let $s$ be the ``semion,'' the unique nontrivial quasiparticle of $U(1)_2$,
and $f$ the transparent fermion (which corresponds to a line operator of $U(1)_{-1}$ in the
approach of section \ref{verysimple}, and
to  $\chi w^2$ in the microscopic
construction based on the $\chi$ field).  The two quasiparticles $s$ and $sf$
have spins $\pm 1/4$ and are exchanged by $\sfT_\sc$.  SF$_+$ and SF$_-$ are distinguished
by the sign of $\sfT_\sc^2$ in acting on these two states.
To analyze this problem in our framework, we first must identify $s$ and $sf$ in our
approach.  Here $s$ is supposed to correspond to the non-trivial
line $\exp(\i \oint b)=\exp(\i\oint(a+2c))$ of $U(1)_2$ (we recall that in eqn.
(\ref{rewrite}), the $U(1)_2$ gauge field is $b=a+2c$).  The coefficient of $c$ in the
exponent on the right hand side is the vorticity
$\v=2$.     Likewise, $sf$ has vorticity 2, since $f=\chi w^2$ has vorticity 0.

For $r=1$, the vortex of $\v=2$
was analyzed in section \ref{quasi}. It has two fermion zero-modes, whose quantization
leads to two states, denoted $\Lambda_\pm$, with spins $\pm 1/4$ and $\k=\pm 1$.  They are
exchanged by $\chi$.  (These two states can be further dressed with powers of $w$, and this
will be important
presently.  In general, these two states have $\k=\pm s$; we set $s=1$.)
 More generally, for any $r$, the $\v=2$ vortex has $2r$ fermion zero-modes.  Quantization
 of these
zero-modes leads again to two quasiparticles $\Lambda_\pm$, which now correspond to the two
spinor representations of $\Spin(2r)$.
They have\footnote{The zero-modes of each of the fermion fields $\chi_1,\dots,\chi_r$
contribute equally, so the spins and the values of $\k$ of $\Lambda_\pm$ are $r$ times what
they are for $r=1$.} spin $\pm r/4$ and $\k=\pm r$.
 However, in topological field theory, we only care about the value of the spin mod 1, and
 similarly
(for $s=1$) we only care about $\k$ mod 4.  So in fact, for odd $r$, the two quasiparticles
obtained by quantizing the fermion zero-modes have spin $\pm 1/4$ and $\k=\pm 1$,
just as for $r=1$.  Moreover, for odd $r$,
the two  spinor representations of $\Spin(2r)$ are complex conjugates,
so $\Lambda_\pm$ are exchanged by any antiunitary symmetry.

  The states $\Lambda_+$ and $\Lambda_-$ are actually part of the T-Pfaffian theory; they
  correspond to $I_2$ and $\psi_2$ in the terminology of \cite{Fidkowski}.
By contrast, the states $s$ and $sf$ of SF have $\k=0$, and correspond to $\bar w\Lambda_+$
and $f\bar w \Lambda_+=\chi w\Lambda_+=w\Lambda_-$.  (We use the fact that the transparent
fermion is $f=\chi w^2$ and that $\chi\Lambda_\pm =\Lambda_\mp$, since the chirality
operator that distinguishes $\Lambda_+$ and $\Lambda_-$
anticommutes with $\chi$. We also use $\bar w w=1$.)   Because of time-reversal symmetry,
there is an arbitrary choice of which of $s$ and $fs$ corresponds to $\bar w\Lambda_+$ and
which to $w\Lambda_-$.  We will
set $s=w\Lambda_-$, $sf=\bar w\Lambda_+$.
 We will determine the action of $\Tsc^2$ on both
pairs $\Lambda_+$, $\Lambda_-$ and $w \Lambda_-$, $\bar w\Lambda_+$.

To determine how $\Tsc^2$ acts on $\Lambda_+$ and $\Lambda_-$, we need to know whether the
representation
of a Clifford algebra of rank $2r$ (which we get by quantizing the $2r$ fermion zero-modes)
is real or pseudoreal.  Because of the mod 8 periodicity of the Clifford algebra,
the answer depends only on the value of $r$ mod 4.  For $r\cong 1$ mod 4, the representation
is real (as we have noted above for $r=1$), leading to $\Tsc^2=1$,
and for $r\cong 3$ mod 4, it is pseudoreal,\footnote{It suffices to take $r=3$,
corresponding to a Clifford algebra of rank 6.  We can represent 6 gamma matrices by
$\g_1=\sigma_1\otimes 1\otimes 1$, $\g_2=\sigma_3\otimes 1\otimes 1$,
$\g_3=\sigma_2\otimes\sigma_2\otimes 1$, $\g_4=\sigma_2\otimes \sigma_3\otimes 1$,
$\g_5=\sigma_2\otimes\sigma_1\otimes\sigma_1$,
$\g_6=\sigma_2\otimes\sigma_1\otimes\sigma_3$.  An antiunitary symmetry $\Tsc$ that commutes
with
these matrices is, up to an irrelevant phase, $\Tsc={\KK} \g_1\g_2\g_3$ (where $\KK$ is
complex conjugation), satisfying $\Tsc^2=-1$.}  corresponding to $\Tsc^2=-1$.
(In \cite{Fidkowski}, it is asserted that the action of $\Tsc^2$ on $I_2$ and $\Psi_2$ is
ill-defined, but in our framework this seems to be well-defined.)
So in general, on these states, $\Tsc^2=(-1)^{(r-1)/2}$.

Given this, to  understand how $\Tsc^2$ acts on $s=w\Lambda_-$  and $sf=w\Lambda_+$, we just
need to know how it acts  on $w$  and $\bar w$.
This is determined by $\Tsc^2=(-1)^F\K$.  Since $w$ and $\bar w$ are invariant under
$(-1)^F$ but
$\K (w)=\i w$, $\K(\bar w)=-\i\bar w$, we get $\Tsc^2 (s)=\i (-1)^{(r-1)/2} s$, $\Tsc^2
(fs)=-\i (-1)^{(r-1)/2}fs$.  This agrees\footnote{We should note that the overall
sign of $\Tsc^2$ acting on the pair $s,fs$ depends on some arbitrary choices.  We made one
arbitrary choice of whether to identify $s$ with $\Lambda_+$ or $\Lambda_-$.
There is also an arbitrary choice of whether to identify the time-reversal symmetry of
\cite{Fidkowski} with what in our language is $\Tsc$ or the equally good antiunitary
symmetry $\Tsc \K^{-1}$, which obeys $(\Tsc\K^{-1})^2=(-1)^F\K^{-1}$.)}  with
\cite{Fidkowski} if we assume that
(as suggested by the values of $\nu_\sc$ claimed in that paper), SF$_+$ corresponds to
$r=1,5$ and SF$_-$ corresponds to $r=3,7$.

An interesting detail is that the sign with which $\Tsc^2$ acts on $s,sf$ is correlated with
the sign with which it acts on $I_2=\Lambda_+$,
$\psi_2=\Lambda_-$,
even though one pair of states is in SF and one is in T-Pfaffian.  Hence, although the model
under consideration admits  four possible actions of $\Tsc$, corresponding to
$r=1,3,5,7$, it is not clear that this can be understood in terms of two possible actions on
SF and two on T-Pfaffian.  This may be related to the following.
It may be oversimplified to interpret the theory under consideration here as
T-Pfaffian$\times$SF, because as observed in footnote \ref{remark},
in the present context the transparent fermion of SF is actually a quasiparticle $\chi w^2$
of T-Pfaffian.

Unfortunately, our methods do not give a natural
way to study  T-Pfaffian$_\pm$ by itself so we have no good way to directly compute
$\nu_\sc$ for TPfaffian$_\pm$.  Moreover, though SF can be usefully described as
 $U(1)_2\times U(1)_{-1}$ as in section \ref{verysimple},
the fact that off-shell $\sfT^2\not=1$ in this model prevents us from being able to
straightforwardly formulate the model on an unorientable manifold; and this in turn leaves
us
with no straightforward way to directly compute $\nu_\sc$ for SF$_\pm$.

\subsection{More On T-Pfaffian}\label{tpfaff}

T-Pfaffian has also been studied from another point of view, motivated by applications to
topological insulators.  We would like to analyze
this in the context of the present paper.

For our starting point, we simply copy a model from eqn.\ (2) of \cite{MV} (the same model
has also been considered elsewhere, e.g.\ \cite{WaSen}).  The
model contains an emergent $U(1)$ gauge field $a$ and a composite Dirac fermion $\Psi_\cf$.
The Lagrangian density is
\be\label{density}\L = \bar\Psi_\cf\left(\g^\mu(\partial_\mu-\i a_\mu)\right)\Psi_\cf
-\frac{\epsilon^{\mu\nu\lambda}}{4\pi}A_\mu\partial_\nu a_\lambda.
\ee
The model is $\sfT$- and $\sfCT$-conserving at the classical level if $a$ transforms under
$\sfT$ as $A$ transforms under $\sfCT$,
and vice-versa,
and  if similarly the $\sfT$ and $\sfCT$ transformations of $\Psi_\cf$ are opposite from
those that we described in section \ref{actsym}
(which were the standard $\sfT$ and $\sfCT$ symmetries of the electron field).  Thus, in
contrast to
(\ref{waction}) and (\ref{twaction}), which we abbreviate to $\sfT(A)= -A$ and $\sfCT(A)=
A$,
we need
 \begin{align}\label{newwaction}\sfT(a)= a,~~~~~\sfCT(a)= -a. \end{align}
Similarly, we  have to reverse the roles of $\sfT$ and
$\sfCT$ in (\ref{easysign}) and (\ref{ctacts}):
\be\label{reversed} \sfT( \Psi_\cf(t,\vec x))=\g_0\Psi_\cf^\dagger(-t,\vec
x),~~~\sfCT(\Psi_\cf(t,\vec x))=\g_0\Psi_\cf(-t,\vec x). \ee
By contrast, so far in this paper, all gauge fields and fermions have had the standard
$\sfT$ and $\sfCT$
transformations.  The transformations in eqn.\ (\ref{reversed}) correspond to $\nu_\sfT=2$,
$\nu_\sfCT=0$, and these values will
be unaffected by the redefinitions below.  The value of $\nu_\sfCT$ is not important for a
topological insulator, but $\nu_\sfT$ mod
16
is a meaningful invariant of a $\sfT$-invariant condensed matter system, and we observe that
it differs from the value $\nu_\sfT=0$
appropriate to the usual topological insulator.  (This obstruction to interpreting the model
(\ref{density})  as a boundary state of a topological
insulator will not be affected by modifications of the model that we consider shortly.)

If we assume that $a$ obeys standard Dirac quantization, the model has several difficulties.
$\Psi_\cf$ couples to $a$ with odd charge, which
induces the usual $\sfT$ and $\sfCT$ anomaly.  Since $a$ is supposed to be an emergent gauge
field that only lives on the boundary
of a $3+1$-dimensional system, we cannot compensate for this with the help of a bulk
coupling. Likewise the $A\d a$ coupling
is not properly quantized.  We can resolve both of these issues if we assume that the flux
quantum of $a$ is $4\pi$ rather than the usual $2\pi$.
With this in mind, we write $a=2a'$ where $a'$ obeys standard Dirac quantization.  Dropping
the prime, we thus rewrite the Lagrangian as
\be\label{twodensity}\L = \bar\Psi_\cf\left(\g^\mu(\partial_\mu-2\i a_\mu)\right)\Psi_\cf
-\frac{\epsilon^{\mu\nu\lambda}}{2\pi}A_\mu\partial_\nu a_\lambda.
\ee
This version of the model has also been considered previously (see eqn.\ (36) in
\cite{Son}).

This theory is well-defined, but it is not a boundary state for a topological insulator
since it lacks the usual $A\d A$ anomaly in time-reversal
symmetry.   We can get a theory that does have the appropriate anomaly if we simply replace
$2a$ everywhere with $2a+A$.  This gives
the version of the model that we will actually study:
\be\label{threedensity}\h\L = \bar\Psi_\cf\left(\g^\mu(\partial_\mu-\i
(2a_\mu+A_\mu))\right)\Psi_\cf -
\frac{\epsilon^{\mu\nu\lambda}}{4\pi}A_\mu\partial_\nu (2a_\lambda+A_\lambda).
\ee
In perturbation theory, the two models (\ref{twodensity}) and (\ref{threedensity}) are
equivalent, since we can go from the first to the second
by the change of variables $a=a'+\frac{1}{2}A$ (where again we drop the prime from $a'$).
But the models are really not equivalent since
$a=a'+\frac{1}{2}A$ is not a valid change of variables.  On the contrary, in going from
(\ref{twodensity}) to (\ref{threedensity}) we have changed
the Dirac condition on quantization of magnetic flux.  (In fact, if $A$ is a $\spinc$
connection, the coefficient of $A$ in such a change of variables should be an even
integer.)

Actually, because $\Psi_\cf$ now has electric charge 1 with respect to $A$, we now
have a model that can be a boundary state for a $\sfT$-conserving system in which the
electromagnetic theta-angle in bulk is $\theta=\pi$.
We now want to ask two questions about the model: (1) Is it $\sfT$- and $\sfCT$-invariant?
(2) Is it consistent with the usual spin/charge relation of
condensed matter physics?

Concerning the first question, if $A$ is understood as a $U(1)$ gauge field, then the model
is $\sfT$- and $\sfCT$-invariant.
Indeed, the starting point (\ref{density}) was invariant classically
under the $\sfT$ and $\sfCT$ transformations (\ref{newwaction}).
By following through the various changes of variable, one can find the classical $\sfT$ and
$\sfCT$ symmetries of (\ref{threedensity}).
They are obtained by just replacing $a$ by $a+A$ on the right hand side of
(\ref{newwaction}):
\be\label{newwf} \sfT(a)= a+A,~~~~\sfCT(a)=-a-A~.\ee
Although this transformation is satisfactory if $A$ is a $U(1)$ gauge field, it is not if
$A$ is a $\spinc$ connection.  In that case the transformation (\ref{newwf}) is not
well-defined: $\sfT$ or $\sfCT$ would map
 a gauge field $a$ to a $\spinc$ connection $\pm(a+A)$.
 Therefore, as it stands, this model cannot,  while also maintaining $\sfT$-invariance,
 satisfy  the spin/charge relation in the strong sense that $A$ can be understood
 as a $\spinc$ connection.

An interesting manifestation of the difficulty arises if we add a scalar field $w$ of
charges $(0,1)$ under $U(1)_A\times U(1)_a$,
to make possible Higgsing to the standard boundary state of a topological insulator.  Then
$\sfT$ will relate $w$ to another
scalar field of charges (1,1), violating the spin/charge relation.  (If both $w$ and its
$\sfT$ conjugate have expectation values, maintaining
$\sfT$ invariance, then electromagnetic gauge invariance is spontaneously broken.)

 As in section \ref{TFTE}, the low energy topological field theory description of this model
 includes an Ising sector that involves an eta-invariant of the Dirac operator coupled to
 $A+2a$.
It is coupled by a $\Z_2$ quotient to an Abelian sector that can be constructed as in
section \ref{Abelians}.  In fact, the only difference from the
$s=1$ case of eqn.\ (\ref{Zfsgs}) is that in the action of the Abelian sector,
we have to include the Chern-Simons terms that are already present in $\h\L$.
The action of the Abelian sector is thus
\be\label{delf}\frac{1}{8\pi}(8c+2a-A)\d (2a+A).\ee
If we set $b=a+2c$, this becomes
\be\label{celf}\frac{1}{4\pi}\left( -8 c\d c+2 b\d b \right)+\frac{1}{2\pi} 2 A\d c
-\frac{1}{8\pi}A\d A.\ee
After including the Ising sector, this differs from the familiar T-Pfaffian model by a decoupled $U(1)_2$ of $b$.

The coupling to $A$ is not relevant for applications to a topological superconductor, but it
is certainly relevant for the topological insulator.
As we have seen, with this coupling, the  theory cannot simultaneously be $\sfT$-invariant
and consistent with the spin/charge relation.
Its low energy description differs from the T-Pfaffian theory by a $U(1)_2$ factor that
suffers from a
similar problem, and by the value of $\nu_\sfT$.   It may well be that, as proposed in the
literature, T-Pfaffian by itself is a satisfactory boundary state for
a topological insulator.  Unfortunately, in the context of the present paper, we do not have
a natural procedure to generate T-Pfaffian without the $U(1)_2$ factor.

\vskip5cm
\centerline{\bf Acknowledgements}

The work of NS was supported in part by DOE grant DE-SC0009988 and the work of EW was
supported in part by NSF Grant PHY-1314311. We would like to thank P.~Etingof, D.~Freed,
P.-S.~Hsin, C.~Kane,
 and M.~Metlitski for comments and discussions.

\appendix

\section{Normalization of Some Couplings}\label{norm}

\subsection{Some $3+1$-Dimensional Terms}\label{fourdc}

Let $X$ be a compact, smooth four-dimensional spin manifold without boundary, and
let $\slashed{D}_0$ be the Dirac operator acting on a spin $1/2$ fermion on  $X$, coupled to
gravity only.  Its index is
\be\label{index}\I(\slashed{D}_0)=\int_X\h A(R)=\frac{1}{48}\int_X\frac{\mathrm{tr}\,R\wedge
R}{(2\pi)^2}, \ee
where $R$ is the Riemann tensor. In four dimensions, $\h A$ is related to the signature:
\be\label{indsig}\int_X\h A(R)= {1\over 8}\sigma~. \ee
The index $\I(\slashed{D}_0)$ is an integer and in fact an even integer, as follows from a
Euclidean version of Kramers doubling.
So the right hand side of eqn.\ (\ref{index}) is an even integer on a spin manifold, and
hence $\sigma$ is divisible by 16.

Now introduce a $U(1)$ gauge field $A$, with field strength $F=\d A$, and let $\slashed{D}$
be the Dirac operator for a spin $1/2$ fermion on $X$ that couples
to $A$ with charge 1 (as well as coupling to gravity).  Its index is
\be\label{zindex}\I(\slashed{D}) =\int_X\h A(R)\exp(F/2\pi) =\int_X\left(\h A(R)
+\frac{1}{2}\frac{F\wedge F}{(2\pi)^2}\right). \ee
The right hand side of this formula is therefore again an integer (not necessarily even, in
this case).

Subtracting these two formulas to cancel the $\h A(R)$ term, we learn that on a spin
manifold $X$, if $F$ is a closed two-form  that  obeys Dirac quantization
(so that it is the field strength of some $U(1)$ gauge field $A$), then
\be\label{halfintegral}\frac{1}{2}\int_X \frac{F\wedge F}{(2\pi)^2}\in \Z. \ee
What we have gained by assuming that $X$ is spin is the factor of $1/2$ in front of this
formula.
If $X$ is not necessarily spin, then $\int_X F\wedge F/(2\pi)^2$ is an integer but not
necessarily even.
Simple examples show that on a spin manifold, $\int_X F\wedge F/(2\pi)^2$ has no
divisibility beyond what is claimed in eqn.\ (\ref{halfintegral}).

Next, suppose $X$ is not necessarily a spin manifold, but a more general $\spinc$ manifold
with $\spinc$ connection $A$.  In this case, $\int_X\h A(R)$ is no
longer an integer, but $\sigma$ is still an integer, so we deduce from eqn.\ (\ref{indsig})
that
\be\label{spincc} \int_X\h A(R)\in \frac{1}{8}\Z.\ee  The Dirac operator $\slashed{D}$ for a
spin $1/2$ particle coupled to $A$ with charge 1  is still defined; its index
$\I(\slashed{D})$ is still an integer and is still given by eqn.\ (\ref{zindex}).
So also
\be\label{spincctwo}\frac{1}{2}\int_X\frac{F\wedge F}{(2\pi)^2}\in {1\over 8}\Z.\ee
This can also be proved by writing the left hand side as $\frac{1}{8}\int_X (2F/2\pi)\wedge
(2F/2\pi)$, and observing that as $2F/2\pi$ is an integral class,
the integral $\int_X (2F/2\pi)\wedge (2F/2\pi)$ is an integer.

A canonical example of a four-manifold that is not a spin manifold is $X=\CP^2$.  The second
Betti number of $X$ is 1, so $X$ has essentially only one interesting two-cycle,
which is a copy of $\CP^1\subset \CP^2$.  Let $G$ be a closed two-form on $X$ normalized to
have unit Dirac flux on $\CP^1$: $\int_{\CP^1}G/2\pi=1$.
Then a standard topological argument shows that
\be\label{standard}\int_{\CP^2}\frac{G\wedge G}{(2\pi)^2}=1. \ee
$\CP^2$ is not a spin manifold, but it has a $\spinc$ structure with a $\spinc$ connection
$A$ whose field strength $F=\d A$ satisfies $F=G/2$.  From this and eqn.\ (\ref{standard}),
we learn that
\be\label{fractional}\frac{1}{2}\int_{\CP^2}\frac{F\wedge F}{(2\pi)^2}=\frac{1}{8}~. \ee
We see that the quantization in eqn.\ (\ref{spincctwo}) cannot be strengthened.  We can
learn the same from the fact that $\CP^2$ has
$\sigma=1$, which implies that the denominator in eqn.\ (\ref{spincc}) cannot be reduced.

To summarize, a four-dimensional manifold $X$ has
\be\label{sigmas}\sigma \in \begin{cases} 16\Z &\mbox{ if $X$ is spin}\\  \Z &\mbox{ for any
$X$ .}\end{cases}\ee
For an ordinary $U(1)$ gauge field $A$, the instanton number is
\be\label{instn}{1\over 2} \int_X\frac{F\wedge F}{(2\pi)^2} \in \begin{cases} \Z &\mbox{ if
$X$ is spin}\\  {1\over 2}\Z &\mbox{ for any $X$ .}\end{cases}\ee
If $A$ is a $\spinc$ connection, then
\be\label{spincins}\frac{1}{2}\int_X\frac{F\wedge F}{(2\pi)^2}\in\frac{1}{8}\Z~, \ee
but
\be\label{spincinsa}{1\over 8}\sigma+\frac{1}{2}\int_X\frac{F\wedge F}{(2\pi)^2}\in\Z~.\ee

\subsection{Theta-Angles}\label{thetangles}

It follows from what we have said that
 a four-dimensional field theory on a $\spinc$ manifold  $X$ has two $\theta$-like
 parameters, corresponding to a topological
 interaction $\theta_1 I_1+\theta_2I_2$, with
\be\label{twothe}I_1= \int_X\left(\hat A(R)+\frac{1}{2}\frac{F\wedge
F}{(2\pi)^2}\right),~~~I_2= 4\int_X\frac{F\wedge F}{(2\pi)^2} .\ee
We have normalized $I_1$ and $I_2$ to be integer-valued,
so $\theta_1$ and $\theta_2$ both have  $2\pi $ periodicity.   $\sfT$- and
$\sfCT$-invariance hold when both $\theta_1$ and $\theta_2$ equal 0 or $\pi$.

 In equation (\ref{twothe}), we have written the topological interactions in one useful
 basis. Another useful basis is given by $I_1$ and $I_2'=8I_1-I_2$:
 \be\label{anotherbasis} I_2'=8\int_X\hat A(R)=\sigma.\ee
 This has the advantage  that $I_2'$ is a purely gravitational coupling.  A coupling
 $\pi\sigma$ is the appropriate gravitational analog of $\theta=\pi$
 for a system that obeys the usual spin/charge relation.  We note that in the $\sfT$- and
 $\sfCT$-conserving case, with both $\theta_1$ and $\theta_2$
 being 0 or $\pi$, $\theta_1 I_1 +\theta_2 I_2$ is equivalent to $\theta_1 I_1+\theta_2
 I_2'$.  (This is certainly not true for generic values of the $\theta_i$.)

 The interior of a conventional topological insulator has $\theta_1=\pi$, $\theta_2=0$.  It
 is also possible to conceive a material with
 $\theta_1=\theta_2=\pi$.   Such a material has a bulk electromagnetic $\theta$-angle of
 $\pi$, like a conventional topological insulator,
 but the fact that $\theta_2\not=0$
 means that its allowed boundary states are different.

The topological interaction at $\theta_1=\theta_2=\pi$ is
\be\label{twothree}\pi I_1+\pi I_2=\pi\int_X\left(\hat A(R)+\frac{9}{2}\frac{F\wedge
F}{(2\pi)^2}\right).\ee
This is precisely $\I(\slashed{D}_3)$, the index of the Dirac operator $\slashed{D}_3$ for a
spin 1/2 particle of electric charge 3.  This observation leads to an easy construction
of a gapless boundary state for a material with $\theta_1=\theta_2=\pi$.  We simply assume
existence on the boundary of a massless Dirac fermion
of electric charge 3.  This leads to a $\sfT$-anomaly by the same logic as applies for
 a boundary fermion of charge 1.  Following the reasoning of section \ref{topins}, to
 restore
$\sfT$-invariance we need in the path integral a bulk factor
$(-1)^{\I(\slashed{D}_3)}=\exp(\i\pi \I(\slashed{D}_3))$.  But this
coincides with $\exp(\i\pi I_1+\i\pi I_2)$, so the charge 3 fermion
is a possible boundary state for a $\sfT$-invariant system with
$(\theta_1,\theta_2)=(\pi,\pi)$.   Similarly, for any integer $k$, a massless Dirac fermion
of charge $8k\pm 1$ is a possible boundary state at $(\theta_1,\theta_2)=(\pi,0)$, and a
massless Dirac fermion of charge $8k\pm 3$ is a possible
boundary state at $(\theta_1,\theta_2)=(\pi,\pi)$.  Symmetry-preserving gapped boundary
states appropriate for $(\theta_1,\theta_2)=(\pi,\pi)$ are
described in section \ref{moregeneral}.

What happens if we specialize to spin manifolds?  Then $I_2$ is divisible by 8, so $\pi I_2$
is a multiple of $2\pi$ and $\theta_2=\pi$ cannot
be distinguished from $\theta_2=0$.  Moreover, $\int_X \h A$ is an even integer and that
term can be dropped in $\pi I_1$.  So in the context of a spin
manifold, $\theta_1=\pi$ just means that the ordinary electromagnetic theta-angle equals
$\pi$.

On the other hand, on a spin-manifold there is an integer-valued coupling
\be\label{spincoup} I_2''=\frac{1}{2}\int_X \h A(R)=\frac{1}{2}\I(\slashed{D}_0). \ee
Accordingly, it is possible to have on a spin manifold a $\sfT$-invariant gapped system with
partition function $\exp(\i\pi I_2'')=(-1)^{\I(\slashed{D}_0)}$.
Such a system cannot be defined on a $\spinc$ manifold in a $\sfT$-invariant fashion, since
$I_2''$ is in general not integer-valued on a $\spinc$
manifold.  Thus, in a $\sfT$-conserving condensed matter system that conserves electric
charge and
satisfies the usual spin/charge relation, one will not encounter the interaction
$\pi I_2''$.

\subsection{$2+1$-Dimensional Chern-Simons Terms}\label{csc}

Naively, the Chern-Simons functional of a $U(1)$ gauge field $A$ on an oriented
three-manifold $W$ is defined by the familiar integral:
\be\label{zelf} \CS(A)=\frac{1}{4\pi}\int_W A\wedge \d A. \ee
This formula is satisfactory in a topologically trivial situation, but in general it is
difficult to interpret as $A$ may have Dirac string singularities.
A procedure that avoids this problem is as follows.   There always exists an oriented
four-manifold $X$ of boundary $W$ such that $A$ extends
over $X$.  Moreover, if $W$ has a chosen spin structure,\footnote{Every orientable
three-manifold $W$ admits a spin structure.  But in general
it may admit inequivalent spin structures and the Chern-Simons action defined by the
procedure in the text will in general depend on the choice of spin structure.} then $X$ can
be chosen so that the spin structure of $W$ extends over $X$.  Picking such an $X$ and  $A$,
and defining $F=\d A$ as before, we
attempt the definition
\be\label{welf}\CS(A)=2\pi \cdot \frac{1}{2}\int_X \frac{F\wedge F}{(2\pi)^2}. \ee
This is manifestly gauge-invariant, but in general it depends on the choice of $X$ and of
the extension of $A$ over $X$.  Had we used in this procedure
another spin manifold $X'$, again with some
chosen extension of $A$, then eqn.\ (\ref{welf}) would be replaced by
\be\label{welof}\CS'(A)=2\pi \cdot \frac{1}{2}\int_{X'} \frac{F\wedge F}{(2\pi)^2}. \ee
The difference between the two definitions is
\be\label{elf}\CS(A)-\CS'(A)=2\pi \cdot \frac{1}{2}\int_{X^*} \frac{F\wedge F}{(2\pi)^2},
\ee
where $X^*$ is a compact spin-manifold without boundary made by gluing $X$ to $X'$ after
reversing the orientation of $X'$ (fig. \ref{gluing} of section \ref{exts}).
According to eqn.\ (\ref{halfintegral}), the right hand side of this formula is an integer
multiple of $2\pi$.  In other words, our procedure for defining $\CS(A)$
has given a result that is independent of the choices mod $2\pi\Z$.

Now let us repeat this in the $\spinc$ case.  The operator $\slashed{D}_0$ acting on a
neutral spin 1/2 fermion is not available, but the operator $\slashed{D}$ acting on a spin
1/2
fermion of charge 1 still exists.  Its index is still given by the formula (\ref{zindex}),
so the integral on the right hand side of that equation is still $\Z$-valued.   This
formula
contains the gravitational term $\h A(R)$ as well as the gauge theory term $F\wedge
F/8\pi^2$, and there is no longer a convenient way to remove
the gravitational term by subtraction.  So if we use integrality of the index to define a
Chern-Simons-like functional,
this functional will have to have a gravitational contribution.
Picking again a $\spinc$ manifold $X$ with boundary $W$ and an extension of $A$ over $X$, we
define
\be\label{orelf} \CS(A)+\Omega(g)=2\pi\int_X \left(\h A(R)+\frac{1}{2}\frac{F\wedge
F}{(2\pi)^2}\right).\ee
Here $\Omega(g)$ is a sort of gravitational Chern-Simons term.  The same argument as before
shows that $\CS(A)+\Omega(g)$ is independent of the choices mod $2\pi$.
Moreover, the normalization of $\CS(A)$ is the same as it was in the spin case (for example,
in a topologically trivial situation, $\CS(A)$ is still given by the naive formula
(\ref{zelf})).
However, only the sum $\CS(A)+\Omega(g)$, and not either of the two terms separately, is
independent of the choice of $X$ mod $2\pi\Z$.

In section \ref{spincharge}, a simple argument was given to show that $\CS(A)$ is
well-defined mod $2\pi/8$, or equivalently that $8\CS(A)$ is well-defined mod $2\pi$.
The discussion leading to eqn.~(\ref{spincins}) shows that this is the sharpest result of
its type: we cannot replace 8 by any smaller integer.
We see that in the $\spinc$ case, $\CS(A)$, defined by the usual formula, is well-defined
only modulo $2\pi/8$.

\subsection{Unorientable Spacetimes}\label{unorientable}

Throughout this paper, we have considered orientable spacetimes only. Without aiming to
explain the generalization of all statements in this paper
to unorientable spacetimes, we will here sketch the generalization of the bulk couplings
corresponding to $\theta_1=\pi$ and $\theta_2=\pi$.

 The unorientable analog of a $\spinc$ manifold is called a $\pinc$ manifold, and (in a
 theory that satisfies
 $\sfT^2=(-1)^F$, so that fermions are in Kramers doublets) the unorientable analog of a
 spin manifold is a $\pinp$ manifold.
 In a theory that satisfies the spin/charge relation, the more general $\pinc$ concept is
 natural, but we will also run into a variant of $\pinc$ below
 (footnote \ref{variant}).

An important preliminary is that either $\sfT$ or $\sfCT$ can be used to define a theory on
an unorientable manifold $X$.  In one case, fields undergo
a $\sfT$  or $\sfCR$ transformation in going around an unorientable loop, and in the other
case they undergo a $\sfCT$ or $\sfR$
transformation.\footnote{We here use the $\sf{CRT}$ theorem of relativistic field theory
(more commonly called the $\sf{CPT}$ theorem), via
which $\sfT$ is related to $\sf{CR}$ and $\sfCT$ to $\sfR$. In Euclidean signature, it is
more natural to think about $\sf{CR}$ and $\sfR$
rather than $\sfT$ and $\sfCT$.}   Concretely, if we use $\sfCT$ symmetry to define a theory
on $X$, then the electromagnetic
field strength $F$ is an ordinary two-form on $X$, while if we use $\sfT$, then it is a
twisted two-form that changes sign in going around an
orientation-reversing loop.

It is more straightforward to generalize the interaction $\pi I_2$ to a $\pinc$ manifold, so
we begin with this.  On a $\spinc$ or $\pinc$ manifold of any
dimension, the integral class $2 F/2\pi$ is congruent mod 2 to $w_2$, the second
Stieffel-Whitney class.  (This is true whether $F$ is an ordinary
two-form or a twisted one.)  So $\exp(\i\pi I_2)$ is equivalent to $(-1)^{\int_X w_2^2}$,
and this generalizes $\theta_2=\pi$ to a $\pinc$ manifold.   On a
$\pinp$ manifold, $w_2=0$ and this interaction is trivial.

Now let us consider the opposite case of a gapped theory with $\theta_1=\pi$, $\theta_2=0$.
On an orientable manifold, its partition function is
$(-1)^{\I(\slashed{D})}$.  But the index of the Dirac operator is only defined on an
orientable manifold.  How can we generalize $(-1)^{\I(\slashed{D})}$
to an unorientable manifold?  The answer to this question is that we should use $\etta$, the
eta-invariant of the Dirac operator, rather than the index.
On an orientable manifold of even dimension, the spectrum of the Dirac operator is symmetric
under $\lambda\leftrightarrow-\lambda$.  Recalling the definition
(\ref{ruff}) of $\etta$, this means that modes with $\lambda\not=0$ do not contribute to
$\etta$, which just receives a contribution of $+1$ from each
zero-mode.  The index $\I$ likewise receives no net contribution from non-zero eigenvalues;
on the other hand, a zero-mode contributes $\pm 1$
to $\I$, depending on its chirality.  So on an orientable manifold of even dimension, $\I$
and $\etta$ are congruent mod 2, and therefore $(-1)^\I
=(-1)^\etta$.

This can be used as the starting point to generalize $\theta_1=\pi$ to an unorientable
manifold, but some subtleties arise.  First, let us
assume that $\sfT$ or equivalently $\sf{CR}$ has been used to define a theory on an
unorientable manifold $X$.  This means that in going around
an orientation-reversing loop in $X$, a fermion field $\chi$ of  charge 1 is  exchanged with
its adjoint $\bar\chi$ of  charge $-1$.
Given this, one cannot define a Dirac operator that acts solely on $\chi$; only a Dirac
operator that acts on the pair $\bp \chi\cr\bar\chi\ep$ can be
defined.\footnote{\label{variant} The fact that $A$ is odd under $\sfCR$ also means that it
is not what is usually called a $\pinc$ connection but a variant of this
with the group $O(2)$ replacing $SO(2)=U(1)$.  $\sfCR$ takes values in the disconnected
component of $O(2)$.}
In eqn.\ (\ref{ruff}), the invariant $\etta$ was defined on an orientable manifold
in terms of a sum over eigenvalues of $\D=\i\slashed{D}$ acting on $\chi$ only.  By $\sfC$
symmetry, the Dirac operator has the same eigenvalues
acting on $\bar \chi$ as it does on $\chi$.  So if we define an eta-invariant $\eta$ for a
Dirac operator that acts on the pair $\bp\chi\cr\bar\chi\ep$,
then on an orientable manifold where both are defined, the relation between $\eta$ and
$\etta$  is just
\be\label{relan}\eta=2\etta.\ee
Unlike $\etta$, $\eta$ is still defined when we use $\sfT$ symmetry to go to an unorientable
manifold.
When we do that,  $\eta$ is always an even integer (as it is in the orientable case) and is
a topological invariant mod 4.\footnote{\label{notice}
An unorientable manifold $X$ has a canonical oriented double cover $\h X$.
Instead of computing $\eta$ in terms of all fermion modes of either charge on $X$, we can
compute it in terms of modes of charge 1 on $\h X$.  When
we do this, because $\h X$ is orientable, only zero-modes contribute.  The number of those
zero-modes is even because,
as $\h X$ has an orientation-reversing symmetry, the index of the charge 1 Dirac operator on
$\h X$ is 0.   To show that $\eta$ is a topological invariant
mod 4, we may reason as follows on the original unorientable manifold $X$. (For background,
see \cite{WQ}, especially the discussion of eqns.
(2.47)
and (B.16).)
In general, in even dimensions, because there is no Chern-Simons function,
 $\eta/2$ is the index of the Dirac operator with APS boundary conditions and hence is a
 topological invariant except for jumps that occur when eigenvalues pass through 0.
In four dimensions,
even on an unorientable manifold, the eigenvalues of the hermitian Dirac operator with
values in a real representation (such as we have here because
we include both signs of the charge) all have even multiplicity because of a version of
Kramers doubling.  When a pair of eigenvalues passes through 0,
$\eta$ jumps by $\pm 4$, so it is a topological invariant mod 4.}  Therefore, in this
situation,
a suitable generalization of $(-1)^\I$ is $(-1)^{\eta/2}$.  This gives a generalization of
$\theta_1=\pi$ when we use $\sfT$ to go to an unorientable manifold.

However, using $\sfT$ to go to an unorientable
manifold in a theory that satisfies the spin/charge relation has not really given us a new
probe of the physics. The mod 2 invariant $(-1)^\I$
has been generalized to a mod 2 invariant $(-1)^{\eta/2}$, but since $\eta$ is always an
even integer in this situation and is only a topological
invariant mod 4, this mod 2 invariant is all that we get.

If instead $\sfCT$ or equivalently
$\sfR$ symmetry is used to define a theory on an unorientable spacetime $X$, matters are
different.   There is no mixing between modes of opposite
charge, so $\etta$ is naturally defined.  But
in this situation, $\etta$ is not necessarily an integer.  So before generalizing to an
unorientable spacetime, we should first rewrite $(-1)^\I$ as $\exp(-\i\pi \I)$ (the sign in
the exponent is an arbitrary choice here), and then
we can generalize  this to an unorientable spacetime as $\exp(-\i\pi\etta)$.  On a $\pinc$
manifold,
 $\etta$ is a topological invariant mod\footnote{\label{jumping} As in footnote
 \ref{notice},
  $\etta$ is a topological invariant except for jumping that occurs when an eigenvalue
  passes through 0.  When this happens, $\etta$ jumps
 by $\pm 2$.  So it is a topological invariant mod 2.  There is no further restriction in
 general.  The difference from footnote \ref{notice} is that we are dealing
 with fermions in a complex representation, so there is no Kramers doubling.} 2, so
 $\exp(-\i\pi\etta)$
is a topological invariant.
But what are the possible values of $\etta$ mod 2 on a four-dimensional $\pinc$
manifold?  The answer is that $\etta$
is always an integer multiple\footnote{This minimum value is assumed for the $\pinc$
manifold $\mathbf{RP}^4$.  See Appendix C of \cite{WQ}. The argument
there is presented for $\pin^+$ manifolds but the same reasoning applies for $\pinc$.}
 of $1/4$.   Hence, there are actually 8 classes of $\sfCT$-conserving theory that can be
 defined on a general $\pinc$ manifold,
 with the partition function
 being $\exp(-\i\pi\rho\etta)$, $\rho=0,\dots,7$.   If we specialize these 8 classes to a
 $\spinc$ manifold, we can only detect the value of $\rho$
 mod 2, since on an orientable manifold, $\exp(-\i\pi\rho\etta)=(-1)^{\rho\I}$.   On an
 orientable manifold, these 8 classes of theory
 have $\theta_1=\pi\rho $ mod $2\pi$.

 If we consider a charge-conserving theory that lacks the conventional spin/charge
 relationship, then a few things are different.   We should work only
 on a $\pin^+$ manifold, not a more general $\pinc$ manifold.   The basic eta-invariant to
 consider is that of a Majorana fermion coupled to gravity
 only.  We will denote it as $\eta$.  It is a topological invariant mod\footnote{The
 reasoning is the same as in footnote \ref{jumping}, except that
 on a $\pin^+$ manifold, the eigenvalues have a two-fold Kramers degeneracy, and hence the
 jumps in $\eta$ are by $\pm 4$.}   4.  So
 $\exp(-\i\pi\nu\eta/2)$ is a topological invariant for any integer $\nu$.  The values of
 $\eta$ are always integer multiples of $1/4$ (the minimum
 value again occurs for $\mathbf{RP}^4$), so $\nu$ is an invariant mod 16.

 In a theory that conserves both $\sfT$ and $\sfCT$ and does not have the standard spin
 charge relation, we should distinguish two  mod 16 invariants
 $\nu_\sfT$ and $\nu_\sfCT$, depending on which symmetry is used to define the theory on an
 unorientable manifold.  Thus we consider
 a theory whose partition function on a $\pinp$ manifold $X$ is $\exp(-\i\pi
 \nu_\sfT\eta/2)$ or $\exp(-\i\pi\nu_\sfCT\eta/2)$, depending on which
 symmetry is used when the fields traverse an orientation-reversing cycle in $X$.  If $X$
 happens to be orientable (and thus spin), these
 factors must be equal since in that case there was no need to make a choice of $\sfT$ or
 $\sfCT$.  Since on an orientable manifold, $\eta/2$ is an arbitrary
 integer ($\eta$ receives contributions only from zero-modes, and there are an even number
 of them because of Kramers doubling), it follows that $\nu_\sfT$ and $\nu_\sfCT$ are
 congruent to each other mod 2.  This is the only general relationship between them, as one
 can see by considering examples constructed from massless free fermions with suitable
 transformations under $\sfT $ and $\sfCT$.  Each massless
 Majorana
 fermion can independently contribute $\pm 1$ to $\nu_\sfT$ and to $\nu_\sfCT$.

Now let us consider a gapped, charge-conserving theory that does obey the usual spin/charge
relation. Suppose that, when
placed on an unorientable manifold using $\sfCT$, this theory has
 partition function
$\exp(-\i\pi\rho\etta)$.  If we specialize to a $\pinp$ manifold, then $\etta$ reduces to
$\eta$ and the partition function
is supposed to be $\exp(-\i\pi\nu_\sfCT\eta/2)$.   Evidently, the relation between $\rho$
and $\nu_\sfCT$ is
\be\label{reln}\nu_\sfCT=2\rho.\ee
In particular, $\nu_\sfCT$ is even for theories that obey the usual spin/charge relation.
This is consistent with the fact that $\rho$ is a mod 8
invariant and $\nu_\sfCT$ is a mod 16 invariant.

\section{The Callias Index Theorem}\label{calliastheorem}

In this appendix, we will use the Callias index theorem \cite{CalliasIndex} to study
time-independent solutions of the Dirac equation for $\chi$.
(This theorem has been elucidated in \cite{BottSeeley} and is based in part on analytical
foundations established in  \cite{Seeley}.  See also  \cite{CalliasAp,JRo,EWeinberg}.)

It is helpful to write the Dirac equation in a completely real form.  To this end, we first
introduce a basis of real gamma matrices:
\be\label{realgamma}\g_0=\bp 0&-1\cr 1&0\ep,~~\g_1=\bp 1&0\cr 0&-1\ep,~~\g_2=\bp 0&1\cr 1&0
\ep.\ee
We also write $\chi$ in terms of Majorana fermions:  $\chi=(\chi_1+\i\chi_2)/\sqrt 2$.  It
is convenient to arrange these Majorana fermions
as a column vector $\bp \chi_1\cr \chi_2\ep$, and to introduce a set of real gamma matrices
that act on this column vector:
\be\label{realgammas}\t\g_0=\bp 0&-1\cr 1&0\ep,~~\t\g_1=\bp 1&0\cr 0&-1\ep,~~\t\g_2=\bp
0&1\cr 1&0 \ep.\ee
We take the matrices $\t\g_i,~i=0,1,2$ that act on the pair $\bp \chi_1\cr \chi_2\ep$ to
commute with the matrices $\g_\mu,~\mu=0,1,2$ that act on the spinor
indices carried by $\chi_1$ and $\chi_2$.

The covariant derivative of $\chi$ (on a flat spacetime) is usually written
$D_\mu=(\partial_\mu+\i A_\mu)\chi$.  On the column vector $\bp \chi_1\cr \chi_2\ep$,
this is equivalent to
\be\label{herm}D_\mu \bp \chi_1\cr \chi_2\ep=(\partial_\mu+\t\g_0 A_\mu)\bp \chi_1\cr
\chi_2\ep.\ee

According to eqn.\ (\ref{ctacts}), $\sfCT$ acts on $\bp \chi_1\cr \chi_2\ep$ simply as
multiplication by $\g_0$ (along with a reversal of the time, which will not
be important here as we will be considering time-independent solutions).
However, according to eqn.\ (\ref{effct}), the appropriate unbroken
symmetry  in the  gapped phase is not $\sfCT$ but $\slCT=\sfCT \K^{1/2}$, where $\K^{1/2}$
is a gauge transformation
that acts on $\chi$ as multiplication by $\i$.  On the column vector $\bp \chi_1\cr
\chi_2\ep$, $\K^{1/2}$ acts as multiplication by $\t\g_0$.  Hence the action of $\slCT$
is
\be\label{tello}\slCT \bp \chi_1\cr \chi_2\ep=\g_0\t\g_0 \bp \chi_1\cr \chi_2\ep. \ee

Setting $\phi=(\phi_1+\i\phi_2)/\sqrt 2$, the  Dirac equation obeyed by $\chi$ is
\be\label{zello}\biggl(\g^\mu(\partial_\mu+\t\g_0 A_\mu)
+\t\g_1\phi_1+\t\g_2\phi_2\biggr)\bp \chi_1\cr \chi_2\ep =0. \ee
We drop the time-derivative (since we want to study time-independent solutions) and we
multiply by $\i\g_0$ to define a hermitian Dirac operator:
\be\label{mello}\D=\i\left(\sum_{\mu=1,2}\g_0 \g^\mu(\partial_\mu+\t\g_0
A_\mu)+\g_0\sum_{i=1,2} \t\g_i\phi_i\right). \ee
$\D$ is imaginary and antisymmetric.

We observe that the matrix $M=\g_0\t\g_0$ that represents the action of $\slCT$ obeys
$M^2=1$ and anticommutes with $\D$:
\be\label{polly} M\D=-\D M. \ee
(The full $\slCT$ operation, which is the product of $M$ with complex conjugation, commutes
with $\D$, since $\D$ is imaginary.)
This is the situation in which one can define an integer-valued {\it index} $\I$, the number
of $M=1$ zero-modes of $\D$ minus the number of $M=-1$ zero-modes of $\D$.
Actually, the ability to define an index depends on the fact that the operator $\D$ has a
discrete spectrum near zero.  This is true as long as $\phi$ is bounded away
from zero near spatial infinity, giving $\chi$ a mass and ensuring that any low-lying modes
of $\chi$ must be localized near the origin in space.

This integer is a topological invariant, in the sense that it is invariant under
deformations of the operator $\D$ that do not change its behavior at infinity in either
real space or
momentum space.  For example, although $A_\mu$ is nonzero in the standard vortex solution,
it vanishes at infinity, and therefore $\I$ would be unchanged if we simply
set $A_\mu$ to 0.  The index only depends on the winding at spatial infinity of the field
$\phi$ that determines the fermion mass.

If explicit $\slCT$-violating terms are added to the action, then $\D$ will be replaced by
an operator that no longer anticommutes with $M$ (or with any similar matrix),
and zero-modes of $\chi$ will no longer be governed by an integer-valued index.  However,
fermi statistics imply that fermion zero-modes can only be lifted in pairs.
So even if $\slCT$ is explicitly violated, the number of $\chi$ zero-modes will always be
equal mod 2 to the index $\I$ that one can define in the $\slCT$-conserving case.
Generically, the number of $\chi$ zero-modes will always be as small as possible subject to
the constraints implied by the index, so the total number of zero-modes
is generically $|  \v|  $ in the $\slCT$-conserving case, and 0 or 1 in the
$\slCT$-violating case.

According to the Callias index theorem, $\I$  is given by a sort of winding number that
should be computed at infinity in phase space.
In other words, we consider a phase space $\R^4=\R^2\times \R^2$ where the first factor is
parameterized by position components $x_1,x_2$ and the second factor
by momentum components $p_1,p_2$.  We let $\D_+$ be the part of $\D$ that maps states of
$M=1$ to states of $M=-1$.  We undo the passage to quantum
mechanics, replacing derivatives by momentum
components, $\partial_\mu\to \i p_\mu$. $\D_+$ thus becomes a function of the $x$'s and
$p$'s.
 Because $\chi$ is gapped near spatial infinity (and the operator $\D$ is ``elliptic,''
 meaning that it is invertible for large
Euclidean momenta), the operator $\D_+$ is invertible when restricted to a large sphere
$S^3$ at infinity in $\R^4$.  The winding number of this invertible
operator, integrated over $S^3$, is equal to the index.

To implement this program, we want to write $\D_+$ explicitly as a map from a space $V_+$ of
modes with $\chi=1$ to a space $V_-$ of modes with $\chi=-1$.
We use a notation in which $| \neg  \up\up\rangle$ represents a joint eigenstate of $\g_0$
and $\t\g_0$ with both having eigenvalue $\i$,
$|  \neg \up\dn\rangle$ represents a joint eigenstate with
respective eigenvalues $\i$ and $-\i$, etc.  So $| \neg  \up\dn\rangle$ and
$|\neg\dn\up\rangle$ give a
basis of $V_+$, while $|\neg\up\up\rangle$ and $|\neg\dn\dn\rangle$ give  a basis of $V_-$.
To write $\D_+:V_+\to V_-$ as a matrix, we represent the basis vectors $|  \neg
\up\dn\rangle$ and $|  \neg \dn\up\rangle$ of $V_+$ as $\bp 1\cr 0 \ep$ and $\bp 0 \cr
1\ep$,
respectively, and similarly we represent basis vectors $| \neg  \up\up\rangle$ and $|\neg
\dn\dn\rangle$ of $V_-$ as $\bp 1\cr 0\ep$ and $\bp 0\cr 1\ep$.  After setting $A_\mu$
to zero (since it will not affect the index), a
calculation gives
\be\label{twobytwo}\D_+=\i\bp \phi_1+\i\phi_2  &  p_1+\i p_2\cr p_1-\i p_2 &
-\phi_1+\i\phi_2\ep. \ee

In a suitable gauge, the  standard vortex solution of vorticity $\v=1$
is invariant under rotations of $\R^2$ together with constant gauge transformations.  The
scalar field $\phi$ is
\begin{align}\label{scalarf} \phi_1 & = x_1 f(|\vec x|) \cr
                                             \phi_2& = x_2 f(|\vec x|). \end{align}
The function $f(|x|)$ vanishes at infinity as $|\langle\phi\rangle|/|\vec x|$.  However, it
will not affect the index if we just set $f=1$ (which means that the fermion mass
grows linearly at infinity), since we can interpolate from a realistic
$f$ to $f=1$ in such a way that the fermions are always gapped near spatial infinity.
With $f=1$, $\D_+$ is simply linear in the $x_i$ and $p_j$ and the winding number
at infinity is easy to calculate.   After relabeling $(x_1,x_2,p_1,p_2)$ as
$(P_0,P_1,P_2,P_3)$ in a fairly obvious way, we find
\be\label{woby}\D_+=P_0+\i \vec \sigma\cdot \vec P, \ee
where $\vec \sigma$ are the Pauli sigma matrices.  If we restrict $\D_+$ to the sphere $S^3$
defined by $\sum_{s=0}^3 P_s^2=R^2$ for large $R$, then up to a factor of $R$,
$\D_+$ is just the identity map from $S^3$ to $SU(2)\cong S^3$.  This identity map is the
basic example of a map of winding number 1, and that is the value of the Callias
index in this situation.

For an example of a field of vorticity $\v>0$, we can just take $\phi$ at infinity to be
defined by
\begin{equation}\label{multiv} \phi_1+i\phi_2=(x_1+\i x_2)^\v.\ee  This just multiplies the
winding number by $\v$.
To get vorticity $\v<0$, we take
\begin{equation}\label{nultiv}\phi_1+\i \phi_2=(x_1-\i x_2)^{-\v}.\ee
Again the winding number if $\v$.  So in all cases, the winding number of the map $\D_+$,
and therefore the index predicted by the Callias index theorem, is equal
to the vorticity $\v$.

For any value of the vorticity $\v$, it is possible to have a classical field that is
rotation-invariant, or more precisely, invariant under a combined rotation plus gauge
transformation.
Indeed, if $\j$ is the standard angular momentum generator and $\k$ is the gauge charge of
the emergent gauge field, then the ansatz (\ref{multiv}) or (\ref{nultiv}) is invariant
under the combination
\be\label{combo}\j'=\j+\v \frac{\k}{4s}. \ee
The factor of $1/4s$ arises because $\phi$ has $\k=4s$.  We set $\k^*=\k/4s$ so
\be\label{ombo}\j'=\j+\v \k^*.\ee
$\chi$ and $\bar\chi$ have $\k=\pm 2s$ and so $\k^*=\pm 1/2$.

A stable quasiparticle with $|\v|>1$ does not necessarily exist in the theory, since the
interaction between vortices may be repulsive, as in a Type II superconductor.
However, if a stable vortex of given $\v$ exists, one may expect that  classically it is
described by a $\j'$-invariant solution.  (Even if this is not the case, the vortex is
deformable to a $\j'$-invariant
situation, so computing in the $\j'$-invariant case  should suffice for determining the
deformation-invariant properties of the spectrum.)  To understand the quantization of the
vortex states,
it is necessary to know not just the number of fermion zero-modes in the vortex field, but
their $\j'$ quantum numbers.

The framework to determine those quantum numbers conceptually -- as opposed to attempting to
directly solve the Dirac equation -- is to compute what is known as the character-valued
index of the Dirac operator.  Define the positive operator $H=\D^2$.      For an angle
$\alpha$, define
\be\label{charindex} F(\alpha)=\Tr\, \exp(\i \alpha\j') M \exp(-\beta H).\ee
The idea behind this definition is that (as in the standard definition of the index of the
Dirac operator)
 nonzero eigenvalues of $H$ appear in pairs with the same $\j'$ but opposite signs of $M$,
 and so cancel out of  the trace.
Hence $F(\alpha)$ is independent of $\beta$, and it can be computed effectively for small
$\beta$ in a high temperature expansion, as we discuss below.
On the other hand, since the nonzero modes of $H$ make no net contribution to $F(\alpha)$,
that function
can be computed just in the space of fermion zero-modes, and therefore captures the desired
information about the $\j'$ eigenvalues of  those modes.

The result that comes from the high temperature expansion is
\be\label{result}F(\alpha)={\mathrm{sign}}(\v)\cdot \bigl(\exp(\i
(\v-1)\alpha/2)+\exp(\i(\v-3)\alpha/2)+\dots+\exp(\i(-\v+1)\alpha/2)\bigr), \ee
meaning that the zero-modes have spins $\j'=(\v-1)/2,(\v-3)/2,\dots, -(\v-1)/2$.  Before
explaining how this formula is obtained, we look at the first few cases.

For $\v=1$, there is only one zero-mode, and it has  $\j'=0$.  Indeed, if the operator $\D$
has only
one zero-mode, it must have $\j'=0$, since complex conjugation (or $\slCT$ symmetry) ensures
that the spectrum of zero-modes is symmetric under $\j'\leftrightarrow -\j'$.

For $\v=2$,  the two zero-modes have spins $\pm 1/2$.  Upon quantization of these modes, one
gets a pair of
states of spins $\pm 1/4$, as asserted in section \ref{quasi}.

For $\v=3$, the three zero-modes have spins $1,0,-1$.  Quantization of the modes of spins
$\pm 1$ gives two states of spins $\pm 1/2$, and existence of the third zero-mode
of spin zero means that the $\v=3$ vortex satisfies non-Abelian statistics.

Finally, for $\v=4$, the four zero-modes have spins $3/2,1/2,-1/2,-3/2$.  Quantization gives
a pair of states of spins $\pm 1$ (which in topological field
theory is equivalent to 0) and another pair of states of spins $\pm 1/2$.  Since the fermion
zero-modes anticommute with the operator $(-1)^F$ that distinguishes
bosons and fermions, one pair of states is bosonic and one is fermionic.  Which is which is
discussed in section \ref{quasi}.

The actual computation of the character-valued index $F(\alpha)$ uses a high temperature
expansion, valid for small $\beta$.
 For a proof of the character-valued index theorem for the Dirac operator using path
 integrals,
see \cite{goodman}.   Here we will describe an equivalent computation in a Hamiltonian
approach.

For small $\beta$, when we compute $F(\alpha)=\Tr\,M e^{\i\alpha\j'} \exp(-\beta H)$ via an
integral over particle orbits, we have to consider
orbits that are periodic up to a rotation by an angle $\alpha$.  Let $R_\alpha$ be the
operator that rotates the $\vec x$ plane by angle $\alpha$.
It has a unique fixed point at $\vec x=0$.  An orbit that propagates from $\vec x$ to
$R_\alpha \vec x$ in imaginary time $\beta$ has a very large
action unless $|(1-R_\alpha)\vec x|\lesssim\beta$.  For fixed nonzero $\alpha$, as $\beta\to
0$, the condition is equivalent to $|\vec x|\lesssim\beta$.  On
such short distances, a particle can be treated as free; the background gauge fields and
scalar fields play no role.
When we set the background field to zero, $H=\D^2$ becomes simply the Laplacian
 $H=-\sum_{i=1,2}\frac{\partial^2}{\partial x_i^2}=\vec p^2$.  This Laplacian acts on a
 Hilbert space $L^2(\R^2)\otimes \H$, where $L^2(\R^2)$ is just
 the space of scalar wavefunctions on $\R^2$, and $\H$ is a four-dimensional Hilbert space
 obtained by quantizing the four real components of $\chi$.
 The trace that we want factorizes as the product of a trace in $L^2(\R^2)$ and a trace in
 $\H$.

 We start with the first factor.  This means that we consider $H$ to act on scalar
 wavefunctions on $\R^2$, and $\j'$ reduces to
  the standard
angular momentum generator $-\i(x_1\partial_2-x_2\partial_2)$. The operator $M$ acts
trivially on $L^2(\R^2)$ and can be dropped. We compute
\begin{align}\label{compute}\Tr\,e^{\i\alpha\j'}\exp(-\beta H)&=\int\d^2x\, \langle R_\alpha
\vec x|\exp(-\beta H)|\vec x\rangle
=\int\d^2x \int\frac{\d^2p}{(2\pi)^2}\exp(-\beta\vec p^2)\exp(\i \vec p\cdot (\vec
x-R_\alpha\vec x))\cr &
=\frac{1}{\det (1-R_\alpha)}=\frac{1}{|1-e^{\i\alpha}|^2}.\end{align}
To compute the trace in $\H$, we observe the following.  As operators on $\H$, the usual
angular momentum is $\j=\frac{\i}{2}\g_0$ and
the charge generator is $\k^*=\frac{\i}{2}\t\g_0$.  So $\j'=\frac{\i}{2}(\g_0+\v\t\g_0).$
The operator $M$ is $\g_0\t\g_0$.  The Hamiltonian acting on $\H$ is 0.
So we need to evaluate
\be\label{ompute}\Tr\,M \exp(\i\alpha\j')=\Tr\,\g_0\t\g_0
\exp(-\alpha\g_0/2-\alpha\v\t\g_0/2).\ee
Since $\g_0$ and $\t\g_0$ act in different two-dimensional spaces and commute with each
other, this is a product
\be\label{mpute}\Tr\,\g_0\exp(-\alpha\g_0/2)\cdot
\Tr\,\t\g_0\exp(-\alpha\v\t\g_0/2)=-(e^{\i\alpha/2}-e^{-\i\alpha/2})(e^{\i\v\alpha/2}-e^{-\i\v\alpha/2}).\ee
So
\be\label{pute}F(\alpha)=-\frac{(e^{\i\alpha/2}-e^{-\i\alpha/2})(e^{\i\v\alpha/2}-e^{-\i\v\alpha/2})}{|1-e^{\i\alpha}|^2}=\frac{e^{\i\alpha\v/2}-e^{-\i\alpha\v/2}}
{e^{\i\alpha/2}-e^{-\i\alpha/2}},\ee
which is equivalent to (\ref{result}).

Finally, in discussing eqn.\ (\ref{neffham}), we claimed that $\chi$ zero-modes can be
lifted in groups of 8 by a quartic Hamiltonian
\be\label{genform}\Delta H=\sum_{ijkl}p_{ijkl}\g_i\g_j\g_k\g_l,\ee
in a way that preserves angular momentum.  More specifically, we want to verify that we can
do this in a way that reduces the spectrum of a vortex
of vorticity $\v+8$ to that of one of vorticity $\v$.
 The  group theoretic fact  that we will use is the following.  In the group $SO(8)$ that
 acts on 8 gamma
matrices $\g_1,\dots,\g_8$, pick a $\mathrm{Spin}(7)$ subgroup, so that the $\g_i$ transform
in the spinor representation $\mathbf 8$ of $\mathrm{Spin}(7)$.
In this representation, there is an invariant fourth order antisymmetric tensor.  Take
$p_{ijkl}$ to be a multiple of this tensor.  The group $SO(8)$ has
two spinor representations, which we denote as $\mathbf{8}'$ and $\mathbf{8}''$.  One of
them, say the $\mathbf{8}''$, is irreducible under $\mathrm{Spin}(7)$
while the other decomposes as $\mathbf{1}\oplus \mathbf{7}$.  If the sign of the Hamiltonian
is chosen correctly, then the $\mathbf{1}$ of $\mathrm{Spin}(7)$
is its ground state and in particular is nondegenerate.  It remains to show that this can be
done in such as way that the ground state is an angular momentum
eigenstate with $\j'=0$.  For this, we first consider the case of a vortex of vorticity 8,
which has precisely 8 zero-modes of angular momenta
\be\label{eivals}\j'=\pm 7/2,\pm 5/2, \pm 3/2,\pm 1/2.\ee  The ground state of $\Delta H$
will certainly be $\j'$-invariant if $\j'$ is a generator of $\mathrm{Spin}(7)$.
For this, we take for $\j'$ a generator of $\mathrm{Spin}(7)$ that in the $\mathbf 7$ has
eigenvalues $0, \pm 1, \pm 2,\pm 4$.  This generator
has in the $\mathbf{8}$ of $\mathrm{Spin}(7)$ the desired eigenvalues of eqn.\
(\ref{eivals}).  More generally, we can in a similar fashion reduce
a vortex $V_+$ of vorticity $\v+8$ to a vortex $V_-$ of vorticity $\v$.  $V_+$ has 8 extra
zero-modes that have no counterpart in $V_-$; their
angular momenta are $\j'=\pm (\v+7)/2, \pm (\v+5/2), \pm (\v+3/2),\pm (\v+1/2)$.  To
construct a quartic Hamiltonian that will lift these modes in a $\j'$-invariant
way, we proceed as before, embedding $\j'$ in $\mathrm{Spin}(7)$ so that in the
representation $\mathbf 7$, it has eigenvalues $0,\pm 1, \pm 2, \pm (4+\v)$.

\section{Some Useful Facts About Chern-Simons Gauge Theories}\label{CSGT}

The purpose of this appendix is to review and clarify some known facts about
$2+1$-dimensional Chern-Simons gauge theories.  In the last subsection we will use this
information to exhibit an explicit Chern-Simons Lagrangian that describes the non-Abelian
statistics of the Read-Rezayi states \cite{ReadRezayi}, which generalize the Moore-Read
state \cite{MR}.  The analysis in that subsection is similar to the analysis in section
\ref{ChernSimons}.

\subsection{$U(1)$}\label{uone}

Consider  a $U(1)$ Chern-Simons theory with the Lagrangian
\be\label{UoneCS}{k \over 4\pi} a  \d a +{1\over 2\pi} A  \d a~,\ee
where $a$ is a $U(1)$ gauge field and $A$ is a classical background $U(1)$ field, which
couples to the current
\be\label{Uoc}j={1\over 2\pi}\d a~.\ee
In order not to clutter the notation, we will first take $k$ positive.  We denote this
theory as $U(1)_k$.  For even $k$ the theory is well defined on any manifold.  For odd $k$
we need the manifold to be spin and the answers depend on the choice of spin structure.
Also, as we discussed in section \ref{spincharge}, even for odd $k$ we can use a $\spinc$
classical gauge field $A$ to define the theory as a non-spin theory.

For $k$ even there are $k$ distinct Wilson lines $W_n=e^{\i n\oint a}$ labeled by
\be\label{nUone}n= 0, \pm 1\dots,\pm {k-2\over 2}, {k\over 2}~;\ee
i.e.\ the lines labeled by $n$ and by $n+k$ are identified.
The lines with $n$ and $-n$ differ by their orientation and the line with $k\over 2$ is
independent of orientation.
The line labeled by $n$ has spin $S$ (modulo one) and it induces holonomy $H$ around it,
where
\be\label{Uones}S={n^2\over 2k}\qquad , \qquad H=e^{2\pi in \over k} ~.\ee
This is consistent with the identification of the lines labeled by $n$ and by $n+k$.  Using
that and the expression for the current (\ref{Uoc}) we learn that $A$ couples to the charge
\be\label{Uonech}Q=-{n\over k}~.\ee
Note that because of the identification of the lines, only $Q \mod 1$ is meaningful.   If we
want to regard $Q$ as a real number, we cannot claim any identifications of lines.

Some of the $2+1$-dimensional Chern-Simons theories have a corresponding $2d$ rational
conformal field theory (RCFT) interpretation.  In the case of $U(1)_k$,
the relevant theory is that of a rational boson.  Its chiral algebra includes the current
$\partial\phi$ and is extended by the operator $e^{\i\sqrt k \phi}$ of dimension $k\over 2$.
This is indeed an integer for $k$ even.\footnote{What is usually called $k$ in  the RCFT
literature  is in our notation $k_{\mathrm{RCFT}}=k/2$.}

For $k$ odd, we can still use the expressions (\ref{Uones}) and (\ref{Uonech}) for the spin,
the holonomy, and the charge.  However, now we can no longer identify the lines labeled by
$n$ and by $n+k$.  These two lines induce the same holonomy, but their spins differ by
${1\over 2}\mod 1$.  For example, the line $E=e^{\i k \oint a}$ does not induce a nontrivial
holonomy, but its spin is half-integer.  Therefore, this line is nontrivial and cannot be
ignored; its expectation value will depend on the spin structure of the three-manifold.
Correspondingly, there are $2k$ inequivalent lines labeled by
\be\label{nUoneo}n= 0, \pm 1...,\pm (k-1), {k}~.\ee
Unlike the case of $k$ even, here $Q=-n/k$ is meaningful mod 2.

In the fractional quantum Hall application, the various Wilson lines represent the
world-lines of quasiparticles.  The special line $E$ represents the underlying electrons.
It has spin $1\over 2$ and charge $1$.  And since it does not induce any holonomy, it has
trivial braiding with all quasiparticles (only Fermi statistics).

Let us consider some examples:
\begin{description}
\item[$k=1$] has two lines, the trivial line with vanishing spin and $E$ with
    half-integer spin.  From the $2d$ RCFT perspective, this theory corresponds to the
    rational boson at the free fermion radius with the fermion $e^{i\phi}$ associated
    with the line $E$.
\item[$k=2$] has two lines with spins modulo one $S=0, {1\over 4}$. This theory $U(1)_2$
    is the same as $SU(2)_1$.
\item[$k=3$] has six lines with spins modulo one $S=0,{1\over 6},{1\over 6}, {1\over 2},
    {2\over 3}, {2\over 3}$.  From the $2d$ RCFT perspective this theory is the first
    member of the $\N=2$ supersymmetric minimal models.  The two supercharges are
    associated with $E$, whose $2d$ dimension is $3\over 2$.
\item[$k=4$] has four lines with spins $0,{1\over 8},{1\over 8}, {1\over 2}$.  This
    theory is similar to the $k=1$ theory.  It is a  free fermion theory, but now it
    includes also two complex conjugate representations of dimension $1\over 8$.  These
    are spin fields.
\end{description}

Finally, we would like to point out that the $U(1)_m$ theory can be viewed as a $\Z_2$
quotient of the $U(1)_{4m}$ theory.  Of course, as groups $U(1)/\Z_2 = U(1)$.  But since the
quotient changes the normalization of the gauge fields, the map is nontrivial.  We start
with $U(1)_{4m}$ with the Lagrangian ${4m\over 4\pi} ada$.  The $\Z_2$ quotient is
implemented by stating that $a$ is not a good gauge field, but $b=2a$ is.  In terms of it
${4m\over 4\pi} a\d a={m\over 4\pi} b\d b$, which describes $U(1)_m$.

This $\Z_2$ quotient can also be described as gauging a one-form global symmetry
\cite{KS,GKSW}.  It can be implemented by identifying a line in the original theory whose
braidings will be trivial in the quotient theory \cite{MSN,MST}.  In this case, the relevant
line is $e^{2\i m \oint a}$, with dimension
$m\over 2$.  Therefore, the quotient theory is a spin-TQFT for odd $m$ and a non-spin-TQFT
for even $m$.  Imposing that this line has trivial braidings projects out all the lines $
e^{\i n\oint a}$ with odd $n$.  When $m$ is even we should also identify the lines $ e^{\i
n\oint a} \sim e^{\i (n+2m) \oint a}$, whose spins differ by an integer \cite{MSN}.  When
$m$ is odd the spins of these two lines differ by half an integer and they should not be
identified.  As a check, the fact that the line with $n=1$ is projected out is equivalent to
the statement in the last paragraph that $a$ is not a good gauge field in the quotient
theory.

\subsection{$U(1)^n$}\label{uonen}

More generally, if the gauge group is $U(1)^n$, the theory is characterized by a $K$ matrix,
a $q$ vector, and a Chern-Simons contact term $k_c$.  In terms of them, the Lagrangian is
\be\label{KqLag}{1\over 4\pi} K_{ij} a^i   \d a^j + {1\over 2\pi} q_i a^i   \d A +{1\over
4\pi} k_c A  \d A~.\ee
All the coefficients in the Lagrangian must be integers. If all the diagonal elements of $K$
as well as $k_c$ are even, the theory does not need a spin structure.  If some of these are
odd, the theory is a spin Chern-Simons theory; it can be defined on a spin manifold and the
results depend on the spin structure.

In the special case where
\be\label{Kqcond}\begin{aligned}
&K_{ii}- q_i \in 2\Z  \qquad {\rm for\ all\ } i \\
&k_c \in 8\Z
\end{aligned}\ee
we can place the theory on an arbitrary manifold with a choice of a $\spinc$ structure,
letting $A$ be a $\spinc$ connection.  The first restriction guarantees that all the
monopole operators of $a$ satisfy the proper spin/charge relation, which is needed for the
$\spinc$ structure.  And the condition on $k_c$ is needed for the Chern-Simons contact term
for $2A$ to be well defined on an arbitrary $\spinc$ manifold.  See also the discussion in
section \ref{spincharge}.

The independent lines are $e^{\i n_i\oint a^i}$ whose spin and charge are $S={1\over 2}
K^{ij}n_in_j$ and $Q=-K^{ij}q_i n_j$, where $K^{ij}$ is the inverse matrix of $K_{ij}$.

\subsection{$\Z_N$}\label{zn}

Here we give a continuum description of $\Z_N$ Chern-Simons gauge theory \cite{MMS,BaS}.

We consider a special case of the theories in section \ref{uonen} based on the gauge group
$U(1)_a\times U(1)_c$ and the Lagrangian
\be\label{uouol}{N\over 2\pi} a  \d c +{k\over 4\pi} a  \d a.\ee
We can also couple the model to $A$.    The field $c$ acts as a Lagrange
multiplier constraining the holonomy $e^{\i\oint a}$ to be an $N^{\mathrm{th}}$ root of
unity.
Therefore, this theory is a $\Z_N$ gauge theory with $e^{i\oint a}$ the $\Z_N$ Wilson line.
The line $e^{\i\oint c}$
creates a holonomy for the $\Z_N$ Wilson line.  The term with $k$ is easily identified as
the Dijkgraaf-Witten
term \cite{DW} of the $\Z_N$ gauge theory.  Here it is given a continuum description in
terms of a $U(1)_a\times U(1)_c$ gauge theory.

For $k\in 2\Z$ the theory is not spin.  But for odd $k$ it depends on the spin structure.
Using $c\to c+a$ we can identify $k \sim k+2N$ and
hence the inequivalent non-spin theories are labeled by $k=0,2,\dots ,2N-2$.  And if we
allow dependence on the spin structure we can have $k=0,1,\dots,2N-1$.

The theory has line operators
\be\label{lineo}W_{n_a,n_c}(\gamma)=e^{\i n_a\oint_\gamma a + \i n_c\oint_\gamma c}   ~,\ee
with spins
\be\label{ZNspins}S=\left({n_an_c\over N}-{kn_c^2\over 2N^2}\right)\mod 1.\ee
Their correlation functions are
\be\label{linecor}\left\langle W_{n_a,n_c}(\gamma) W_{n_a',n_c'}(\gamma') \right\rangle \sim
\exp\left({2\pi i\over N^2}\#(\gamma,\gamma')(Nn_an_c'+Nn_cn_a'- k n_cn_c')\right)~,\ee
where $\#(\gamma,\gamma')$ is the linking between the two lines $\gamma$ and $\gamma'$.
From equations (\ref{ZNspins}) and (\ref{linecor}),  we see that the lines
$W_{n_a,n_c}$ and $W_{n_a+N,n_c}$ have the same correlation functions and the same spins.
They should be identified.  The lines $W_{n_a,n_c}$ and $ W_{n_a+k,n_c+N}$ have the same
correlation functions, but their spins satisfy $S(W_{n_a+k,n_c+N})= S(W_{n_a,n_c})-{k\over
2}$.  Therefore, for even $k$ they should be identified.
For odd $k$, the theory needs a spin structure.  For odd $k$, the line $E=W_{k,N}$ has
half-integer spin and  is nontrivial.

We conclude that for $k$ even
there are only $N^2$ lines and for $k$ odd there are $2N^2$ lines.  In other words
the lines satisfy
\be\label{linerel}\begin{aligned}
&W_{1,0}^N=1\\
&W_{1,0}^kW_{0,1}^N =E ~,\end{aligned}\ee
where for $k$ even we can set $E=1$ and for $k $ odd we have $E^2=1$.

\subsection{$U(2)$}\label{utwo}

An $SU(2)$ Chern-Simons theory has the Lagrangian
\be\label{SUtwoCS}{k \over 4\pi}\Tr\ (b  \d b + {2\over 3}b  b   b)~,\ee
where $b$ is an $SU(2)$ gauge field.  Here $k\in \Z$ and we take it to be positive.  The
Wilson lines are labeled by $j=0,{1\over 2},\dots,{k\over 2}$ and their spins are
$S={j(j+1)\over k+2}\mod 1$.

An $SU(2)\times U(1)$ Chern-Simons theory has the Lagrangian
\be\label{SUtwoUCS}{k_1 \over 4\pi} a  \d a+ {k_2 \over 4\pi}\Tr\ (b  \d b +{2\over 3} b  b
b)~. \ee
Here $k_{1,2}\in \Z$.  In order not to clutter the equations we take $k_2$ positive, but we
allow $k_1$ to be negative. For $k_1\in 2\Z$ the theory is non-spin and the lines are
labeled by $j=0,\dots,{k_2\over 2}$ and $n=0,\pm1,\dots, {|k_1|\over 2}$.  On a spin
manifold with a choice of spin structure (or by coupling to a $\spinc$ classical gauge field
$A$) we can also have odd $k_1$ and then $n=0,\pm1,\dots, {|k_1|}$.

The $U(2)= (SU(2)\times U(1))/\Z_2$ theory is characterized by two levels $k_{1,2}$ and is
denoted by $U(2)_{k_1,k_2}$.  One way to describe it is to combine $a$ and $b$ in eqn.\
(\ref{SUtwoUCS}) to a conventionally normalized $U(2)$ gauge field $c=b+a{\bf 1}$ with the
Lagrangian
\be\label{UtwoCSm}{k_2 \over 4\pi}\Tr\ (c  \d c +{2\over 3} c  c   c)+ {k_1-2k_2\over4\cdot
4\pi}(\Tr\ c)   \d(\Tr\ c) ~.\ee
As a three-dimensional gauge theory it differs from the $SU(2)\times U(1)$ theory of
equation (\ref{SUtwoUCS}) by having additional bundles.  Therefore, there might be
additional conditions on $k_{1,2}$.  One way to find them follows from focusing on a
specific $U(1)\subset U(2)$ gauge field of the form $\begin{pmatrix} c_{11} & 0 \\ 0& 0
\end{pmatrix}$, whose ``effective Lagrangian'' is ${2k_2 + k_1 \over 4\cdot 4\pi}c_{11} \d
c_{11}$.  Therefore, the theory is consistent only when
 \be\label{rela}2k_2 +k_1  \in 4 \Z \ee
and it depends on the spin structure unless $2k_2+k_1 \in 8\Z$.  In particular, $k_1$ must
be even. It can be shown by adapting
to $U(2)$ the arguments given for $U(1)$ in Appendix \ref{norm} that there are no further
conditions.

Let us understand this in more detail.   We start with the $SU(2)_{k_2}\times U(1)_{k_1}$
theory and perform the quotient.  From the
three-dimensional point of view this means that we gauge a $\Z_2$ one-form global symmetry
\cite{KS,GKSW}. This is achieved by summing over additional bundles that are not
$SU(2)\times U(1)$ bundles.  One way to think about it to start with the original
$SU(2)\times U(1)$ theory and to identify in it a Wilson line that induces a nontrivial
holonomy in $SU(2)\times U(1)$, which is trivial in $U(2)$.  This holonomy should be $-1$ in
each of the factors.  Then we should impose that this line has trivial correlation
functions.  In the $SU(2)$ factor this line should have $j={k_2\over 2}$.  The holonomy
around it is in the center $\Z_2 \subset SU(2)$.  Looking at equation (\ref{Uones}) it is
clear that from the $U(1)$ factor we need $n={|k_1|\over 2}$ and hence $k_1$ should be even.
The spin of the full line is
\be\label{konetwo}{{k_2\over 2}({k_2\over 2} +1)\over k_2+2} + {k_1\over 8}= {2 k_2+
k_1\over 8} ~. \ee
When this spin is an integer, i.e.\ when $2k_2+k_1=0\mod 8$, the quotient makes sense on
every manifold.  Alternatively, we can have $2k_2+k_1=4\mod 8$ and then the quotient makes
sense only on a spin manifold.

In general, when we perform such a quotient the spectrum of lines is modified by three rules
\cite{MSN,MST}.  First, of the original lines labeled by $(j,n)$ we should keep only those
that have trivial braiding with the special line $(j={k_2\over 2}, n={|k_1|\over 2})$.  This
leads to a selection rule
\be\label{Utwoselection}j +{n\over 2} \in \Z~.\ee
Indeed, this is the condition that $(j,n)$ is a $U(2)$ representation rather than a faithful
$SU(2)\times U(1)$ representation.  Second, we should identify some lines.  It is easy to
see that
\be\label{Utwoiden}(j,n)\qquad {\rm and}\qquad   \left({k_2\over 2}-j, ({k_1\over 2}+n )\mod
|k_1|\right)\ee
induce the same holonomy.  Their spins differ by
\be\label{spind} \left({({k_2\over 2}-j)({k_2\over 2}-j+1)\over k_2+2} + {\left({k_1\over 2}
+n\right)^2\over 2k_1} \right)-\left({j(j+1)\over k_2+2} + {n^2\over 2k_1}\right)= {k_1+2k_2
\over 8} -(j-{n\over 2}) ~.
\ee
For $k_1 +2k_2 = 0\mod 8$, when the theory is not a spin theory, we use the selection rule
(\ref{Utwoselection}) to see that these two representations have the same spin modulo one
and therefore they should be identified.  For $k_1 +2k_2 = 4\mod 8$, when the theory is a
spin theory these two representations induce the same holonomy, but their spins differ by
half an integer and therefore they should not be identified.

The third rule of a quotient \cite{MSN,MST} applies to lines that are fixed points of the
identification. Such lines should be treated more carefully.  But it is easy to see from
(\ref{Utwoiden}) that no such fixed point exists in this case.

Let us consider the special case $k_2=k$, $k_1=-2k$.  Since $2k_2 + k_1=0$, this is not a
spin theory. Since $k_1$ is negative, this particular Chern-Simons
theory can be coupled on the boundary of a three-manifold to a two-dimensional field theory
that has both left movers realizing $SU(2)_k$ and right movers realizing $U(1)_{-2k}$.  But
in this particular case, there is also another option for the boundary theory
\cite{MSN,MST}.  Since $U(1)_{2k}$ is conformally embedded in $SU(2)_k$, this theory
describes also the GKO coset $SU(2)_k \over U(1)_{2k}$ \cite{GKO}.  This theory is known as
the parafermion theory.  (Recall that our normalization of $k_1$ differs by a factor of $2$
from the standard RCFT normalization.) The special case of $k=2$ corresponds to the Ising
model.  Here the possible Wilson lines are labeled by $j=0,{1\over 2},1$ and $n=0,\pm 1,2$.
The selection rule (\ref{Utwoselection}) and the identification (\ref{Utwoiden}) show that
the distinct lines are the trivial line with $(j=0,n=0)$, $W_\sigma$ with $(j={1\over
2},n=1)$, and $W_\psi$ with $(j=1,n=0)$. They correspond to the identity operator (spin
$0$), the spin field (spin $1\over 16$), and the free fermion (spin $1\over 2$).  Note that
even though this is a theory of $1+1$-dimensional fermions, the corresponding
$2+1$-dimensional Chern-Simons theory is not a spin theory.  In the language of section
\ref{ising}, this theory is the Ising TQFT, not the Ising spin-TQFT.

We can quotient the $U(2)$ theory again by $\Z_2$ and turn the gauge group into $U(2)/\Z_2 =
SO(3)\times U(1)$.  This is possible only when $k_2$ is even and $k_1=0\mod4$.  One way to
see that is based on identifying the special lines in the $SU(2)_{k_2}\times U(1)_{k_1}$
theory that have trivial braiding in the quotient.  They are $(j={k_2\over2}, n=0)$, whose
spin is ${k_2\over 4}$ and $(j=0, n={|k_1|\over 2})$, whose spin is $|k_1|\over8$.  The
product of these lines is the special line in the quotient leading to $U(2)$ and should not
be considered separately.  The  theory that results from the additional quotient  is
$SO(3)_{k_2} \times U(1)_{k_1/4}$.  It is non-spin when $k_2=0\mod 4$ or $k_1=0\mod 8$.
Otherwise the theory is spin.  Note that when the two factors of the gauge group are both
spin, i.e.\ when $k_2=2\mod 4$ or $k_1=4\mod 8$, we do not need to double the number of
lines twice.

  Above we described the Ising TQFT as the non-spin Chern-Simons theory $U(2)_{2,-4}$.  Its
  $\Z_2$ quotient is the spin-TQFT $SO(3)_2\times U(1)_{-1}$.  This theory
  is the Ising spin-TQFT, in the language of section \ref{ising}.  It has only one
  nontrivial line, $W_\psi$, with spin  ${1\over 2}\mod 1$.  The line in the $U(2)_{2,-4}$
  theory that is associated with the spin field $W_\sigma$ is projected out.

\subsection{Lifting a Spin-TQFT to a Non-Spin-TQFT}\label{LSTQFT}

In Appendix \ref{utwo}, we discussed examples of a $\Z_2$ quotient of a non-spin-TQFT
leading to a spin-TQFT.  Here we discuss the opposite process.  Starting with a spin-TQFT we
try to lift it to a non-spin-TQFT.  Unlike the $\Z_2$ quotient, this process is not unique.
(For a related mathematical discussion see e.g.\ \cite{DNO}.)

The partition function $Z_s$ of a spin-TQFT depends on the spin structure $s$.  One way to
turn it into a non-spin-TQFT is by summing over $s$.  When we do that we have freedom in
inserting $s$ dependent coefficients
\be\label{totalZ} {\cal Z} =\sum _s z_s Z_s ~.\ee
In order not to add degrees of freedom, we take $z_s$ to be the partition function of an
almost trivial spin-TQFT, sometimes called an invertible TQFT.  This means that for a closed
manifold $z_s$ is a phase and the theory has only two line operators, the identity operator
with vanishing spin and a fermion line with spin $1\over 2$.

Examples of such theories are $SO(n)$ Chern-Simons theories with level one, denoted
$SO(n)_1$.  These theories have $c={n\over 2}$ and therefore, up to an overall
$s$-independent phase, we can identify\footnote{The 2d version of this statement is that 16
chiral fermions, upon summing over spin structures, give
a modular invariant theory.} the theory with $n$ and the theory with $n+16$.  We can also
extend this to negative $n$ by $SO(-n)_1 \equiv SO(n)_{-1}$.  Special cases are the
following.\begin{description}
\item[$n=1$] is the Ising spin-TQFT.
\item[$n=2$] is usually denoted by $U(1)_1$.
\item[$n=3$] is usually denoted by $ SO(3)_2=SU(2)_2/\Z_2$.  Denoting it as $SO(3)_2$ does
    not fit our general expressions, where it should better be denoted by $SO(3)_1$.
    Therefore, below we will denote it by $ SU(2)_2/\Z_2$.
    \end{description}

This leads to $16$ distinct ways, labeled by $n$, of lifting a given spin-TQFT to a
non-spin-TQFT
\be\label{sixteenT}{\cal Z}^n(\mbox{spin-TQFT})=\sum_s
Z_s(SO(n)_1)Z_s(\mbox{spin-TQFT})~.\ee

We will also be interested in the non-spin-TQFT of $\mathrm{Spin}(l)_1$.  Like the spin
theory based on\footnote{Note the confusing terminology.  The TQFT based on
$\mathrm{Spin}(l)_1$ is non-spin, while the TQFT based on $SO(l)_1$ is a spin-TQFT.}
$SO(l)_1$, it also has $c={l\over 2}$.  For $l$ odd it has three lines: the identity, the
fermion line, and a spin field line with spin ${l\over 16}$.  For $l$ even it has four
lines: the identity, the fermion line, and two spin field lines with spin ${l\over 16}$.
Special cases are the following.\begin{description}
\item[$l=1$] is the non-spin Ising TQFT.
\item[$l=2$] is usually denoted by $U(1)_4$.
\item[$l=3$] is usually denoted by $SU(2)_2$.
\end{description}

Let us lift the spin-TQFT of $SO(m)_1$ to a non-spin TQFT.  Tensoring it with $SO(n)_1$ and
summing over the spin structures the partition function is
\be\label{liftsom}{\cal Z}^n(SO(m)_1) = \sum_s Z_s(SO(n)_1)Z_s(SO(m)_1)=\sum_s
Z_s(SO(n+m)_1) = {\cal Z}(\mathrm{Spin}(n+m)_1) ~.\ee
The resulting theory is $\mathrm{Spin}(n+m)_1$.  The simple case
$n=0$ lifts $SO(m)_1$ to the non-spin theory $\mathrm{Spin}(n)_1$.

A more interesting case corresponds to $n=-2$.  It lifts $SO(m)_1$ to
$\mathrm{Spin}(m-2)_1$.  For $m=3$ this lifts $SU(2)_2/\Z_2$ to the non-spin Ising theory.
We can also write this as
\be\label{liftIsing}{SU(2)_2\over \Z_2} \longrightarrow {SU(2)_2\over \Z_2} \times U(1)_{-1}
\longrightarrow {SU(2)_2\times U(1)_{-4} \over \Z_2}~,\ee
where the first term represents $Z_s$, the second is $z_sZ_s$ and the third is the result of
the sum over $s$.  In the third step the numerator is the result of independently summing
each factor over $s$ and the $\Z_2$ quotient guarantees that $s$ in the two sectors is the
same.  The last expression is the Chern-Simons description of the RCFT coset $SU(2)_2\over
U(1)_4$ that describes the non-spin Ising theory.  As in Appendix \ref{utwo}, it can also be
written as $U(2)_{2,-4}$.

\subsection{Chern-Simons Gauge Theories for Non-Abelian FQHE}\label{CSFQHE}

Here we use the analysis in the previous subsections to find a Chern-Simons description of
the series of models of non-Abelian statistics of Read and Rezayi  \cite{ReadRezayi}, which
generalize the famous Moore-Read state \cite{MR}.

The Read-Rezayi construction starts with the chiral algebra
\be\label{Ntcog}{{\rm Parafermions}_k \times U(1)_{kL} \over \Z_k}
= {{SU(2)_k \over U(1)_{2k}}\times  U(1)_{kL} \over \Z_k} ~.\ee
The first factor, the parafermion theory, was analyzed above.  It is not a spin theory.  The
level of the $U(1)$ factor is taken to be a multiple of $k$ so that we can mod out by
$\Z_k$.  More explicitly, this guarantees that the $U(1)$ factor has a global $\Z_k$
one-form symmetry, which can be gauged.  The quotient by $\Z_k$ is achieved by ensuring that
the line $(j=0, n=2, l=L)$ has trivial braiding. (Here the labels of the line are the
$SU(2)_k\times U(1)_{-2k}\times U(1)_{kL}$ representations.) Its spin is $S={L-2\over 2k}
\mod 1$ and therefore we need
\be\label{valueL}L=kM +2  \qquad, \qquad M\in \Z    ~.\ee
For $M$ odd it is a fermion and the theory is spin and for $M$ even it is a boson.

Let us describe the $2+1$-dimensional Chern-Simons theory.  The parafermion theory is
described by a $U(2)_{k,-2k}$ theory (\ref{UtwoCSm}) with gauge field $c$ and the
$U(1)_{k(kM +2)}$ theory by a gauge field $b$.  The Lagrangian of the $U(2)_{k,-2k} \times
U(1)_{k(kM +2)}$ theory is
\be\label{UtwoCSmp}{k \over 4\pi}\Tr\ (c  \d c +{2\over 3} c  c   c)-{k\over 4\pi}(\Tr\ c)
\d(\Tr\ c) + {k(kM +2)\over 4\pi} b  \d b + {k \over 2\pi} A  \d b~.\ee
As we will soon see, the coupling to $A$ is such that we can mod out by $\Z_k$.

Next, we should mod out by $\Z_k$.  One way to do that is to include additional bundles.  An
easier way to do that is to find an appropriate change of variables to conventionally
normalized $U(2)\times U(1)$ gauge fields.  We change variables from $c$ and $b$, which are
no longer good gauge fields, to
\be\label{bcchage}\begin{aligned} \tilde c=c- b{\bf 1} \\ \tilde b = kb~.\end{aligned}\ee
These combinations were chosen such that the $\Z_k$ one-form global symmetry does not act on
them.  Equivalently, all the bundles of the original $U(2)\times U(1)/\Z_k$ theory
(\ref{UtwoCSmp}) can be described as ordinary $U(2)\times U(1)$ bundles with the gauge
fields $\tilde c$ and $\tilde b$.
In terms of these variables the Lagrangian (\ref{UtwoCSmp}) is
\be\label{UtwoCSmpt}{k \over 4\pi}\Tr\ (\tilde c  \d\tilde c +{2\over 3} \tilde c  \tilde c
\tilde c)-{k\over 4\pi}(\Tr\ \tilde c)   \d(\Tr\ \tilde c) -{1\over 2\pi} (\Tr\ \tilde c)
\d\tilde b + {M\over 4\pi} \tilde b  \d\tilde b + {1 \over 2\pi} A  \d\tilde b~.\ee
The fact that the coefficients here are properly normalized is an independent derivation of
the condition (\ref{valueL}).

The equations of motion of the Abelian factors are
\be\label{abeom}\begin{aligned}
k\d(\Tr\ \tilde c) + 2\d\tilde b &= 0\\
-\d(\Tr\ \tilde c )+ M\d\tilde b +\d A &= 0 \end{aligned}\ee
and therefore, the response to a background $A$ is
\be\label{jgen}j={\delta S \over \delta A} = {1\over 2\pi} d\tilde b = -{k\over 2\pi L} dA
~.\ee
This allows us to identify the filling factor
\be\label{nug}\nu={k\over L}={k\over kM+2}~.\ee

Various models are included in this series:
\begin{description}
\item[$k=1$] Here the parafermions theory is trivial and we have $U(1)_{M+2}$.  For
    $M=2q-1$ this is the standard Abelian FQHE. The charges are multiples of $1\over
    2q+1$ and $\nu={1\over 2q+1}$.
\item[$k=2$] Here the parafermion theory is a Majorana-Weyl fermion (the Ising model)
    and the chiral algebra is $\left({SU(2)_2 \over U(1)_{4}}\times  U(1)_{4M+4}
    \right)/\Z_2$.  Here $\nu={1\over M+1}$.  For $M=1$ this is the Moore-Read state
    \cite{MR}.  The models with higher values of $M$ were analyzed by \cite{Fendley}.
\item[$M=1$] Here we find the $\N=2$ supersymmetric minimal models \cite{ReadRezayi},
    where the charges are multiples of $1\over k+2$ and $\nu={k\over k+2}$.  In this
    case we can shift $\tilde b\to \tilde b +\Tr\ \tilde c$ and write eqn.\
    (\ref{UtwoCSmpt}) as \be\label{LagCStcs}{k \over 4\pi} \Tr\ (\tilde c   \d\tilde c
    +{2\over 3} \tilde c  \tilde c  \tilde c ) - {k+1\over 4\pi}(\Tr\ \tilde c )
    \d(\Tr\ \tilde c)+ {1 \over 4\pi} (\tilde b +A)   \d(\tilde b +A) +{1 \over 2\pi } A
    \d(\Tr\ \tilde c) - {1\over 4\pi} A  \d A~.\ee                                 In
    this form the gauge group is $U(2)_{k,-2(k+2)}\times U(1)_1$ and the $U(1)_1$ part
    decouples.  Note that we did not shift $\tilde b$ by $A$ in order to preserve the
    spin/charge relation.  An interesting aspect of the theory (\ref{LagCStcs}) is the
    role of the decoupled $U(1)_1$ factor.  The theory without it is still spin.  So its
    presence does not add additional lines; the theory without it has all the necessary
    lines.  However, keeping the $U(1)_1$ factor, the theory has a purely left moving
    chiral algebra ($\N=2$ supersymmetry).  But if we remove the $U(1)_1$ factor, the
    corresponding two-dimensional theory needs both left-moving $SU(2)_k$ and
    right-moving $U(1)_{-2(k+2)}$ modes.  Finally, in the form (\ref{LagCStcs}) the
    theory is $U(2)_{k,-2(k+2)}$ and it is straightforward use the expressions in
    section \ref{utwo} to work out the spins and charges of the quasiparticles.
\item[$k=2$, $M=-3$]  Here eqn.\ (\ref{Ntcog}) becomes $\left({\rm Ising} \times
    U(1)_{-8}\right)/\Z_2$.  This is the topological field theory of the T-Pfaffian
    state \cite{TPfaffian,TPfaffiantwo}, which is expected to play a role both in
    topological insulators and topological superconductors
    \cite{Fidkowski,MetlitskiTSC}, and which
we have discussed in section \ref{app}.  Substituting these values in eqn.\
(\ref{UtwoCSmpt}) we find the Lagrangian
        \be\label{UtwoTpf}{2 \over 4\pi}\Tr\ (\tilde c  \d\tilde c +{2\over 3} \tilde c
        \tilde c   \tilde c)-{2\over 4\pi}(\Tr\ \tilde c)   \d(\Tr\ \tilde c) -{1\over
        2\pi} (\Tr\ \tilde c)  \d\tilde b - {3\over 4\pi} \tilde b  \d\tilde b + {1
        \over 2\pi} A  \d\tilde b~.\ee
\end{description}

\end{document}